\documentclass[a4paper,fleqn,usenatbib]{mnras}
\pdfoutput=1

\usepackage{mathptmx}

\usepackage[T1]{fontenc}
\usepackage{ae,aecompl}

\usepackage{graphicx}	
\usepackage{amsmath}	
\usepackage{amssymb}	
\newcommand{\angstrom}{\textup{\AA}} 
\usepackage{hyperref}

\newcommand{\lya}{Ly$\alpha$}
\newcommand{\unitcgssb}{erg s$^{-1}$ cm$^{-2}$ arcsec$^{-2}$}
\newcommand{\unitfluxdcgs}{erg s$^{-1}$ cm$^{-2}$ \AA$^{-1}$}
\newcommand{\civ}{\ion{C}{iv}\ }
\newcommand{\heii}{\ion{He}{ii}\ }
\newcommand{\siiv}{\ion{Si}{iv}\ }
\newcommand{\nv}{\ion{N}{v}\ }
\newcommand{\ciii}{\ion{C}{iii}]}
\newcommand{\cii}{\ion{C}{ii}]}

\def \qso {SDSS~J1020$+$1040}

\def \cgssb {{\rm\,erg\,s^{-1}\,cm^{-2}\,arcsec^{-2}}}

\title[Inspiraling Halo Accretion around a $z\sim3$ Quasar.]{Inspiraling Halo Accretion Mapped in Lyman-$\alpha$ Emission around a $z\sim3$ Quasar}

\author[F. Arrigoni Battaia et al.]{
Fabrizio Arrigoni Battaia$^{1}$\thanks{E-mail: farrigon@eso.org}, 
J. Xavier Prochaska$^{2,3}$, Joseph F. Hennawi$^{4,5}$, \newauthor
Aura Obreja$^{6,7}$, Tobias Buck$^{4}$,
Sebastiano Cantalupo$^{8}$, Aaron A. Dutton$^{7}$, \newauthor 
and Andrea V. Macci\`{o}$^{7,4}$
\\
$^{1}$European Southern Observatory, Karl-Schwarzschild-Str. 2, D-85748 Garching bei M\"unchen, Germany\\
$^{2}$Department of Astronomy and Astrophysics, University of California, 1156 High Street, Santa Cruz, California 95064, USA\\
$^{3}$University of California Observatories, Lick Observatory, 1156 High Street, Santa Cruz, California 95064, USA\\
$^{4}$Max-Planck-Institut f\"ur Astronomie, K\"onigstuhl 17, D-69117 Heidelberg, Germany\\
$^{5}$Department of Physics, Broida Hall, University of California, Santa Barbara, CA 93106-9530, USA\\
$^{6}$University Observatory Munich, Scheinerstr. 1, D-81679 Munich, Germany\\
$^{7}$New York University Abu Dhabi, PO Box 12988, Saadiyat Island, Abu Dhabi, United Arab Emirates\\
$^{8}$Institute for Astronomy, Department of Physics, ETH Zurich, CH-8093 Zurich, Switzerland
}

\date{Accepted XXX. Received YYY; in original form ZZZ}

\pubyear{2017}

\begin{document}
\label{firstpage}
\pagerange{\pageref{firstpage}--\pageref{lastpage}}
\maketitle

\begin{abstract}
In an effort to search for \lya\ emission from circum- and intergalactic gas on scales of hundreds of kpc around $z\sim3$ quasar, and 
thus characterise the physical properties of the gas in emission, we have initiated an extensive fast-survey with the Multi Unit Spectroscopic Explorer (MUSE): 
{\it Q}uasar {\it S}napshot {\it O}bservations with {\it MU}se: 
{\it S}earch for {\it E}xtended {\it U}ltraviolet e{\it M}ission (QSO MUSEUM). 
In this work, we report the discovery of an enormous \lya\ nebula (ELAN) around the quasar SDSS~J102009.99+104002.7 at $z=3.164$, 
which we followed-up with deeper MUSE observations. 
This ELAN spans $\sim297$~projected kpc, has an average \lya\ surface brightness ${\rm SB}_{\rm Ly\alpha}\sim
6.04\times10^{-18}$\unitcgssb (within the $2\sigma$ isophote), and 
is associated with an additional four, previously unknown embedded sources: 
two \lya\ emitters and two faint active galactic nuclei (one Type-1 and one Type-2 quasar).
By mapping at high significance the line-of-sight velocity in the entirety of the observed structure, 
we unveiled a large-scale coherent rotation-like pattern spanning $\sim300$~km~s$^{-1}$ with a velocity 
dispersion of $<270$~km~s$^{-1}$,  
which we interpret as a signature of the inspiraling accretion of substructures within the quasar's host halo.
Future multiwavelength data will complement our MUSE observations, and are definitely needed to fully characterise such a complex system.
None the less, our observations reveal the potential of new sensitive 
integral-field spectrographs to characterise the dynamical state of diffuse gas on 
large scales in the young Universe, and thereby witness the assembly of galaxies.
\end{abstract}

\begin{keywords}
quasars: general, quasars: emission lines, galaxies: high-redshift, (galaxies):intergalactic medium, cosmology: observations, galaxies:haloes
\end{keywords}



\section{Introduction}
\label{sec:intro}

It is predicted that the universe's initial conditions have zero net angular momentum.
Yet spin is a fundamental property of galaxies, especially in disky spirals like our Milky Way.
The accepted model of galaxy formation explains the required angular momentum build-up as follows.
During the Universe's lifespan, baryons collapse into the potential well of dark matter (DM) haloes from the ``cosmic web'', i.e. 
the diffuse intergalactic medium (IGM) tracing the large-scale structure in the universe. 
In this process, the gas is shock-heated to the virial temperature of the DM haloes, 
and subsequently cools down and settles in galaxies, where it is partially transformed into stars (\citealt{WR1978}).
In the initial phases of the collapse, the gravitational forces exerted between neighbouring DM haloes 
produce torques which generate a net angular momentum within these systems (\citealt{hoyle51}).
In an expanding universe (\citealt{hubble29}), this initial phase of angular momentum build-up ends when 
the DM haloes become bound structures in themselves, sufficiently far away from their neighbours such 
that the large scale gravitational torques stop being the dominant evolutionary force.
As a result, the angular momentum acquired at earlier epochs forces the baryons 
to assemble in rotating structures inside haloes (\citealt{FE1980}). 
Following this theory, analytic calculations predicted spins for modern galaxies 
(\citealt{peebles69}), which have been 
confirmed by numerical simulations within our modern
cosmological paradigm (e.g. \citealt{bullock01,porciani02}).

In the past decade, an important element has been added to this picture.
Hydrodynamic simulations of galaxy formation have started to show
the accretion of ``streams'' of cool gas into
dark matter haloes (\citealt{kkw+05}).  These streams are predicted
to fuel star-formation within the central galaxy (\citealt{db06}), 
funnel dwarf galaxies into the surrounding dark matter halo, and 
provide a reservoir of cool halo gas (\citealt{fg11,fumagalli11a}).
Recent numerical work has found that the baryons within these streams
exhibit a significantly higher angular momentum than the matter
within the inner regions of the dark matter haloes.  This suggests that
streams contribute to the net angular momentum of the system
(\citealt{stewart11,danovich+15}).
These same models predict that large regions of the halo may contain
inspiraling gas or structures with a projected velocity shear qualitatively 
similar to rotation (\citealt{stewart+16}).  
Therefore, both these channels of accretion available to baryons would lead to the 
presence of cool gaseous structures within galaxy haloes  
carrying a net angular momentum.

Unfortunately, the halo gas, i.e. 
the so-called circumgalactic medium (CGM), 
is expected to have very low densities, and thus to be typically too faint to detect 
directly.  Observations using the light from bright background 
sources to probe haloes along individual sightlines have revealed a high 
covering fraction of cool gas around high-$z$ galaxies (\citealt{rudie12,QPQ5,qpq6, QPQ7}).
This cool CGM is manifest around galaxies
of essentially all luminosity and mass, and across all of cosmic time
(e.g., \citealt{pwc+11,tumlinson+13}).
The absorption-line experiment, however, cannot resolve the `morphology' 
of the CGM, and given that it is an inherently one-dimensional probe, 
it does not uniquely constrain kinematics relevant to
the angular momentum of the system.  The past few years, however,
have witnessed the discovery of enormous \lya\ nebulae (ELANe) --
cool, emitting hydrogen gas that extends hundreds of kpc around 
$z>2$ quasars (\citealt{cantalupo14,hennawi+15,Cai2016}).  
On these scales, the \lya\ emission
probes gas throughout the dark matter halo and even beyond into the surrounding IGM. 
Furthermore, the emission is sufficiently bright (\lya\ surface brightness ${\rm SB}_{\rm Ly\alpha}\sim10^{-17} \cgssb$ at 100~kpc from the AGN) 
to measure line-of-sight velocities throughout the ELAN.  In turn, one may search for
signatures of inspiraling gas predicted by models of galaxy formation.

However, ELANe seems to be extremely rare, and one would need a statistical sample to 
find such bright and thus ``easy-to-observe'' systems. Indeed, 
the reminder of the studies 
in the literature show (i) fainter \lya\ emission at these large projected distances (100~kpc), 
${\rm SB}_{\rm Ly\alpha}\sim 10^{-18}$\unitcgssb\ (\citealt{Borisova2016, Fumagalli2016}), (ii) detections only 
on smaller scales ($R<50$~kpc) in 50-70\% 
of the cases (\citealt{HuCowie1987, heckman91a, heckman91b, Christensen2006, North2012, qpq4, Roche2014, fab+16}), or (iii) non-detections 
(e.g., \citealt{Herenz2015, fab+16}). 
By conducting a stacking analysis of the narrow-band data targeting the \lya\ emission of 15 $z\sim2$ quasars, 
\citet{fab+16} shows that the average \lya\ profile for 
typical quasars at this redshift should be very low (${\rm SB}_{\rm Ly\alpha}\sim 10^{-19}$\unitcgssb\ at $\approx$100~kpc) 
and thus quite difficult to be detected routinely around each 
object. On the other hand, using MUSE, \citet{Borisova2016} show, on average, higher azimuthal 
\lya\ profiles around $z\sim3$ quasars, but still lower than the observed profiles for the ELANe.
Thus, ELANe are indeed the best target to map line-of-sight velocities at high significance. Nevertheless, 
given their recent discovery and the low-number statistic the ELAN phenomenon is still poorly constrained.

In this study, we report the discovery of an additional ELAN: an enormous (maximum projected distance of 297~kpc) bright 
(${\rm SB}_{\rm Ly\alpha}\sim 6.04\times10^{-18}$\unitcgssb, average within the $2\sigma$ isophote) \lya\ 
nebulosity around the quasar SDSS~J102009.99+104002.7 
at $z=3.164\pm0.006$ (for details on the redshift determination see Appendix~\ref{appZ}).
This new discovery together with the other ELANe reported so far (\citealt{cantalupo14, hennawi+15, Cai2016}) 
seems to hint to a scenario in which such bright extended \lya\ emission around quasars is linked to 
dense environments or to the presence of companions (e.g., \citealt{hennawi+15}).
Further, the high flux of the 
\lya\ emission from the large-scale structure 
together with the unprecedented capabilities of MUSE/VLT, 
allows us to map the velocity field in 
the entirety of such a large-scale nebula, revealing a rotation-like pattern.

This work is structured as follows. In \ref{sec:obs}, we describe our observations and data reduction.
In \ref{sec:results}, we present the observational results. In particular, 
we show the quasar's companions, the maps (surface brightness, velocity, and sigma) for the \lya\ emission, 
and the constraints on the extended emission in the \civ and \heii lines.
In \ref{sec:disc} we discuss our observations in light of current models for galaxy formation and 
predictions from different powering mechanisms. Finally, \ref{sec:summ} summarises our conclusions.  

Throughout this paper, we adopt the cosmological parameters $H_0=70$~km~s$^{-1}$~Mpc$^{-1}$, $\Omega_M =0.3$ 
and $\Omega_{\Lambda}=0.7$. In this cosmology, 1\arcsec\ corresponds to about 7.6 physical kpc at $z=3.164$.
All magnitudes are in the AB system (\citealt{Oke1974}), and all distances are proper, unless specified.

\section{Observations and Data Reduction}
\label{sec:obs}

We targeted the quasar SDSS~J102009.99+104002.7 (henceforth \qso) during the program 094.A-0585(A) with 
the Multi Unit Spectroscopic Explorer (MUSE; \citealt{Bacon2010}) on the 
VLT 8.2m telescope YEPUN (UT4), as part of the survey QSO MUSEUM: {\it Q}uasar {\it S}napshot 
{\it O}bservations with {\it MU}se: {\it S}earch for {\it E}xtended {\it U}ltraviolet e{\it M}ission (Arrigoni Battaia et al. in prep.). 
This ``snapshot survey'' has been designed to target the population of $z\sim3$ quasars with fast observations 
of 45 minutes on source, with the aim of (i) uncovering additional ELANe, similar to \citet{cantalupo14} and \citet{hennawi+15},  
(ii)  conducting a statistical census to determine the frequency of the ELAN phenomenon, (iii) studying the size, luminosity, 
covering factor of the extended \lya\ emission, and any relationship with the quasar luminosity or radio activity, 
(iv) looking for any evolutionary trend by comparing this sample with the $z\sim2$ quasar population (e.g., \citealt{fab+16}).
At the moment of writing, this survey consists of $48$ radio-quiet quasars and $11$ radio-loud objects, with $i$-band magnitude 
in the range $17.4 < i < 19.0$ (median 18.02), and the  redshift spanning $3.03 < z < 3.46$ (median 3.17). This survey thus expands on
the work by \citet{Borisova2016}, both in number of targeted sources and by encompassing fainter sources. 
A preliminary analysis of this sample shows that ELANe (i.e. showing a surface brightness of $10^{-17}\, \cgssb$ at 100 projected kpc from the quasar) 
have been discovered in only 2\% of the sample in agreement with previous statistics reporting $\lesssim 10\%$ at $z\sim2$ (\citealt{hennawi+15}).

The observations of \qso\ were carried out in service-mode during UT 18-19 of February
2015 and consisted of 3 exposures of 900~s each, with the 
exposures rotated with respect to each other by 90 degrees. 
These observations resulted in the discovery of
a bright ELAN around the quasar \qso, 
with a maximum extent of $199.3$~kpc (or $26.2\arcsec$), 
determined from the 2$\sigma$ isophote. 
After this initial discovery in the ``fast-survey'' data, 
we followed up this target with MUSE for an 
additional 4.5 hours on source on UT 8-9 December 2015 and 10-11 January 2016, as part of the service program 096.A-0937(A). 
The exposure time was split into 18$\times$900~s exposures, with 90 degree rotations and a small dither between each exposure. 
The average seeing for these observations was $0.66\arcsec$ (FWHM of the Gaussian at 5000~\AA, 
measured from the combined 4.5 hours datacube), and
ranged between $0.59\arcsec-0.82\arcsec$ in
different observation blocks (OBs).
In this follow-up program 096.A-0937(A), on which we focus in the reminder of this work, we offset 
the pointing centre so that the quasar moved
about $15\arcsec$ to the East 
with respect to program 094.A-0585(A) 
(where the quasar was centred in the MUSE field of view), so that
our observations would cover two bright radio sources located $\sim30\arcsec$ away from the quasar. As the 
nature of these radio sources was unknown, our aim 
was to determine if they were linked to the system studied here. The spectral-imaging capabilities of MUSE, together with the radio data from the literature, allowed us to
assess that these sources are actually the
radio lobes of a previously unidentified radio galaxy at $z=1.536\pm0.001$, and are thus not physically related to the system considered  here
(see appendix~\ref{appB}).

The data were reduced using the MUSE pipeline recipes 
v1.4 (\citealt{Weilbacher2014}). In particular, each of the individual
exposures has been bias-subtracted, flat-fielded, twilight and illumination corrected, and wavelength
calibrated. The flux calibration of each exposure has been obtained using a spectrophotometric standard star
observed during the same night of each of the six observing blocks.
Sky-subtraction was not performed with the ESO pipeline, but instead using the {\tt CubExtractor} package (Cantalupo in prep., \citealt{Borisova2016}) 
developed to enable the detection of very low surface brightness signals in MUSE datacubes.
Specifically we apply the procedures {\tt CubeFix} and {\tt CubeSharp} described in \citet{Borisova2016} to apply a
flat-fielding correction and to subtract the sky, respectively. Finally, the datacubes of the individual
exposures are combined using an average $3\sigma$-clipping algorithm. As described in \citet{Borisova2016},
we applied another iteration of the aforementioned procedures to improve the removal of self-calibration
effects. The products of this data reduction are a final science datacube and a variance datacube,
which takes into account the propagation of errors for the MUSE pipeline and the correct propagation during
the combination of the different exposures. 
Our final MUSE datacube (4.5 hours on source) results in a $2\sigma$ surface brightness limit of SB$_{\rm Ly\alpha}=3.7\times10^{-19}\cgssb$ (in 1 arcsec$^2$ aperture) 
in a single channel (1.25\AA) at $\approx 5066$\AA\, ($z=3.167$, \lya\ line for the nebula; see section~\ref{sec:cubex}).

\section{Observational Results}
\label{sec:results}

\begin{figure*}
\begin{center}
\includegraphics[width=0.95\textwidth]{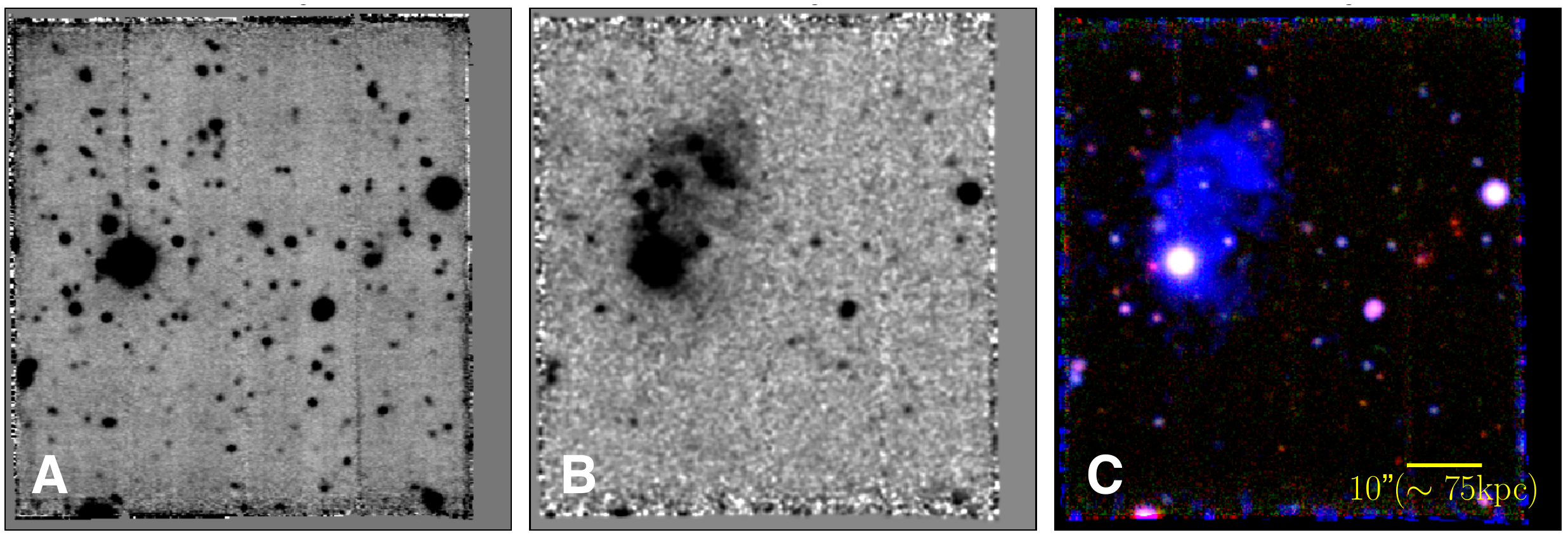}
\end{center}
\caption{{\bf Discovery images of the ELAN around the quasar \qso.} 
    {\bf (A)} white-light image of the combined exposures showing the whole FOV of our observations. Vertical stripes with slightly lower
    signal-to-noise are visible in this image due to our observing strategy (see section~\ref{sec:obs} for details). 
    {\bf (B)} NB image of 40~\AA\ centred at the redshift $z=3.167$ of the extended \lya\ emission. 
    The image has been smoothed with a gaussian kernel of $0.6\arcsec$ (or 3 pixels). 
    {\bf (C)} RGB image composed of the NB image, 
    and two intermediate continuum bands free of line emissions at the redshift of \qso\ of about $200$~\AA\ 
    centred at $5400$~\AA\ and $8237$~\AA, respectively. 
North is up, East is to the left.}
\label{Fig1}
\end{figure*}

In Fig.~\ref{Fig1} we show the whole $1.02'$~$\times$~$1.11'$ field of view (FOV) 
of our combined MUSE observations (i) as a ``white-light'' image , 
(ii) as a narrow-band (NB) image ($40$~\AA) centred on
the \lya\ line at $z=3.167$ (spatially smoothed by 3 pixels or $0.6\arcsec$),
and (iii) a RGB image composed of 
the NB image, and two intermediate continuum-bands free of line emission at 
the redshift of \qso, with a width of about $200$~\AA\
and centred at $5400$~\AA\ and $8237$~\AA, respectively.  
The $40$~\AA\ narrow-band image has a $2\sigma$ surface brightness limit of SB$_{\rm Ly\alpha}=1.9\times10^{-18}\cgssb$ (in 1 arcsec$^2$ aperture).
Note that our observational strategy inevitably results in vertical stripes with higher noise because 
we only performed three 90 degree rotations instead of four, making the vertical stripes under-sampled in comparison to the rest of the field. This artefact
is more 
significant in the white-light image than in NB
(or single channel) images, and has no effect
on the 
results presented in this work.
Fig.~\ref{Fig1} clearly shows the presence of a bright ELAN around \qso. Before analysing it, we focus on the characterisation of the 
compact sources associated with \qso. 

\subsection{Search for Companion Sources: Hints of an Overdensity}
\label{sec:compact}

\begin{figure}
\centering
\includegraphics[width=0.95\columnwidth, clip]{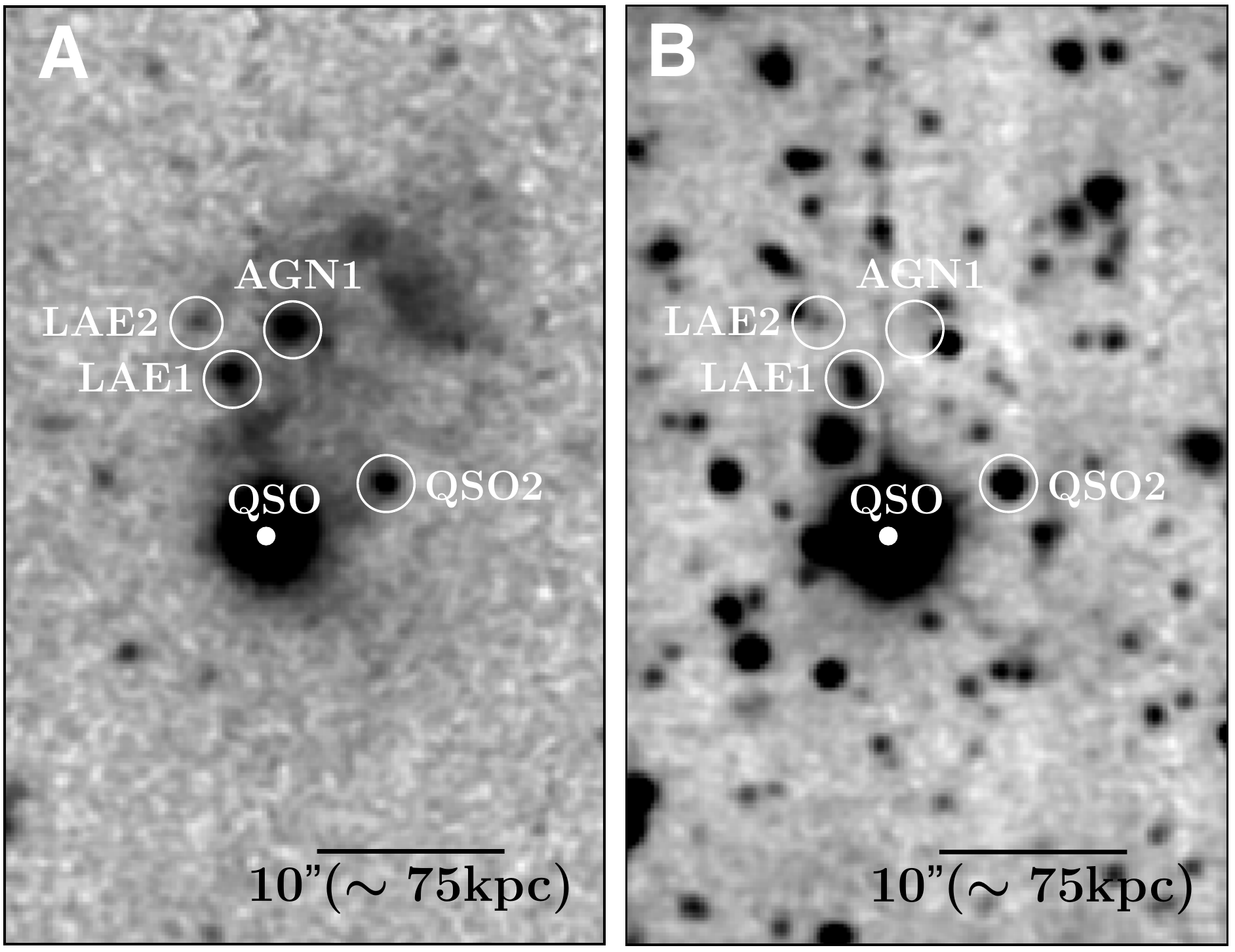} 
\caption{{\bf Position of the compact sources associated with the ELAN.} {\bf (A)} Portion of the $40$~\AA\ 
NB image matching the \lya\  line of the quasar \qso\ extracted from the MUSE datacube. 
{\bf (B)} Portion of the white-light image obtained by collapsing the MUSE datacube. In both panels, 
the positions of the objects associated with the ELAN are indicated. North is up, East is to the left.}
\label{figsources}
\end{figure}

\begin{figure*}
\begin{center}
\includegraphics[width=0.85\textwidth]{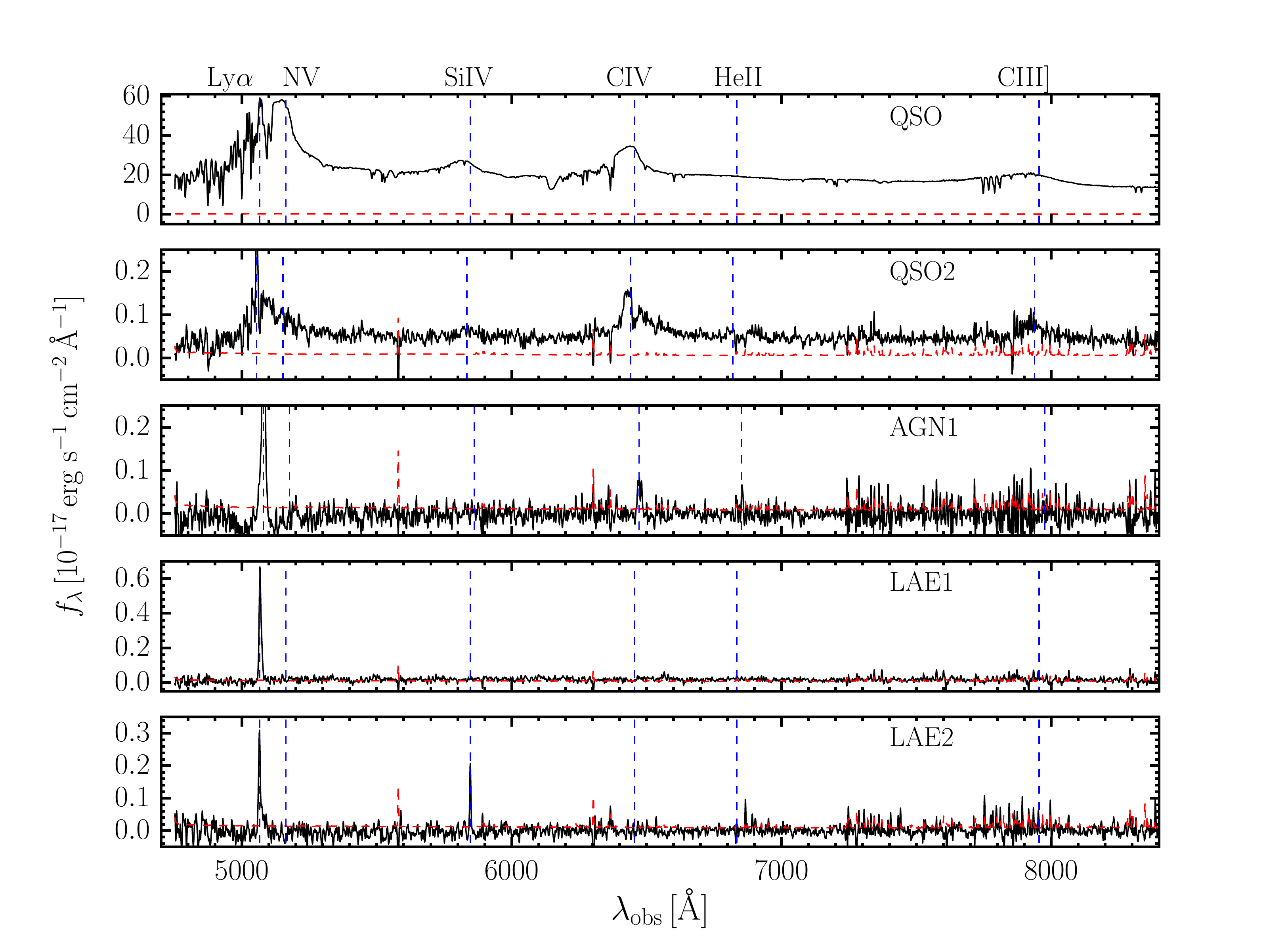}
\end{center}
\caption{ {\bf One-dimensional spectra of the strong \lya\ emitters 
associated with the ELAN.} 
The 1D spectrum of \qso\ has 
been extracted in a $1.5\arcsec$ radius circular aperture, while the 
other four spectra have been extracted within the 2$\sigma$ isophote 
of each source (see section~\ref{sec:compact}). 
For each object, the red dashed line shows the noise spectra extracted within the same aperture. 
At long wavelengths the MUSE spectra become 
noisier due to the presence of residuals of stronger and numerous sky-lines (more 
evident for the fainter sources). 
The blue dashed lines indicate 
the location of the typical strong 
ultra-violet emission lines encountered in quasar's spectra. The individual y-axes
have been scaled for presentation.
QSO2 and AGN1 have been classified as AGN. 
Having a strong \ion{Si}{IV} line in emission, LAE2 is a clear outlier in comparison to
typical \lya\ emitters at this redshift (see section~\ref{sec:compact} and Fig.~\ref{LAEcomparison}).} 
\label{figSpectraSources}
\end{figure*}

To identify galaxies at the same redshift of the quasar,
we  use two complementary approaches. We search for i) compact continuum sources and ii) 
compact \lya\ emitters whose continuum might have been too faint to be detected in the former case. 
We then determine the redshifts by looking for lines or continuum breaks (if any). This analysis led to the discovery
of four companions of which 
we show the positions in both the white-light image and in the NB image in Fig.~\ref{figsources}, and list them in Table~\ref{tabPos}.
Further, we show their spectra in Fig.~\ref{figSpectraSources}.

\begin{table}
\caption{Positions and $i$-mag for the associated sources QSO2, AGN1, LAE1, and LAE2.}
\center
\begin{tabular}{lcccc}
\hline
\hline
Source	& RA	& Dec	& Separation	& $i$-mag       \\  
	& (J2000) & (J2000)	& (arcsec)	& (AB)	\\
\hline
QSO2  		       &  10:20:09.58	&  +10:40:02.30   &  6.8  &  $24.30\pm0.02$	\\
AGN1 		       &  10:20:09.92	&  +10:40:13.93   &  11.3 &  $>28.6$	        \\
LAE1 		       &  10:20:10.15	&  +10:40:11.53	  &  9.1  &  $25.45\pm0.05$	\\
LAE2 		       &  10:20:10.24	&  +10:40:14.13	  &  11.9 &  $>27.0$	        \\
\hline
\end{tabular}
\flushleft{\small We quote the projected distances from \qso, and the 
$i$ magnitudes (SDSS filter) extracted from the MUSE datacube.
The magnitudes values for AGN1 and  LAE2 are determined as forced photometry at their locations.
}
\label{tabPos}
\end{table}

More specifically, the companions have been found as follows.
First, we use the white-light image shown in Fig.~\ref{Fig1} as a deep continuum detection image and
construct a catalogue of source candidates\footnote{The continuum sensitivity level in the white-light image 
is $f_{\lambda, 1\sigma}=1.1\times10^{-21}$~erg~s$^{-1}$~cm$^{-2}$~\AA$^{-1}$~pixel$^{-1}$.}. 
Specifically, we run {\tt SExtractor} (\citealt{bertin96}) with a detection area of 5 pixels and a threshold of
2$\sigma$ above the background root-mean-square. We then use the segmentation map generated by {\tt SExtractor} as a
mask to extract 1D spectra for each identified source from the final MUSE datacube.
Next, we inspected the 1D spectra to identify emission or absorption lines at the redshift of \qso. 
Of all the 177 continuum detected sources in the deep white-light image, only two 
are clearly at redshifts similar to \qso. These are a quasar, referred to as QSO2, and a strong \lya\ 
emitter (EW$_{\rm rest}=105$~\AA), referred to as LAE1, at projected distances of $51.7$~kpc ($6.8\arcsec$) 
and $69.2$~kpc ($9.1\arcsec$) from \qso, respectively (Fig.~\ref{figsources}).    
We evaluate the redshift of QSO2 with the same approach as for \qso\  (see appendix~\ref{appZ}), i.e.  following \citet{Shen2016}.
Specifically, we estimate the peak of the \civ line, as specified in \citet{Shen2016}, and obtained $z=3.158$.
With a measured $\lambda L_{1700\angstrom} = (5.0\pm0.9)\times10^{43}$~erg~s$^{-1}$ for QSO2, we estimate an expected shift 
of +173~km~s$^{-1}$ for \civ with respect to systemic. 
We thus obtain a redshift of $z=3.156\pm0.006$ for QSO2, where we took into account the systematic uncertainty 
of $415$~km~s$^{-1}$ in the error estimate (\citealt{Shen2016}). 
Bearing in mind the uncertainties on the redshift determination for both quasars, QSO2 seems to have a blueshift 
of 576~km~s$^{-1}$ with respect to \qso.

For LAE1, we have obtained a redshift of $z=3.168\pm0.001$ by fitting its \lya\ emission 
line with a gaussian (despite the presence of a red tail in its shape). For both QSO2 and LAE1, we have extracted an 
equivalent $i$-band SDSS magnitude from the MUSE datacube, obtaining $24.30\pm0.02$ and $25.45\pm0.05$, respectively.
These sources are therefore 
much fainter than \qso\ ($i$=17.98, fiber magnitude).

Secure redshifts could not be obtained for any of the other continuum-detected sources, e.g. no presence of emission lines, 
no continuum breaks. 
In particular, note that continuum sources that happen to be at the same location of the ELAN, show faint line emission  
at the wavelength of the \lya\ emission of the ELAN itself. Given that these faint continuum sources are not visible as compact \lya\
emitters in the NB image (see Fig.~\ref{figsources}), and that they do not show other emission lines, we ascribe the
emission to the nebula.

\begin{figure*}
\centering
\includegraphics[width=0.99\textwidth, clip]{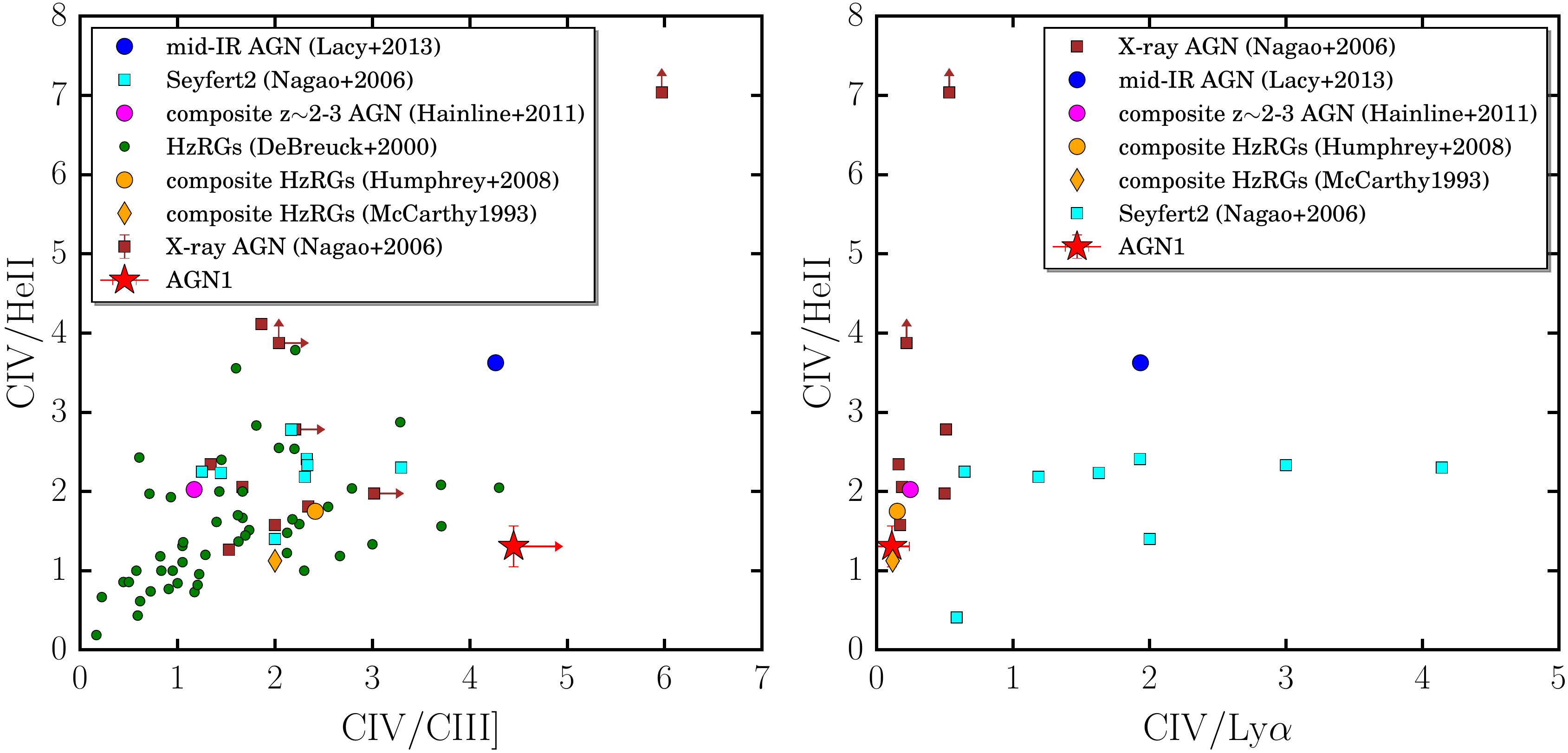} 
\caption{{\bf Comparison between AGN1 and the Type-2 AGN in the literature.} The line-ratios of AGN1 (red star) are compared to other 
Type-2 AGN from the literature in the \ion{C}{iv}/\heii vs. \ion{C}{iv}/\ciii\ diagram (left), and in the 
    \ion{C}{iv}/\heii vs. \ion{C}{iv}/\lya\ diagram (right). We show the values for (i) individual HzRGs (green circles; \citealt{DeBreuck2000}), 
    (ii) individual narrow-line X-ray sources 
    (brown squares; \citealt{Nagao2006}), (iii) individual Seyfert-2s (cyan squares; \citealt{Nagao2006}), (iv) two composites for HzRGs 
    (orange diamond; \citealt{McCarthy1993}; and orange circle,
    \citealt{Humphrey2008}), (v) a composite of narrow-lined AGN at $z\sim2-3$ (magenta; \citealt{Hainline2011}), 
    and (vi) a composite for mid-IR selected Type-2 AGN 
    (blue circle; \citealt{Lacy2013}). 
    To avoid confusion, we only show the errorbars for our data point. The typical uncertainties on the line ratios from the literature
    are of the order of 0.5-1. Overall, AGN1 seems to agree with the type-2 AGN reported in the literature.}
\label{AGNcomparison}
\end{figure*}

\begin{figure*}
\centering
\includegraphics[width=0.99\textwidth, clip]{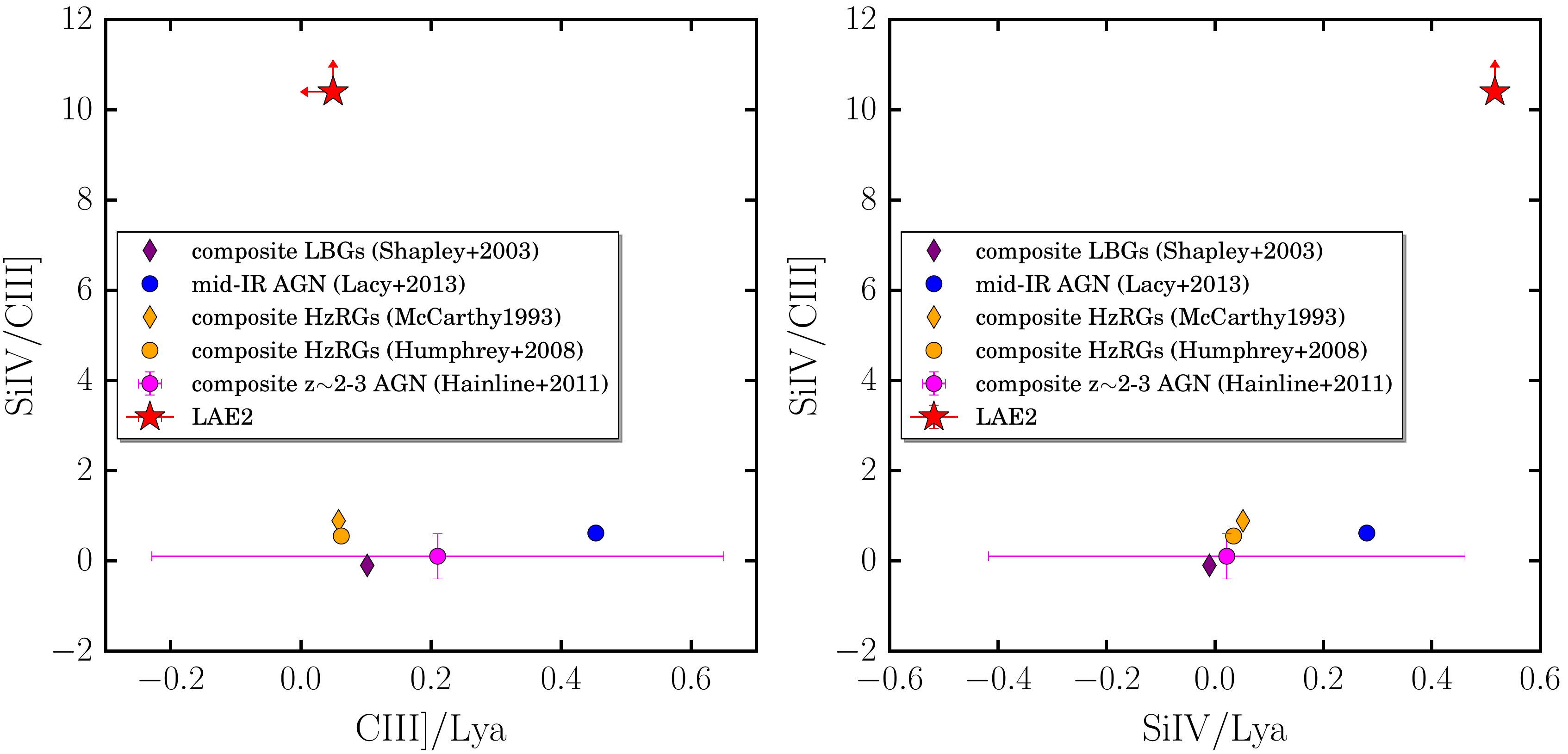} 
\caption{{\bf Comparison between LAE2 and the Type-2 AGN and LBGs in the literature.} The line-ratios of LAE2 (red star) are compared to Type-2 AGN 
and LBGs from the literature in the \ion{Si}{iv}/\ciii\ vs. \ciii/\lya\ diagram (left), and in the \siiv/\ciii\ vs. \ion{Si}{iv}/\lya\ diagram (right). 
We show the values for (i) two composites for HzRGs 
    (orange diamond, \citealt{McCarthy1993}; and orange circle,
    \citealt{Humphrey2008}), (ii) a composite of narrow-lined AGN at $z\sim2-3$ (magenta; \citealt{Hainline2011}), (iii) a composite for mid-IR selected Type-2 AGN 
    (blue circle; \citealt{Lacy2013}), and (iv) a composite for bright \lya\ emitting LBGs (EW$_{\rm Ly\alpha}^{\rm rest}=52.63\pm2.74$, 4th quartile in \citealt{shapley03}). 
    Note that LBGs have \siiv in absorption and thus their ratio on this plot is negative. LAE2 is a clear outlier.}
\label{LAEcomparison}
\end{figure*}

As stated earlier, we also search for  associated compact 
line emitters whose faint continua were not detected in the deep white-light image. 
Specifically, we use the 40~\AA\ NB image presented in Fig.~\ref{Fig1} as a detection image
to construct a catalogue of source candidates. We run {\tt SExtractor} with a detection area of 5 pixels and a threshold of
2$\sigma$ above the background root-mean-square\footnote{To avoid contamination from the 
nebula, the background rms is computed using SExtractor with a large mesh size (i.e. 128 pixels or $\sim26$ arcsec).}. Next we 
matched this line-emitter catalog to our catalog of continuum sources from the foregoing discussion, and removed 
all those with continuum detections.  The segmentation map of the remaining sources\footnote{We were left with six compact sources to analyze, 
but four of them were foreground objects.}
was used as a mask to extract 1D spectra from the final MUSE datacube.
This approach led to the identification of two \lya\ emitters at a redshift similar to \qso, AGN1 
at $z=3.179\pm0.001$ (EW$_{\rm rest}>168$~\AA) at $11.3\arcsec$ ($\sim85.9$~kpc), and LAE2 at $z=3.167\pm0.001$ (EW$_{\rm rest}>29$~\AA) at 
$11.9\arcsec$ ($\sim90.4$~kpc) (Fig.~\ref{figsources}). AGN1 also exhibits a \civ\ emission line, and tentative 
evidence for \heii and \nv\ emission lines, all indicative of a hard-ionizing source, making this source
most likely a type-2 AGN. 
Indeed, its line ratios are in good agreement with type-2 AGN reported in the literature as shown in Fig.~\ref{AGNcomparison} (\citealt{Nagao2006,McCarthy1993, Humphrey2008}).
Given its redshift, AGN1 has a velocity shift of 864~km~s$^{-1}$ with respect to \qso.
Note that AGN1 does not have a fully gaussian \lya\ line, although redshifts determined from both \lya\ and \civ 
agree within their uncertainties. 
On the other hand, LAE2 shows strong \siiv emission, 
also requiring a hard-ionizing source 
(ionization energy of I$_{\rm SiIV}=33.49$~eV; \citealt{Draine2011}).  However, we don't detect other high ionization 
lines (\ion{He}{ii}, \ion{C}{iv}, \ion{N}{v}, \ciii) which typically accompany \ion{Si}{iv} in AGN spectra 
(e.g., \citealt{McCarthy1993,Humphrey2008,Hainline2011,Lacy2013}). 
To our knowledge, there are no sources reported in the literature with only \lya\ and \siiv detected.
In the population of Lyman break galaxies (LBGs; e.g., \citealt{shapley03}) or LAEs characterized by similar 
strong \lya\ emission as LAE2, \siiv is usually observed in absorption (e.g., 4th quartile in \citealt{shapley03}).
We show the discrepancy between the line ratios for LAE2 and the population of type-2 AGN and LBGs 
in Fig.~\ref{LAEcomparison}.  
LAE2 thus seems a clear outlier for both type-2 AGN and LBGs. However, we argue that the strong \siiv emission should be a signature 
of powering mechanisms more similar to an AGN. The nature of LAE2 remains unclear, and follow up studies are needed.
Intriguingly, \citet{Dey2005}, while studying a LAB at $z=2.656$, reported the 
detection of an emission line at $\lambda=5081.6$~\AA, which could have been an improbable blueshifted \siiv 
line, or an interloper as the author suggested.  
The information for the three sources 
with narrow emission lines are summarized in Table~\ref{tabCompactSources}. 

\begin{table*}
\caption{Information on the emission lines detected in the spectrum of AGN1, LAE1, and LAE2.}
\centering
\begin{tabular}{lcccccc}
\hline
\hline
Line	    	           & Line Center      & Redshift 	 & Line Flux 			       & Continuum Flux			       		 & EW$_{\rm rest}$ & Line Width	\\
		           & (\AA)	      & 	 	 & (10$^{-17}$ erg s$^{-1}$ cm$^{-2}$) & (10$^{-20}$ erg s$^{-1}$ cm$^{-2}$ Hz$^{-1}$)   & (\AA)  	 & (km s$^{-1}$)\\
\hline
\multicolumn{7}{c}{AGN1 ($\langle z\rangle =3.179\pm0.001$, d$=85.9$~kpc)} \\
\hline
\lya			 &  5080.51$\pm$0.03  &  3.179$\pm$0.001 & 9.90$\pm$0.08			   & -4.4$\pm$7.0			       & $>$168 	 & 232.4$\pm$1.9   \\
\civ			 &  6474.6$\pm$0.6    &  3.180$\pm$0.001 & 1.09$\pm$0.08		       & -3.2$\pm$3.0				       & $>$42  	 & 213$\pm$26	\\
\heii			 &  6854.2$\pm$0.2    &  3.179$\pm$0.001 & 0.83$\pm$0.04		       & -2.9$\pm$4.0				       & $>$25  	 & 156$\pm$10	\\
\hline
\multicolumn{7}{c}{LAE1 (d$=69.2$~kpc)}\\
\hline
\lya			 &  5067.68$\pm$0.05  &  3.168$\pm$0.001 & 5.49$\pm$0.06			   & 12.6$\pm$9.8			       & 104.96$\pm$0.78 	 & 239.6$\pm$2.8   \\
\hline
\multicolumn{7}{c}{LAE2 ($\langle z\rangle =3.167\pm0.001$, d$=90.4$~kpc)} \\
\hline
\lya			 &  5064.9$\pm$0.1    &  3.166$\pm$0.001   & 2.41$\pm$0.07			   & -1.4$\pm$9.8			       & $>$29	 	 & 186.9$\pm$6.0   \\
\siiv			 &  5846.42$\pm$0.07  &  3.167$\pm$0.001   & 1.24$\pm$0.04		           & -3.7$\pm$3.6			       & $>$41  	 & 90.6$\pm$3.9    \\
\hline
\end{tabular}
\flushleft{For each narrow emitter we quote the estimated average redshfit and the projected distance from \qso.
For each detected emission-line in the spectrum of each source, we report 
the line center, the redshift, the line flux, the continuum flux, the rest-frame equivalent width, 
and the line width as $\sigma_v$ of a Gaussian fit to the line.
}
\label{tabCompactSources}
\end{table*}

The presence of three confirmed AGN within 90 projected kpc makes this system similar to the other two known 
physical quasar triplets discovered at lower redshifts (\citealt{Djorgovski2007, Farina2013}).
In addition, given the presence of unusual compact emitters embedded in the ELAN, 
it might resemble the properties of the so far only known quadruple quasar (\citealt{hennawi+15}), known to be embedded in
a bright ELAN as well.
It is thus interesting to compare the environment of that system with \qso. 
\citet{hennawi+15} show that the bright quasar SDSS~J084158.47+392121.0 (henceforth SDSS~J0841) 
in their system inhabits a clear overdensity of LAEs, 
exceeding the average protoclusters, i.e. high-redshift radio galaxies (HzRGs) and Lyman-Alpha Blobs (LABs; e.g., \citealt{Yang2009}), by a factor of $\gtrsim20$ for $R<200$~kpc and by 
$\sim3$ on scales of $R\simeq 1$~Mpc. To allow a comparison with studies of HzRGs and LABs, the analysis of \citet{hennawi+15} relies 
on the LAEs selected to have $EW_{\rm rest}>20$\AA\, and with $f_{\rm Ly\alpha}>4.0\times10^{-17}$~erg~s$^{-1}$~cm$^{-2}$.
Given our pointing strategy and the MUSE FOV, we are able to fully probe the environment only within $R\lesssim122$~kpc (or
$\lesssim16 \arcsec$) from \qso. Within this region, following the same selection criteria of \citet{hennawi+15}, we find 3 objects 
(2 of which are AGN), while SDSS~J0841 has 4 objects (2 of which are AGN), suggesting that the system studied here inhabits a
similar overdensity on small $\lesssim 100$~kpc scales 
as SDSS~J0841, and making it a rare overdense system as well. A follow-up narrow-band study of a wider field around \qso\ is needed for a full
comparison. However, the similarity on small scales with SDSS~J0841, together with the presence of multiple AGN, 
make \qso\ likely to be the progenitor of a very massive object. 

Finally, we note that the other two ELANe so far discovered are also associated to multiple AGN and overdensities of galaxies (\citealt{cantalupo14,Cai2016}).  
\citet{cantalupo14} report the discovery of an ELAN associated to the quasar UM~287 ($z=2.279$) 
and a fainter companion quasar. Further, \citet{Cai2016} unveil a bright ELAN at $z=2.319\pm0.004$ displaced by $33.1\arcsec$ 
from the quasar SDSS~J144121.66+400258.8  ($z=2.311\pm0.006$) by targeting the density peak of the large-scale structure BOSS1441 (\citealt{Cai2016a}). 
Showing extended \civ and \heii in emission, this ELAN probably hosts an additional obscured AGN which is powering the nebula through an outflow 
and/or photionization (\citealt{Cai2016}). 
More generally, it has been argued that the presence of a giant \lya\ nebula (wether is an ELAN, a LAB or associated to an HzRG) 
is physically connected to the location of overdensities of galaxies and AGN (\citealt{Matsuda2005,Matsuda2009,Saito2006,Venemans2007,Prescott2008,Yang2009,hennawi+15}).

\subsection{The Enormous Lyman-Alpha Nebula}
\label{sec:gb_lya_neb}

In Fig.~\ref{Fig1} we have shown an NB image of the ELAN around the quasar \qso.
In this section we report how we analysed the final MUSE datacube to constrain this extended \lya\ emission, while in appendix~\ref{appC} we explain 
how we tested the reliability of our approach.

\subsubsection{Empirical PSF subtraction and Moments of the Flux Distribution}
\label{sec:cubex}

We extracted the properties of the \lya\ emission using the {\tt CubExtractor} package, as explained in \citet{Borisova2016}.
Briefly, we have first subtracted the quasar PSF to remove any contamination from the unresolved QSO on large scales, 
and then we have characterised the properties of the ELAN by calculating the moments of the flux distribution. 

More specifically, the subtraction  of the PSF has been performed using the {\tt CubePSFSub} algorithm within {\tt CubExtractor}, 
which empirically reconstructs the PSF of the quasar in user-defined wavelength layers
within the MUSE datacube. For each wavelength layer, the empirical PSF image is obtained as a pseudo-NB image, 
it is rescaled to the flux within $5\times5$ pixels (or $1\arcsec\times1\arcsec$) around the quasar position, 
and then subtracted (see \citealt{Borisova2016} for more details).  In our case, we have used a wavelength layer of 150 spectral 
pixels ($187.5$\AA), shown to be optimal for the case of extended emission around quasars (\citealt{Borisova2016}). 
Note that this method is not reliable in the central $1\arcsec \times 1\arcsec$ region used for 
the PSF rescaling, leading to residuals on these scales (\citealt{Borisova2016}). However, in appendix~\ref{appC} we show that the PSF subtraction following a
different algorithm does not change our results (\citealt{Husemann2013}).   
In addition to the PSF subtraction, we have removed all the continuum-detected sources from the datacube using the {\tt CubeBKGSub} algorithm, which uses
a fast median-filtering approach (see \citealt{Borisova2016} for more details).

\begin{figure*}
\centering
\includegraphics[width=0.96\textwidth]{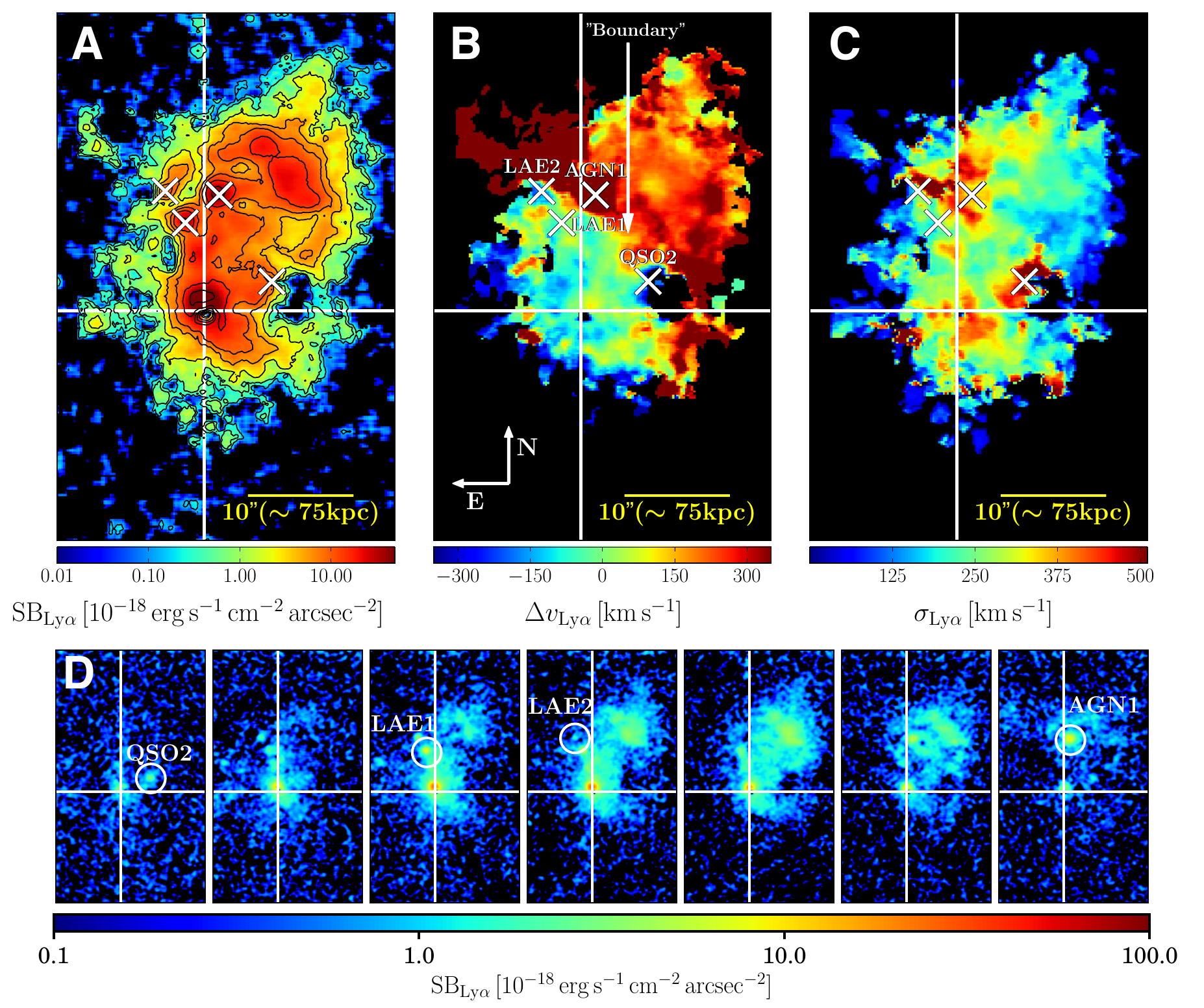}
\caption{{\bf The enormous \lya\ nebula (ELAN) surrounding the quasar \qso.} 
{\bf (A)} ``optimally-extracted'' \lya\ surface brightness map obtained
after subtraction of the quasar point-spread-function (PSF) and continuum in the final MUSE datacube (see section~\ref{sec:cubex}). 
The black contours indicate the isophotes corresponding to a signal-to-noise ratio of ${\rm S/N}=2,\,4,\,10,\,20,\,30,\,50$, and $100$. 
This image reveals an extremely bright nebula (SB$_{\rm Ly\alpha}\sim 10^{-17}\cgssb$) extending on 100 kpc scales on the NW side of the quasar. 
Additional four strong \lya\ emitters (diagonal crosses) are associated with \qso, and the ELAN. Two of these sources have been 
spectroscopically confirmed as AGN (Fig.~\ref{figSpectraSources}), making this system the third known quasar triplet.
{\bf (B)} flux-weighted velocity-shift map with respect to 
the systemic redshift of the quasar \qso\  obtained from the first-moment of the flux 
distribution.  A velocity shear between the SE and NW portion of the nebula is evident. The transition region is referred to as the ``Boundary''.
{\bf (C)} velocity dispersion map obtained from the second-moment of the 
flux distribution. Regions of higher dispersion 
($\sigma_{\rm Ly\alpha}\approx430$~km~s$^{-1}$) 
are visible in proximity of the three AGN, but overall the \lya\ nebula shows 
quiescent kinematics ($\sigma_{v} < 270$~km~s$^{-1}$).
{\bf (D)} Each cut-out image (same size as A, B, and C) shows the surface brightness map of the ELAN within a  3.75~\AA\ layer ($3\times$ MUSE sampling) in the wavelength range  5058~\AA~$\lesssim \lambda \lesssim$~5084~\AA\ (from left to right).
In all of the panels (A, B, C, D) the large white cross indicates the position of the 
quasar \qso\  prior to PSF subtraction.} 
\label{figlya}
\end{figure*}

We then use {\tt CubExtractor} to identify the diffuse \lya\ emission by searching for 
regions with a minimum ``volume'' of 10,000 voxels (volume pixels) above a signal-to-noise ${\rm S/N}>2$ within the MUSE 
datacube\footnote{Such a large minimum ``volume'' has been chosen because we are only interested in selecting the ELAN, i.e. 
a very extended structure around the quasar \qso. Indeed, 10,000 voxels would corresponds to e.g. $100\times100$ pixels 
(or $20\arcsec \times 20\arcsec$) with width of 1 spectral pixel (or $1.25$\AA). 
For reference, the ELAN extracted with this approach has 133,632 connected voxels, and it is the only source found 
by {\tt CubExtractor} above the chosen threshold.}. 
In this way, we are left with a three-dimensional mask which indicates 
the voxels associated with the extended \lya\ emission around the quasar \qso. 
We used this mask to obtain an ``optimally extracted'' NB image by integrating the flux 
along the wavelength direction for only the voxels belonging to the nebulosity. 
Each pixel itself in the obtained 2D-image thus represents a narrow-band filter, whose width is set by the 
${\rm S/N}=2$ threshold. 
We have then added to the ``optimally extracted'' NB image a 'background' layer of 40~\AA\ around $z=3.167$ (central wavelength 
of the nebula for region closer to QSO position) to recreate the noise for a NB image with the wavelength range equal to 
the maximum width of the nebulosity, i.e the maximum width of the 3-D mask previously obtained.  
In panel A of Fig.~\ref{figlya} we show this optimally extracted NB image, which clearly reveals
that this \lya\ nebula is one of the so far brightest and largest known around
radio-quiet quasars. 
The nebula has an average SB$_{\rm Ly\alpha}=6.04\times10^{-18}\, \cgssb$ (within the $2\sigma$ isophote).
Note that an ``optimally extracted'' image does not allow a visual estimate of the noise as it depends on the number of layers at each spatial position. 
Hence, to enable a better interpretation  we show in Fig.~\ref{figlya} the signal-to-noise contours for $S/N=2,4,10,20,30,50$, and $100$, 
estimated through variance propagation accounting for the number of layers along each pixel position (see \citealt{Borisova2016}).

\begin{figure}
\begin{center}
\includegraphics[width=0.9\columnwidth, clip]{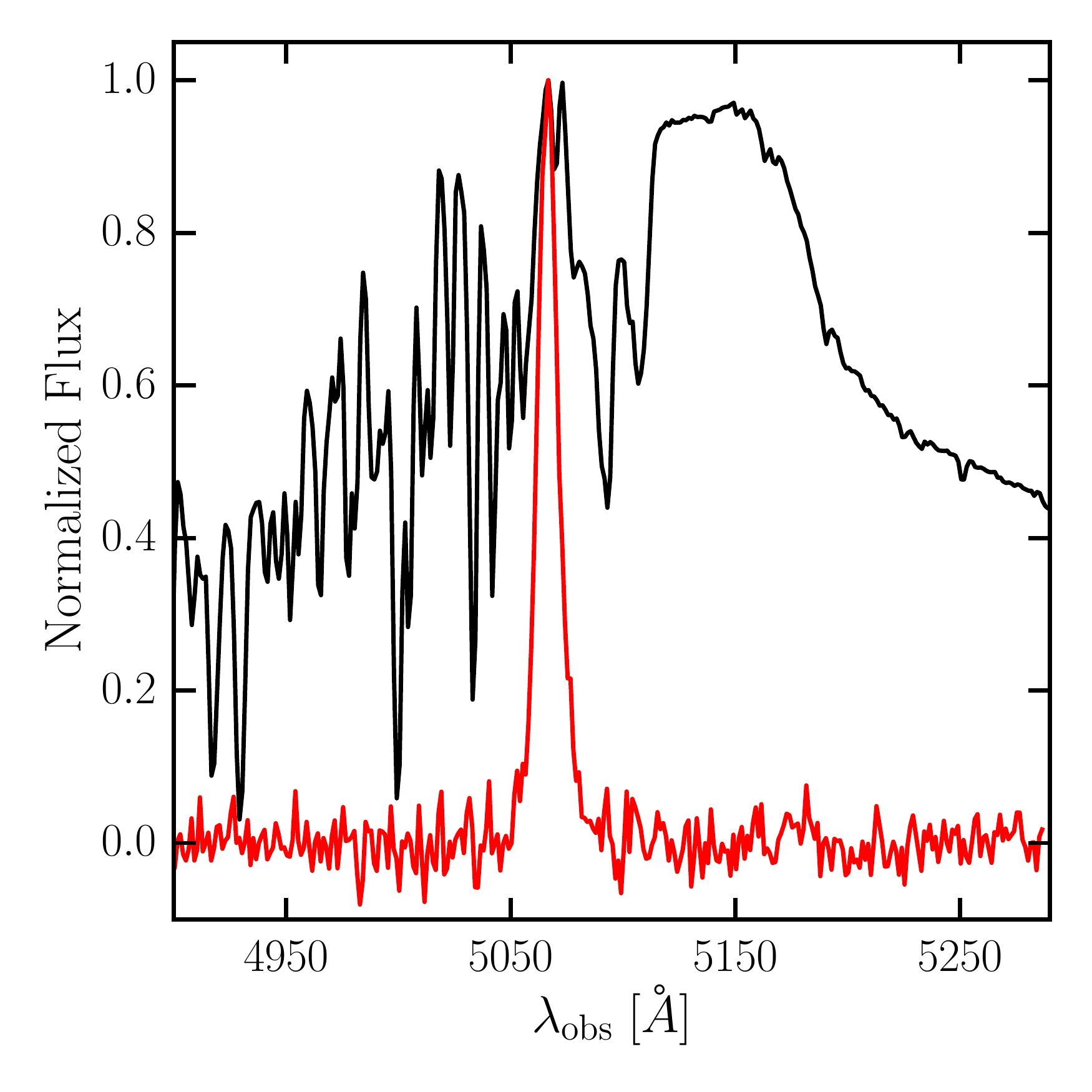} 
\end{center}
\caption{{\bf \lya\ line shape for the quasar \qso\ and the peak of the ELAN.} We compare 
the \lya\ line emission of the quasar (black) and of the peak of the ELAN (red), i.e. at $\approx1\arcsec$ from the quasar. 
The spectrum of the peak of the ELAN has been extracted within a circular aperture of $2\arcsec$ radius centred at $\alpha=155.0428$~deg 
and $\delta=10.6679$~deg (Region 3 in Fig.~\ref{LyaLineShape}). Both spectra have been normalized to their maximum. 
The ELAN emission is much narrower than the complex \lya\ emission  of the quasar \qso. We are thus confident that the peak of the ELAN close 
to the quasar position is not due to PSF residuals.}
\label{QSONebComparison}
\end{figure}

Further, the PSF subtracted image 
reveals a bright peak for the \lya\ emission $\simeq 1\arcsec$ to the North of the quasar \qso\  (Fig.~\ref{figlya}). 
The \lya\ emission at this location show a much narrower profile ($\sigma_{\rm Ly\alpha}=316.4\pm7.9$~km~$s^{-1}$) than the more complex 
quasar's broad \lya\ line blended together with the \ion{N}{V} emission. This comparison is shown in Fig.~\ref{QSONebComparison}. 
Given the great differences in the line shapes, we are therefore confident that the bright emission in close proximity to 
the quasar \qso\ cannot constitute a PSF subtraction residual. 
This bright knot may represent the extended emission line regions (EELR; \citealt{Stockton2006}) associated with quasars, i.e. narrow emission-line 
regions extending for tens of kpc around AGN, 
and thought to be mainly powered by the central bright source or star-formation from the host-galaxy (e.g., \citealt{Husemann2014}).
Interestingly, 
this emission is at the redshift of $z=3.167\pm0.001$, and is connected to the emission on larger scales. This occurrence thus confirm
that the \lya\ nebula and the quasar \qso\ are at the same redshift, once 
the known uncertainties are taken into account (see appendix~\ref{appZ}). 

We then use the aforementioned three-dimensional mask to analyse the kinematics 
of the \lya\ emitting gas.  Specifically, to derive the centroid velocity and the width of the emission line, 
we have computed at each spatial location the first and second moment of the flux distribution only for the voxels selected by the mask.  
The use of only the selected ``volume'' should minimise the effect of the noise for this approach.

Panel B in Fig.~\ref{figlya} shows the flux-weighted centroid
of the \lya\ emission throughout the ELAN with respect to the systemic redshift of \qso.   
The ELAN clearly exhibits a significant velocity shear as one progresses from its SE edge to the NW.  
The southern half of the ELAN shows systematically blueshifted velocities (by $\approx 300$~km~s$^{-1}$)
compared to its northern half, and there is a relatively sharp discontinuity
across the ``boundary''.
Panel C in Fig.~\ref{figlya} shows the flux-weighted standard
deviation of the emission.  The values are relatively small, $\sigma_v < 270$~km~s$^{-1}$ and are nearly consistent with the 
spectral resolution of MUSE. 
One concludes that the motions within this ELAN are highly coherent, and have amplitudes consistent with being gravitational motions within a 
dark matter halo hosting a quasar.
This high coherence of the velocity field can be greatly appreciated by looking at panel D in Fig.~\ref{figlya}. 
This panel dissects the ELAN presented in panel A, B, and C in a sequence of narrow-band images of $3.75$~\AA\ ($3\times$ MUSE sampling) 
in the wavelength range $5058$~\AA~$\lesssim \lambda \lesssim$~$5084$~\AA. 
We will further discuss this observed velocity shear in section~\ref{sec:favInterp}.

\subsubsection{The Enormous Lyman-Alpha Nebula in \heii and \civ}
\label{sec:heiiciv}

To try to mitigate the possible uncertainties  that arise from the use of the \lya\ emission as a unique diagnostic (section~\ref{sec:kinematics}), 
our QSO MUSEUM  survey (Arrigoni Battaia et al., in prep.)
specifically targets $z\sim3$ quasars to be able to cover other strong rest-frame ultraviolet lines beside \lya. 
In particular, it has been shown that the \civ$\lambda1549$ and \heii$\lambda1640$ lines are important 
indicators that may constrain the density, ionization state, metallicity, and the importance of scattering within the emitting gas
(\citealt{Nagao2006,Prescott2009,Humphrey2008,Prescott2015,fab+15a,fab+15b}).  
Further, as described in detail in \citet{fab+15a,fab+15b}, 
these emission lines can disentangle the powering mechanisms for the \lya\ emission, such
as photoionization from AGN or star-formation (\citealt{fab+15a} and references therein), scattering of \lya\ photons (e.g., \citealt{Dijkstra2008, Cen2013}), 
shock-heated gas in superwinds (e.g., \citealt{TanShi2000, Mori2004, Cabot2016}), or cooling radiation (e.g., \citealt{Yang2006,Rosdahl12}).  
Such mechanisms can also act together,  and additional diagnostics might be needed to characterise the different contributions 
(e.g. polarization of the \lya\ line \citealt{Prescott2011}).
Here we report our observations at the \civ and \heii expected wavelengths, while we discuss the powering mechanisms for the ELAN around 
\qso\ in section~\ref{sec:PowMech}.

To assess the presence of extended \civ and \heii in our data, we proceed in two ways.
First, using {\tt CubExtractor}, we have searched for connected voxels above a signal-to-noise ${\rm S/N} > 2$ 
at the wavelength of the two emission lines, on unsmoothed data, 
and leaving the minimum ``volume'' unconstrained, so that 
less extended emission than the enormous \lya\ emission would have also
been detected. This approach led to the detection of compact unresolved emission close to the quasar PSF residuals, 
but no extended emission has been detected at the expected redshift, with only a hint for extended emission in the 
\civ line if the ${\rm S/N} > 2$ constrain is relaxed. 
Secondly, given the possibility of a velocity shift between the \lya\ line emission and the \civ and \heii emission, we have searched the datacube
with the same approach within a window of 2000~km~s$^{-1}$ on both side of the expected location. 
This approach has also not resulted in the detection of the two lines on large scales.
Here, we have thus decided to present our data by simply constructing two narrow-band images of 30~\AA\  
centred at the expected wavelength of the two emission lines at $z=3.167$ (redshift of the peak of the Ly$\alpha$ nebula; section~\ref{sec:cubex}), 
i.e. $\lambda=6454.7$~\AA\, and $\lambda=6833.9$~\AA, respectively for
\civ and \heii\footnote{We select this width for these NB images to confidently include the shift of AGN1 and QSO2 within our images. 
We remind that the three-dimensional mask for the ELAN has a maximum width of $40$~\AA\ down to a $2\sigma$ limit (section~\ref{sec:cubex}). 
Here we thus restrict to higher ${\rm S/N}$ levels for the \lya\ emission.}. The NB images are obtained by collapsing the final datacube, 
after the quasar PSF and the continuum sources have been removed (as explained in section~\ref{sec:cubex}).

\begin{figure*}
\begin{center}
\includegraphics[width=0.9\textwidth, clip]{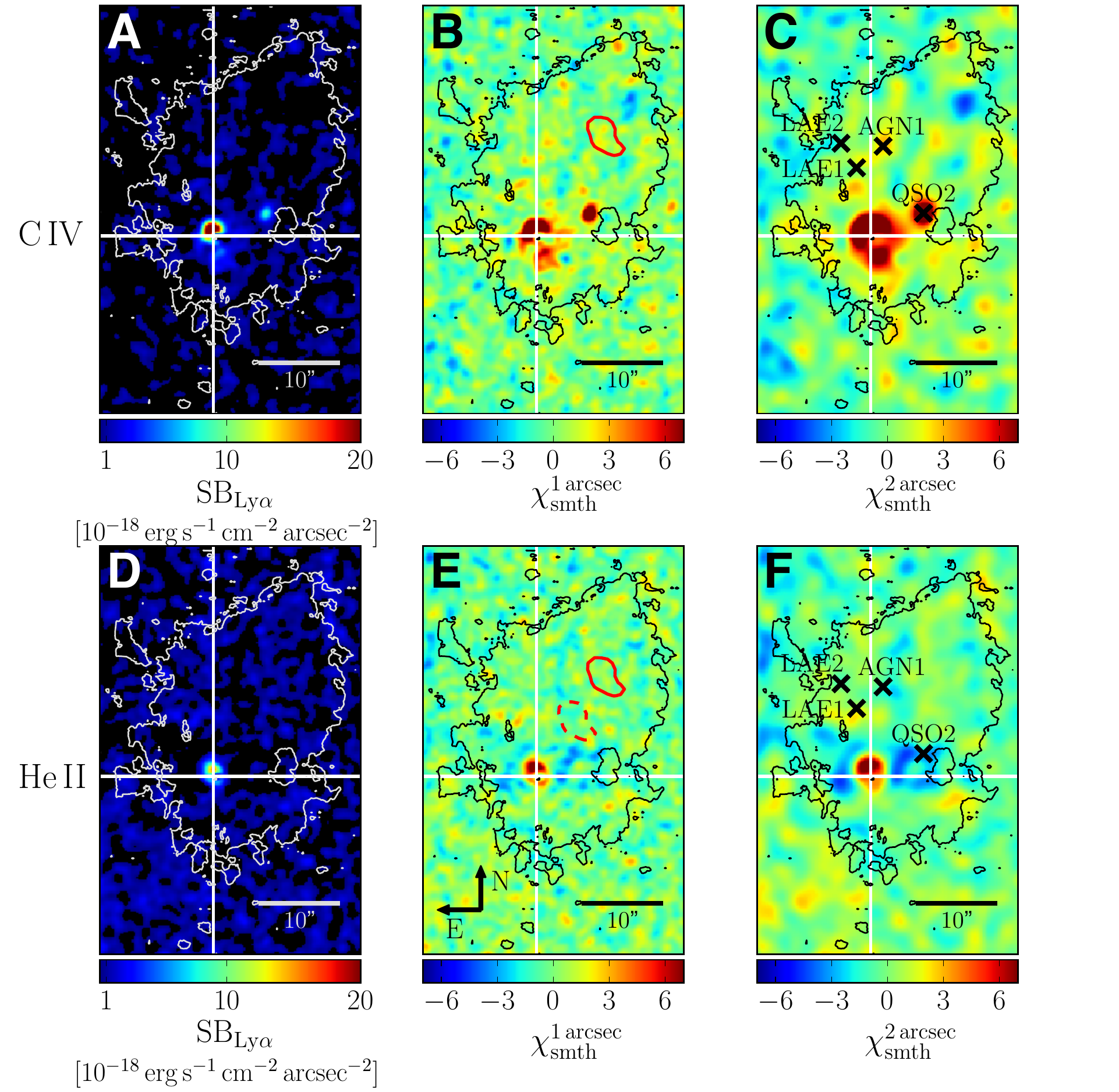}
\end{center}
\caption{{\bf Narrow-band images extracted at the expected wavelengths for} \civ {\bf and} \ion{He}{ii}.
{\bf (A)} \civ surface brightness map for a $30$~\AA\ NB image obtained by collapsing the final MUSE datacube around $z=3.167$ (redshift of the ELAN), 
after quasar PSF and continuum subtraction. {\bf (B)}  $\chi_{\rm smth}$ image for the same wavelength range (expected \civ location) obtained using a 
Gaussian kernel with FWHM$=1\arcsec$ as explained in section~\ref{sec:heiiciv}. The red contour highlights the ${\rm S/N}=50$ isophote for the \lya\ 
emission used to extract the line ratios for the clump at $\approx120$~kpc from \qso. 
{\bf (C)} $\chi_{\rm smth}$ image for the same wavelength range (expected \civ location) obtained using a Gaussian kernel with FWHM$=2\arcsec$ 
as explained in section~\ref{sec:heiiciv}. 
 {\bf (D), (E), (F)} Same as for (A), (B), and (C), respectively, but for the \heii line. In (E), the dashed red contour indicates 
 the portion of the ELAN used in the analysis that resulted in Fig.~\ref{CloudyFourth}.  
No clear extended emission in \civ and \heii is associated to the ELAN down to our sensitivity limits. 
However the aggressive smoothing seems to suggest very faint extended \civ emission in the South-West direction from \qso.} 
\label{civheii}
\end{figure*}

Panels A and D of Fig.~\ref{civheii} show the surface brightness maps for the NB images of the giant \lya\ nebula in the \civ and \heii lines after smoothing the data
with a Gaussian kernel of 1 arcsec. 
These images have a 2$\sigma$ SB limit within a square arcsecond of SB$_{\rm lim}^{\rm CIV}=1.05\times10^{-18} \cgssb$, 
and SB$_{\rm lim}^{\rm HeII}=1.18\times10^{-18} \cgssb$,
while the depth achieved for individual channels at these wavelength is SB$_{\rm lim}=2.25\times10^{-19} \cgssb$, 
and SB$_{\rm lim}=2.69\times10^{-19} \cgssb$, 
respectively for \civ and \heii. 
To indicate the significance of the mild detection of the compact emission and to search 
for evidence of faint emission on large scales, we have computed a smoothed $\chi$ image
following the technique in \citet{qpq4} and \citet{fab+15a}, for two Gaussian kernels with FWHM$=1\arcsec$ and FWHM$=2\arcsec$, respectively.
To obtain these images, we proceed as follows.
First, we obtain an unsmoothed NB image $I$ of 30~\AA\ by collapsing the final MUSE datacube at the wavelength of interest. Then, we smooth this image using the Gaussian 
kernel, obtaining $I_{\rm smth}$. Next, from the variance cube we have computed the unsmoothed variance image $\sigma^2_{\rm unsmooth}$ for each NB image, 
and obtained the smoothed sigma
image $\sigma_{\rm smth}$ by propagating the variance image of the unsmoothed data:
\begin{equation}
\sigma_{\rm smth} = \sqrt{{\rm CONVOL}^2[\sigma_{\rm unsmooth}^2]},
\end{equation} 
where the CONVOL$^2$ operation denotes the convolution of the
variance image with the square of the Gaussian kernel. Thus, the smoothed $\chi$ image is defined by
\begin{equation}
\chi_{\rm smth}=\frac{I_{\rm smth}}{\sigma_{\rm smth}}.
\end{equation}
These $\chi_{\rm smth}$ maps are more effective in visualising the presence of extended emission.

From Fig.~\ref{civheii} it is clear that \civ emission is only definitively detected in close proximity of the two quasars, \qso\ 
and its companion, while the \heii emission is seen clearly only
close to \qso\ (North direction). In the case of \qso, the additional emission lines
have their maxima at the same position of the peak of the \lya\ emission at $\approx1\arcsec$ from \qso. Being narrow 
and at the same redshift of the quasar, this small-scale emission is thus probably due to the so-called EELR around 
the bright quasar (see section~\ref{sec:gb_lya_neb}).   
Deeper data at higher spatial resolution are needed 
to explore in further detail the presence, kinematics, and geometry of the
\civ and \heii line-emissions on these small scales. Such observations will be feasible once 
the adaptive-optic system for MUSE becomes available (GALACSI; \citealt{Stuik2006}).

In addition, there is the hint (very low significance in the unsmoothed data) in the \civ map for extended emission in the South-West 
direction of \qso. 
Given the faint levels for this
emission, we are not able to investigate if it traces the same kinematics as the \lya\ emission in this region. 
Deeper observations are needed to confirm this emission and eventually compare it with the \lya\ emission.

\begin{figure}
\begin{center}	
\includegraphics[width=1.0\columnwidth, clip]{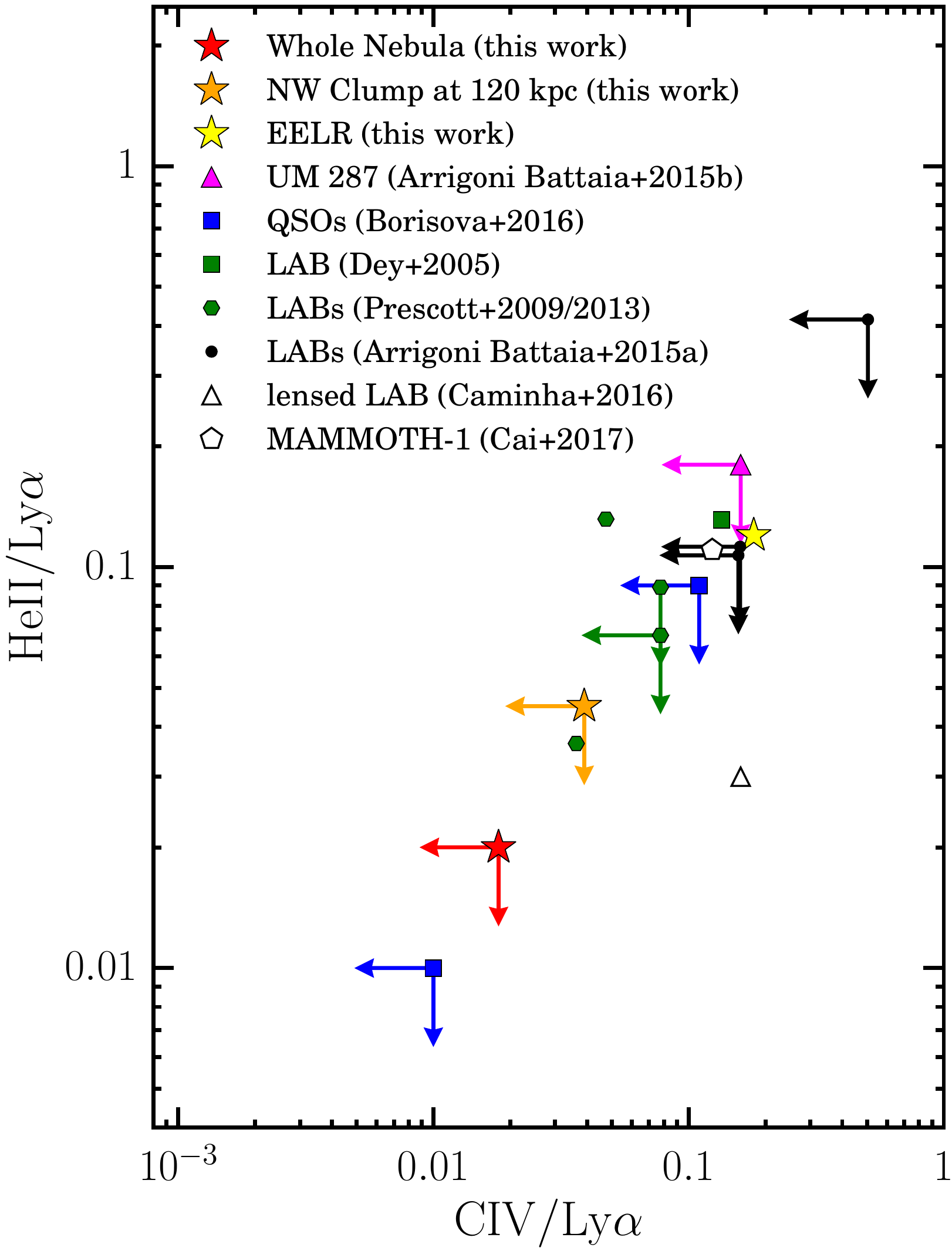}
\end{center}
\caption{\ion{He}{ii}{\bf /\lya\ vs. } \ion{C}{iv}{\bf /\lya\ log-log plot for radio-quiet \lya\ nebulae.} 
Our upper limits of the \ion{He}{ii}/\lya\ and \ion{C}{iv}/\lya\
    ratios for the whole \lya\ nebula (red), the NW clump at ${\approx120}$~kpc (orange) and the datapoint for the EELR (yellow) here studied are compared with 
    data available in the literature for other radio-quiet \lya\ nebulae. Specifically, we show data for (i) LABs from 
    \citet{Prescott2009,Prescott2013,Dey2005}, \citet{Caminha2016}, 
    and \citet{fab+15a} (only the most and less stringent data-points); (ii) \lya\ nebulae around radio-quiet QSOs 
    from \citet{fab+15b} and from \citet{Borisova2016} (only the most and less stringent data-points); (iii) the MAMMOTH-1 ELAN
    (\citealt{Cai2016}). 
    The errorbars for both ratios ($\sigma=0.01$) for our EELR datapoint are smaller than the size of the symbol used.} 
\label{plotCivHeii}
\end{figure}

Finally, as they can be used to infer the physical properties of the gas in emission, we report here the line ratios between \ion{C}{iv}, \ion{He}{ii}, and \lya, 
and discuss in section~\ref{sec:PowMech} their implications. 
Specifically, given that we do {\it not} have a detection in \civ and \heii on large scales, we compute the 5$\sigma$ SB 
limits within the ${\rm S/N}=2$ isophote defined by the \lya\
emission (see black contour in Fig.~\ref{civheii}), which corresponds to an area on the sky of 609.36 arcsec$^2$. 
We obtain SB$_{\rm HeII}<1.2\times10^{-19} \cgssb $, 
and SB$_{\rm CIV}<1.1\times10^{-20} \cgssb$, 
resulting in \ion{He}{ii}/\lya$<0.020$,  
and \ion{C}{iv}/\lya$<0.018$ when using the average SB$_{\rm Ly\alpha}=(6.04\pm0.26)\times10^{-18} \cgssb$ for the whole nebula. 
We compute the $5\sigma$ limits also for the bright clump at a projected distance of $\approx120$~kpc from \qso. In particular, 
we use the isophote within which the \lya\ emission has ${\rm S/N}>50$ 
(region indicated with a red contour in Fig.~\ref{civheii}, area of 13.4 arcsec$^2$) finding SB$_{\rm HeII}<8.1\times10^{-19} \cgssb$, 
and SB$_{\rm CIV}<7.2\times10^{-19} \cgssb$,
resulting in \ion{He}{ii}/\lya$<0.045$, and \ion{C}{iv}/\lya$<0.039$ when using the 
average SB$_{\rm Ly\alpha}=(1.81\pm0.08)\times10^{-17} \cgssb$ within the same region.
Further, we also compute the line ratios for the detected EELR close to \qso\ and 
obtained SB$_{\rm HeII}=(7.24\pm0.35)\times10^{-18} \cgssb$, and SB$_{\rm CIV}=(1.11\pm0.05)\times10^{-17} \cgssb$,
resulting in \ion{He}{ii}/\lya$=0.12\pm0.01$, and \ion{C}{iv}/\lya$=0.18\pm0.01$ when using the 
average SB$_{\rm Ly\alpha}=(6.08\pm 0.27)\times10^{-17} \cgssb$ at this location.
Fig.~\ref{plotCivHeii} summarizes our datapoints and compares them to data in the literature for radio-quiet extended 
\lya\ nebulosities at high redshift. 
Our data for the whole ELAN are consistent with the non-detections 
usually reported in the literature for radio-quiet \lya\ nebulae (e.g., \citealt{fab+15a}). 
While the values of the EELR are in agreement with the detections in the literature 
for nebulae currently explained as powered by an enshrouded AGN (i.e. \citealt{Dey2005,Cai2016}). 

\section{Discussion}
\label{sec:disc}

In this section we first show how the velocity shear observed within the ELAN around \qso\ is remarkably similar to a rotation-like pattern and discuss the favoured interpretation
in light of the current data. Secondly, we consider in turn alternative scenarios for the presence of such a velocity shear.
Finally, we discuss the possible powering mechanisms that could give rise to such bright \lya\ emission on 100~kpc scales, with no extended \heii and \civ emission
down to our current SB limits.

\subsection{The favoured interpretation: witnessing mass assembly around a $z\sim3$ quasar}
\label{sec:favInterp}

\begin{figure*}
\begin{center}
\includegraphics[width=0.8\textwidth]{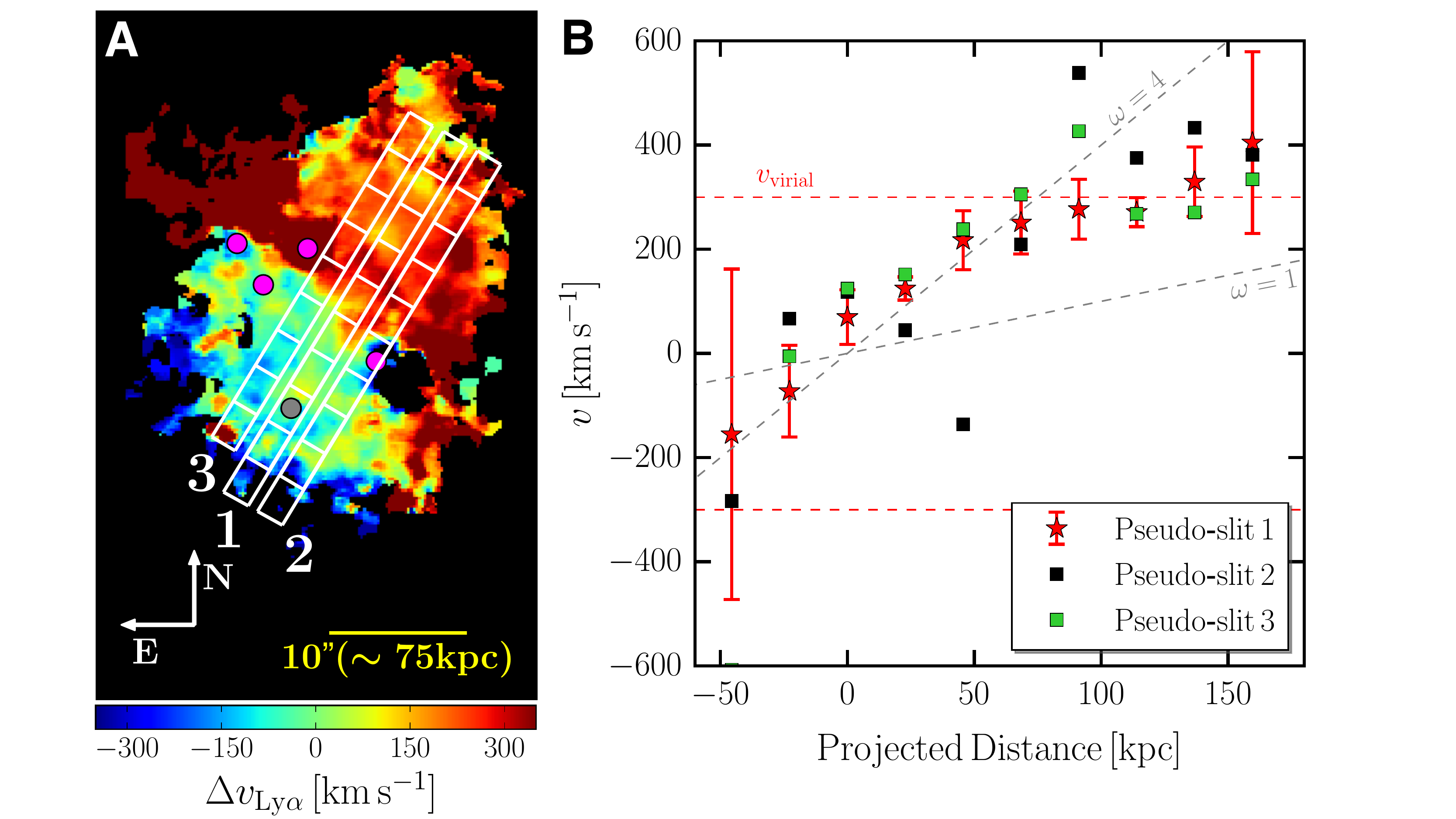}
\end{center}
\caption{{\bf \lya\ velocities along three ``pseudo-slits'' through the ELAN.} 
{\bf (A)} The 
three {$2\arcsec$-wide} pseudo-slits (long white rectangles) used in 
our analysis are overlaid on the flux-weighted velocity-shift map for the ELAN. ``Pseudo-slit 1'' intersects the position of the quasar \qso\ (grey circle), the peak of the ELAN at a distance 
of $1\arcsec$ from the quasar (at $\alpha=155.0428$~deg, $\delta=10.6679$~deg), and a 
secondary peak at 120 projected kpc from \qso.
Two additional pseudo-slits probe the same 
orientation but at symmetrically shifted positions. 
Each pseudo-slit is oriented at $149$\,deg E from N.
Along each pseudo-slit we indicate the $2\arcsec 
\times 3\arcsec$ boxes within which the \lya\ centroids 
have been calculated (see B).
The magenta circles indicate the 
position of the additional associated sources (Fig.~\ref{figsources}).
{\bf (B)} Flux-weighted velocity shift computed along the three 
pseudo-slits shown in A with respect to the position and systemic redshift of \qso. 
The errorbars show the uncertainty on the mean velocity $\sigma/\sqrt{S/N}$. 
To avoid confusion, similar errorbars are not shown for the pseudo-slits 2 and 3.  
The red dashed line indicates the 
virial velocity $v_{\rm vir}=293$~km~s$^{-1}$ for a halo of $M_{\rm 
DM}=10^{12.5}$~M$_{\odot}$.
While the grey dashed lines represent the velocity curve for 
solid-body rotation with $\omega=1$, or $4$ in units of 
$3.1\times10^{13}$~Hz.
The whole \lya\ structure seems to show a rotation-like pattern.
Analogous results are found by using Gaussian-centroided velocities (see appendix~\ref{appC}).}
\label{figchords}
\end{figure*}

We better highlight the velocity shear presented in Fig.~\ref{figlya} and section~\ref{sec:gb_lya_neb} with two complementary views.
First, we present the velocities with respect to the quasar systemic redshift along three parallel ``pseudo-slits'' 
through the ELAN. Fig.~\ref{figchords} show this test.
The central pseudo-slit (``Pseudo-slit 1'') intersects the position of \qso\  (assumed as reference), the flux-weighted centroid 
of the ELAN ($1\arcsec$ from \qso), and also a bright peak in the extended \lya\ emission at $\approx120$ projected kpc NW from \qso.
The other two pseudo-slits are offset by $15.2$~kpc (or $2\arcsec$) to either side.
This configuration has been chosen to cover the brightest parts of the extended 
\lya\ emission, while avoiding contamination from the known compact sources.

Within each pseudo-slit the nebula shows 
small flux-weighted velocity shifts at small projected distances,  
and up to $300-400$~km~s$^{-1}$ shifts at $160$ projected kpc. 
The velocity shift at large projected distances
is remarkably similar to the expected virial velocity 
($293$~km~s$^{-1}$) of the dark matter haloes that host
$z \sim 3$ quasars ($M_{\rm DM}\sim10^{12.5} M_\odot$, \citealt{white12}). 
Outlying points along each velocity curve are due to
substructures in the giant nebula, like the gas associated with the faint QSO2 along ``Pseudo-slit 2''. 
Overplotted on the figure are predictions (dashed lines) for a sphere in solid body rotation centred on the ELAN and with its angular momentum axis oriented perpendicular to the 
pseudo-slits. 
The data-points flatten out at large radii and thus do not follow such simplified models.  

\begin{figure*}
\begin{center}	
\includegraphics[width=0.75\textwidth, clip]{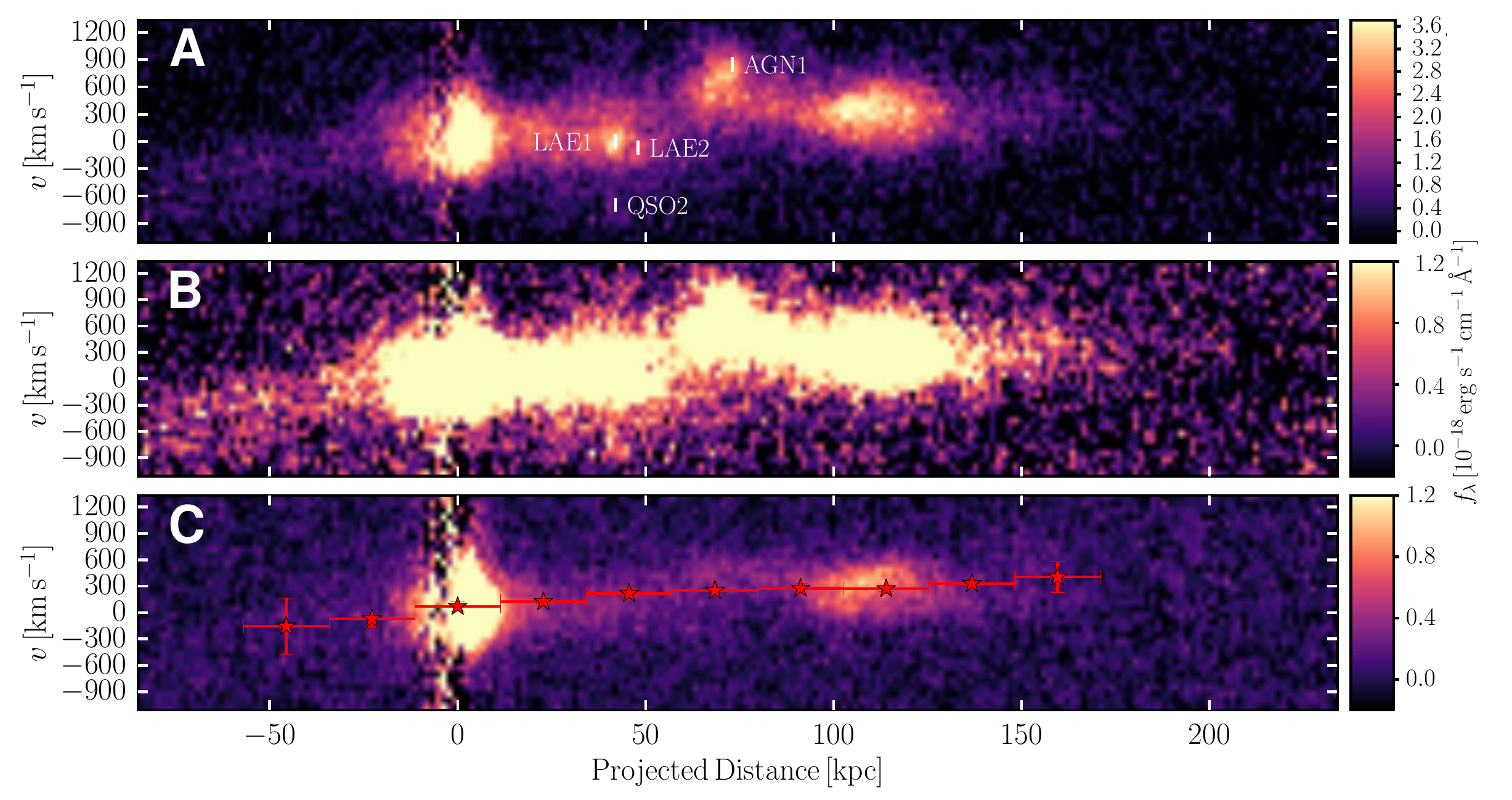}
\end{center}
\caption{{\bf Two-dimensional spectrum of the ELAN along the direction of ``Pseudo-slit 1''.} 
{\bf (A)} Two-dimensional spectrum of the ELAN obtained by collapsing the final MUSE datacube (PSF and continuum subtracted) along the direction of ``Pseudo-slit 1'' 
for a spatial region including the whole \lya\ extended emission. The position for the contribution of the associated compact sources is indicated. 
Note that along this direction LAE1 and QSO2 are at the same projected distance from the quasar \qso\ (used here as reference).  {\bf (B)} same as A, 
but with a different flux scale (same as in panel C) to emphasise the emission at lower levels. {\bf (C)} Two-dimensional spectrum of the ELAN extracted within ``Pseudo-slit 1'' 
with superimposed the data-points shown in Fig.~\ref{figchords}. Within this pseudo-slit, the velocity pattern is not 
contaminated by the contribution from the compact sources, and thus clearly visible. 
The higher noise at 0 kpc is due to the PSF subtraction of \qso\  (see section~\ref{sec:cubex}).} 
\label{TwoDspectrum}
\end{figure*}

Secondly, to fully support our analysis using pseudo-slits, in Fig.~\ref{TwoDspectrum} we show a two-dimensional spectrum obtained by collapsing the 
final MUSE datacube (PSF and continuum subtracted) along the direction of ``Pseudo-slit 1''. 
In particular, panel A and B of Fig.~\ref{TwoDspectrum} report the two-dimensional spectrum for the whole spatial range comprehending 
the ELAN (i.e. like a pseudo-slit with width of $\approx21\arcsec$). In these panels is visible the contribution of the associated 
compact sources (QSO2, LAE1, LAE2, and AGN1) to the \lya\ emission.
At positive distances their intrinsic \lya\ emission hides the signature of the rotation-like pattern on halo scales, 
but at negative distances the shear is visible.  On the other hand,  the shear at positive distances is evident 
when excluding these sources from the extracted two-dimensional spectrum. Indeed, panel C of Fig.~\ref{TwoDspectrum} 
shows the two-dimensional spectrum extracted within ``Pseudo-slit 1''  together with the data-points presented 
in Fig.~\ref{figchords} (first moment of the flux distribution). The velocity shear of the diffuse \lya\ emission extending from negative 
to positive distances is now clearly visible. Note that we change the flux scale between panel A and B to allow a comparison with panel C.

This behaviour of a monotonically increasing velocity with increasing distance, 
and a flattening at larger distances, is suggestive of a ``classical'' galaxy 
rotation curve, impacted by the presence of the underlying dark matter on large scales (e.g., \citealt{Persic1996}).  
Such a large-scale rotation-like pattern is also expected in the current paradigm of 
galaxy formation for inspiraling material within galaxy haloes (e.g., \citealt{stewart+16}).
Specifically, the fraction of cool halo gas 
mapped in \lya\ emission, is likely tracing the overall accretion motions within 
the halo, roughly centred at or near
the position of \qso. This baryonic 'rotation' is 
in turn predicted to trace the motions of the underlying dark-matter.

\begin{figure*}
\begin{center}
\includegraphics[width=0.92\textwidth]{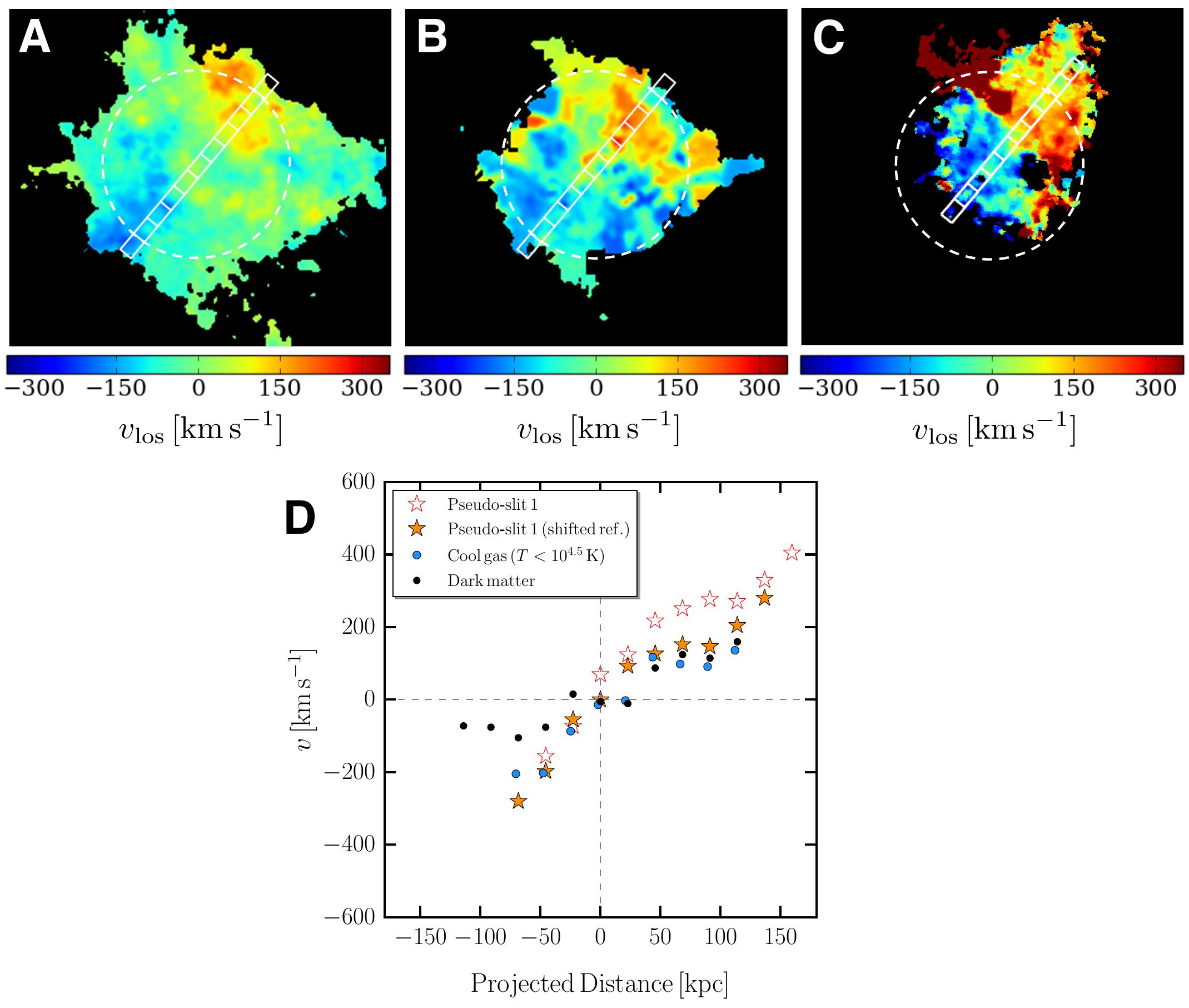}
\end{center}
\caption{{\bf Comparison of the observed \lya\ 
velocity pattern with the prediction of a particular direction to a simulated halo.}
Line-of-sight mass-weighted velocity map for {\bf (A)} the dark matter and  {\bf (B)} the cool gas ($T<10^{4.5}$~K) in our simulation (\citealt{Nihao2015}, appendix~\ref{sec:simulation}), 
smoothed to the same resolution of the observations.
The maps show the halo as seen along a direction perpendicular to the net angular momentum of the cool gas (appendix~\ref{sec:simulation}). 
In both maps, the white rectangle indicates the pseudo-slit used to extract the velocity curves presented in D.
The circles denote the virial radius of the simulated dark matter halo, $R_{200}=[3M_{\rm halo}/(800\pi \rho_{\rm crit}(z))]=101$~kpc. {\bf (C)} 
flux-weighted velocity-shift map of the ELAN adapted from Fig.~\ref{figlya}, now shown with respect to the shifted position (see D).
{\bf (D)} Simulated mass-weighted velocity shift computed along the 
pseudo-slit shown in A and B. The cool gas (blue) and dark matter (black) are compared to the flux-weighted velocity shift along ``Pseudo-slit 1'' 
(red) as from Fig.~\ref{figchords}. In 
orange we show the data-points for ``Pseudo-slit 1'' 
as if the halo centre was not co-spatial with the 
quasar \qso, but shifted by $\approx 23$~kpc (or 
$\approx 3\arcsec$) along ``Pseudo-slit 1'' and by $\approx - 125$~km~s$^{-1}$ (i.e. next
white box along ``Pseudo-slit 1'' from the quasar position in Fig.~\ref{figchords}), as shown in panel C.  
This position is also displaced from the peak of the extended \lya\ emission at $1\arcsec$ from \qso.
To avoid confusion we omit the errorbars. Note that the 
errors on the simulated data-points are much smaller than the uncertainty shown for our observations in Fig.~\ref{figchords}. 
} 
\label{figmodel}
\end{figure*}

To further explore 
this possibility, we compare our observations with a 
cosmological zoom-in simulation centred on a dark-matter halo of 
mass $M_{\rm DM} = 10^{12.29}M_\odot$ at $z\approx3$ (\citealt{Nihao2015}, see appendix~\ref{sec:simulation} for details on the
selection of the halo) close to the 
masses estimated for quasar host haloes (\citealt{white12,fanidakis13}).
In particular, as the observed \lya\ emission traces cool gas ($T\sim10^{4}$~K) (e.g., 
\citealt{hennawi+15}), we directly compare
the observed velocity shear with the velocity patterns of 
the cool gas ($T<10^{4.5}$~K) in our 
cosmological zoom-in simulation (see appendix~\ref{sec:simulation} for details on this temperature cut).
Intriguingly, it is easy to find views of the simulated halo for which 
the velocity shear is similar to what is seen in our data\footnote{We discuss the effect of resonant scattering in section~\ref{sec:kinematics}, concluding 
that they should not be able to introduce in the data a coherent gradient on scales of hundred kpc.}.  
Fig.~\ref{figmodel} shows mass-weighted line-of-sight velocity maps of the dark matter (panel A) and the cool gas (panel B) 
in our simulation. 
These maps are constructed for a direction perpendicular to 
the angular momentum axis of the cool gas in a $400$~kpc box centred on 
the minimum of the halo potential (appendix~\ref{sec:simulation}). 
Although the dark matter and the 
cool gas show similar shears of the order of hundreds km/s, the dark 
matter lags behind the baryons which are inspiraling. 

The similarity between our data (panel C in Fig.~\ref{figmodel}) and the shear present within the simulated gas can be better 
quantified by extracting the velocity curve along a pseudo-slit, analogous
to what was done with the observations. 
In particular, we select the pseudo-slit that passes through the centre of the halo 
and best probes the velocity shear of the cool halo gas as seen in this orientation (appendix~\ref{sec:simulation}).
The bottom panel of Fig.~\ref{figmodel} illustrates that the observed velocity shear (red points) agrees, within the uncertainties, 
with the simulated cool gas velocity shear (blue points)  at small and negative projected distances. 
However, the simulation seems to under-predict the line-of-sight velocities at large positive projected distances.  
This can be explained by the fact that at smaller radial distances from the centre,
the halo potential should dominate the kinematics, 
whereas halo gas at large radii (large projected distances) is expected to be
more influenced by the particular large scale configuration (e.g. mergers, filaments) 
at a given time (e.g. \citealt{More2015}). 
Variations of the signal on $100$~kpc scales are thus expected in a halo to halo comparison. In addition, the quasar \qso\ could sit in a more massive DM halo than selected here. 
Such a halo would be characterised by higher circular velocity. Further, we caution that the comparison between observations 
and simulations suffers from the uncertainty on firmly placing the centre of the halo in observations (see section~\ref{sec:HaloCenter}).
Indeed, the agreement between observations and the simulation would be improved by shifting the reference position 
for the observed data (previously placed  at \qso) towards the companion \lya\ emitters and the \lya\ nebula by $\approx23$~kpc (or $\approx3\arcsec$)
and by $-125$~km~s$^{-1}$, i.e. by one box along ``Pseudo-slit 1'' (orange points in Fig.~\ref{figmodel}). Such a better agreement would imply 
that the centre of mass of the system is not precisely located at the position of \qso.

\begin{figure}
\begin{center}
\includegraphics[width=0.97\columnwidth]{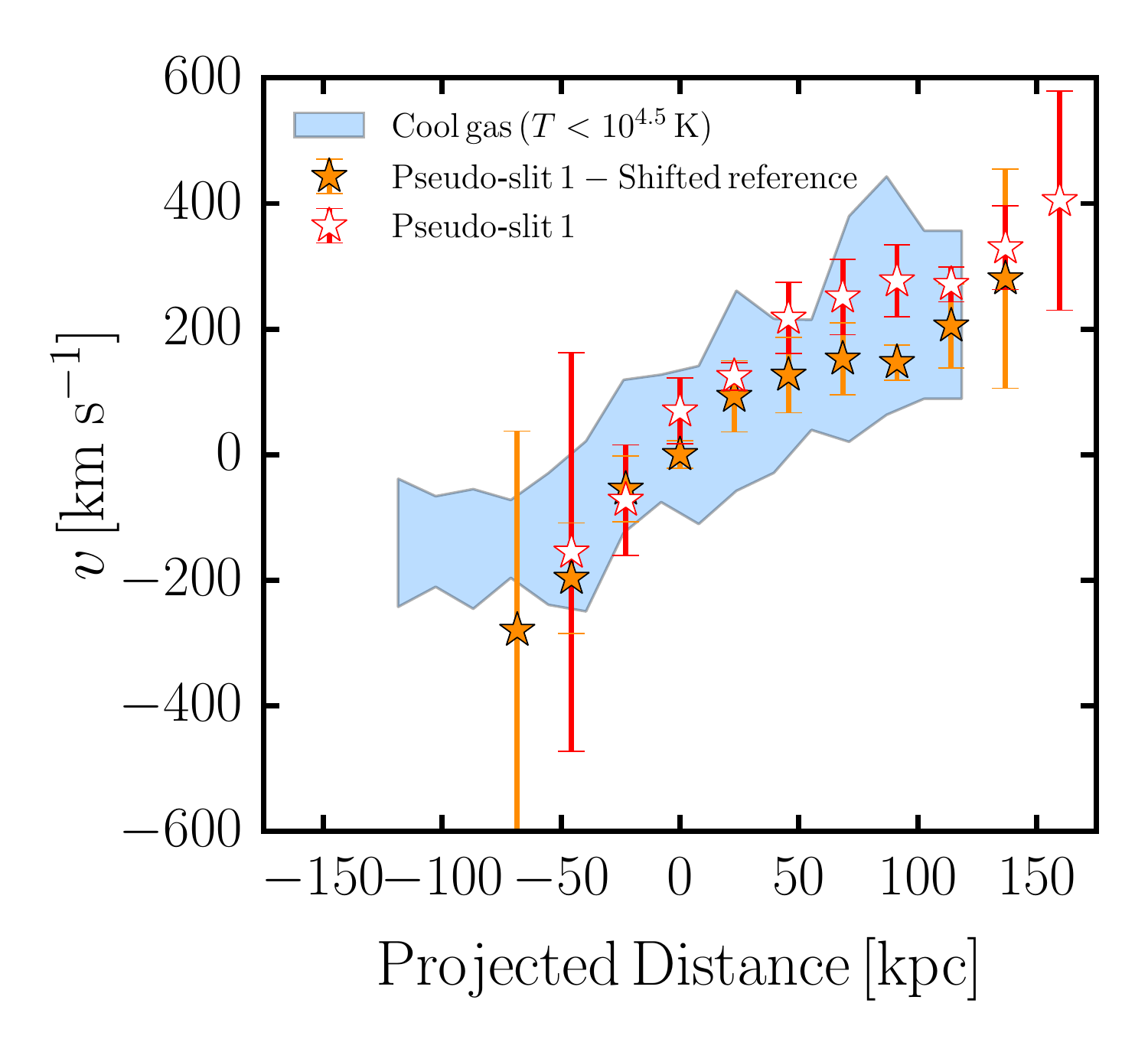}
\end{center}
\caption{{\bf Comparison of the observed \lya\ 
velocity pattern with the overall prediction of a cosmological zoom-in simulation of galaxy formation.}
The data points (red and orange) are the same as in Fig.~\ref{figmodel}. The blue shaded area indicates the region 
covered by the predicted velocity curves for directions to the simulated halo which show a clear rotation-like signal within the virial radius. 
Such directions corresponds to $\sim20$\% of the possible orientations.
Given the remarkable agreement, our data seems to reveal structure assembly onto a massive dark matter halo.}
\label{figmodel_stat}
\end{figure}

We further test this scenario by selecting all the orientations of the simulated halo which show a clear rotation-like signal within
the virial radius, and compute the velocity curves as done for the previous orientation. 
Specifically, we have generated hundred velocity maps for the cool halo gas by sampling the whole sphere. $\sim20$\% of this maps show a clear rotation-like signal
within the virial radius. We have then extracted the velocity curves within a pseudo-slit as done previously. 
In Fig.~\ref{figmodel_stat} we compare the region spanned by these simulated velocity curves drawn from different orientations with our observations, for both the data-points estimated using 
the quasar as reference (red) and the shifted reference position (orange, as in Fig.~\ref{figmodel}). 
The good agreement between the observed and the simulated velocity 
curves confirms the plausibility of our interpretation, namely that the nebular 
\lya\ emission traces motions of inspiraling baryons within the halo hosting the quasar \qso.
It is important to note that such motions can be easily probed by observations only if the bright \lya\ emission extend out to large distances (comparable
with the halo scale) from the quasar.

\subsubsection{The Position of the Centre of the Main Halo}
\label{sec:HaloCenter}

As discussed in the previous section, our interpretation of the velocity pattern as  
the signature of inspiraling structures within the quasar's halo requires the centre of the dark-matter halo to be in close 
proximity to \qso. 
This hypothesis is supported by several pieces of evidence.
First of all, clustering studies place SDSS quasars at these redshifts in very 
massive dark matter haloes $M_{\rm halo}\approx 10^{12.5}\, M_{\odot}$ 
(\citealt{white12}), at least $30\times$ more massive than the haloes hosting LAEs at the same redshift, $M_{\rm halo}=10^{11\pm1}M_{\odot}$ (e.g., \citealt{Ouchi2010}). 
Also, bright quasars at $z=2.7$ seem to sit in $M_{\rm halo} = 10^{12.3\pm0.5}\, M_{\odot}$ and to reside in groups of galaxies (\citealt{Trainor2012}).
In addition, the two-point correlation function of quasars (combining SDSS DR7 quasars and the sample in \citealt{Hennawi2006}) requires only a very small fraction
of quasars to be satellites ($f_{\rm sat}\sim10^{-4}$ at $z=1.4$) to explain their small-scale clustering, when interpreted in the framework 
of the halo occupation distribution (\citealt{richardson12}).
It is thus plausible that the halo of \qso\ acts as 
the main halo in its overdense environment, with the fainter objects being interacting 
structures or satellites. Intriguingly, note that the most favoured position of the 
centre of mass from the comparison with the direction perpendicular to the net angular momentum of the cool gas in the simulation (Fig.~\ref{figmodel}) would place the two companions 
AGN, QSO2 and AGN1, at symmetric velocity shift, i.e. $-701\pm450$~km~s$^{-1}$ and 
$739\pm72$~km~s$^{-1}$ respectively. Keeping in mind the uncertainties on the redshift determination for the two QSOs, these shifts could then reflect the peculiar 
velocities of the two AGN with respect to \qso\ within a massive structure.

Secondly, the \lya\ nebula is clearly associated with \qso, being at its systemic redshift and 
showing its maximum near the quasar position (at $\approx 1\arcsec$, Fig.~\ref{figlya}). It is indeed plausible that the \lya\ emission 
shows its maximum in proximity of the quasar where higher densities are expected. For example,  
the existence of a 
density profile within CGM with higher densities near the central galaxy has been proposed  by
\citet{Stern2016}, and inferred with the absorption technique in \citet{QPQ8}.

Even though the aforementioned hints are compelling, a firm determination of the halo 
centre would require challenging observations. Indeed, an additional evidence could be 
obtained by conducting a deep spectral-imaging survey  (e.g. deeper IFU observations, submm data to estimate the redshift of faint or obscured companions)
to characterise the velocity distribution of 
galaxies around \qso. If such a distribution clearly peaks around the quasar's systemic redshift and 
no comparable massive galaxies are found, 
\qso\ would then be considered as 
dominant.  Obvious to say, this approach would need a statistical sample of satellites to  accurately compute a velocity distribution.
In addition to this approach, if \qso\ is indeed at the centre of a massive group, 
very deep X-ray observations might be able to constrain the emission from the diffuse 
hot-phase. If the peak of such emission (after removal of the intrinsic quasar's X-rays) appears to be close to \qso\ our hypothesis would 
then be definitely strengthened.

Our overall approach reflects the techniques used at low redshift to constrain the 
properties of galaxy clusters, where the most massive galaxy has been often used as the 
centre of 
mass when other information were still not available (e.g., \citealt{KentGunn1982}).
Finally, note that a shift of  the brightest object of a group from the centre of the mass distribution is frequently seen at low redshift.
For example, the brightest cluster galaxy (BCG) in the prototypical strongly lensed massive clusters A383 ($z=0.1887$) 
is displaced from the cluster centre by tens of kpc ($21\pm56$~kpc) and tens of km~s$^{-1}$ ($-11\pm110$~km~s$^{-1}$) (\citealt{Geller2014}).

\subsection{Alternate Scenarios for the Observed Velocity Shear within the ELAN}
\label{sec:scenShear}

In the previous section, we argue that the velocity shear within the ELAN most 
likely traces the kinematics of the cool gas, which is expected to show a rotation-like 
pattern while accreting onto a dark matter halo, as shown in current structure-formation theories and cosmological 
simulations. 
The velocity offsets detected within the gaseous structure and the association of three AGN with the ELAN, 
together with their large velocity offsets from \qso, thus seems to reflect 
the gravitational motions within a massive structure (e.g., \citealt{Miley2006,hennawi+15}).
However, so far, we have neglected other possible mechanisms that might have shaped the \lya\ emission around \qso\ 
 (e.g., resonant scattering of \lya\ photons).

In the next sections we thus discuss the alternate scenarios that could reproduce the observed velocity shear, which we conclude  
seems disfavoured given the current data. 
However, we emphasise that only future follow-up observations compared against future zoom-in cosmological simulations 
would be able to definitively rule out some of these scenarios.

\subsubsection{Resonant Scattering within the ELAN} 
\label{sec:kinematics}

The resonant nature of the \lya\ transition makes challenging, in most of the cases, its use as a
tracer of the kinematics in astrophysical observations (e.g., \citealt{Neufeld_1990}, and references thereafter). 
In particular, a \lya\ photon typically experiences several scatterings before escaping the system in which it is produced because of the high opacity at line centre.
As \lya\ photons must diffuse into the wings of the line to leave the system (e.g., \citealt{Neufeld_1990, Cantalupo2005}), 
double-peaked emission line profiles are thus expected especially for high neutral hydrogen column densities.
In addition, because of the high number of scatterings, the emergent \lya\ line profile could be also affected by the amount of dust and its particular
distribution within the system (e.g., \citealt{Neufeld_1990,Duval2014}). 
Further, infalling or outflowing gas on galaxy scales have been shown to imprint a distinctive feature to the \lya\ line profile, 
with absorption of the red or blue side of the line, respectively (e.g., \citealt{Verhamme2006}).
Therefore, the line profile is not expected to follow a simple analytical function, and additional non-resonant 
diagnostics are usually required to firmly characterise the motions within a system (e.g., \citealt{Yang2014,Prescott2015}). 
Notwithstanding these challenges, the \lya\ line is the brightest and, in most of the cases, the only detected emission from the diffuse CGM and IGM, 
and hence it provides a unique opportunity to study the kinematics of these diffuse gas phases on very large scales, once the resonant scattering effects are taken into account.

\begin{figure}
\begin{center}
	\includegraphics[width=1.02\columnwidth, clip]{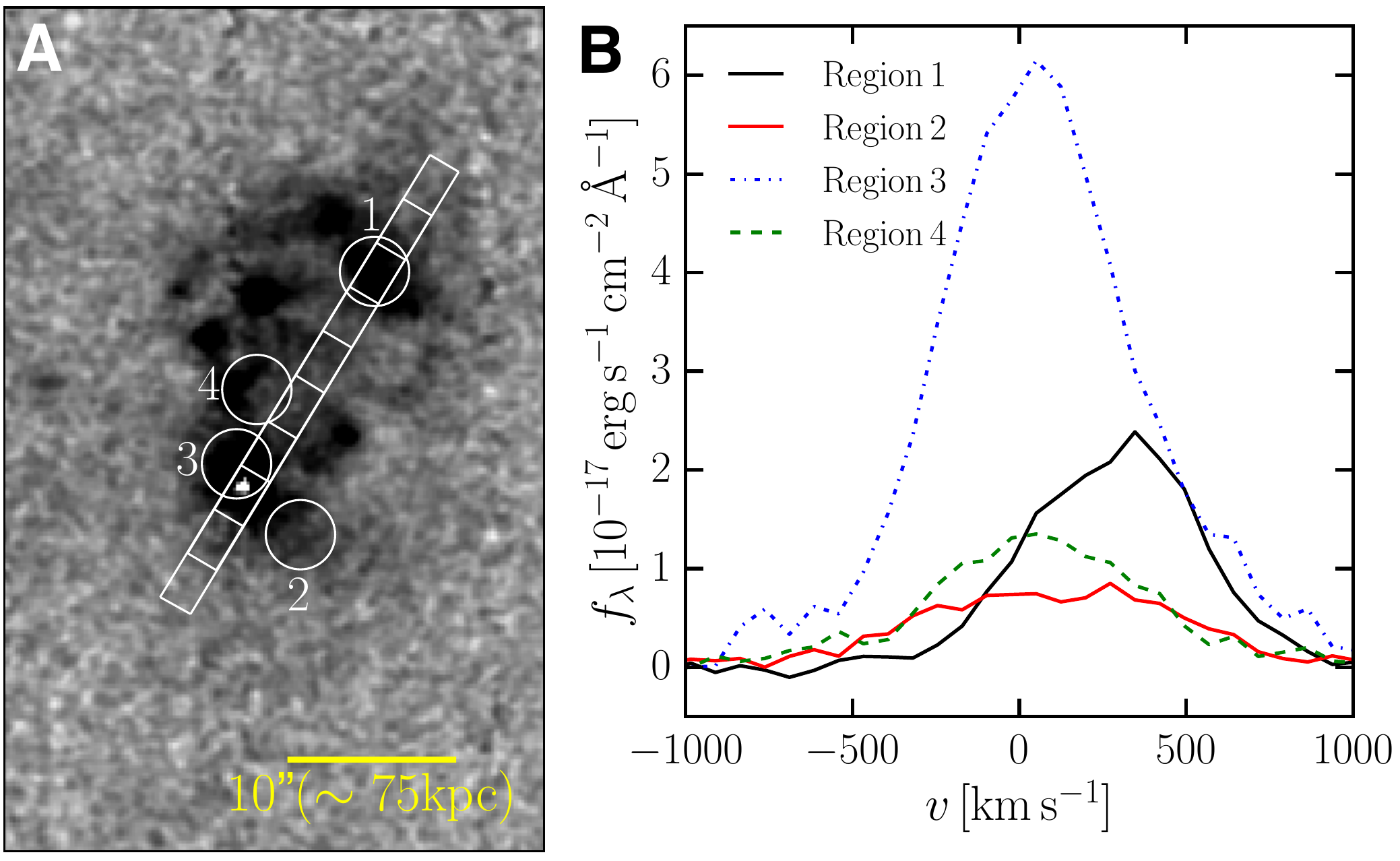}
\end{center}
\caption{{\bf \lya\ emission-line shapes for four representative regions within the ELAN.} {\bf (A)} NB image of the \lya\ nebula around \qso\ after PSF subtraction. 
The position of 4 regions (circles with $2\arcsec$ radius) and of the slit used in Fig.~\ref{figchords} are shown for comparison. {\bf (B)} \lya\ line 
spectra of the nebula extracted within the region shown in the left panel. Note that at the resolution of our
    observations (FWHM~$\approx170$~km~s$^{-1}$ at 5000\AA), there is no evidence for double peak profiles expected for resonantly trapped \lya\ emission. 
    Note also the velocity shift of $\approx300$~km~s$^{-1}$ from the quasar systemic redshift for Region 1, as expected from our analysis.}  
\label{LyaLineShape}
\end{figure}

Not being able to rely on non-resonant emission lines (\heii is not detected, section~\ref{sec:heiiciv}), to assess the importance of resonant scattering of \lya\ photons within the ELAN here studied, 
we have thus carefully inspected the \lya\ line shape.
First, our test in appendix~\ref{sec:kubeviz} shows that the moment analysis presented in section~\ref{sec:cubex} is in complete agreement 
with a Gaussian fit for all the extent of the ELAN, resulting in fully compatible maps (Fig.~\ref{QdebKub}). This agreement has been further demonstrated in Fig.~\ref{figchordsSup} 
for the emission within ``Pseudo-slit 1''. As an additional check, in Fig.~\ref{LyaLineShape} we show the \lya\ emission line shape in four circular regions of radius $2\arcsec$.  
We choose these regions because they are representative of the different velocity dispersions that are present in Fig.~\ref{figlya}, 
spanning more quiescent (Region 1 and 4) and active (Region
2 and 3) portions, while covering the range of distances from \qso\ within the ELAN. 
Indeed, from Fig.~\ref{LyaLineShape} it is clear that all regions show in first approximation Gaussian lines (detected at high significance), 
with regions 2 and 3 presenting wider emission (FWHM$\approx1000$~km~s$^{-1}$), 
while regions 1 and 4 more quiescent kinematics (FWHM$\lesssim600$~km~s$^{-1}$). 
Therefore, we conclude that our data are well approximated by a Gaussian down to the MUSE spectral resolution of FWHM$\approx2.83$\AA, 
or $\approx170$~km~s$^{-1}$ at 5000\AA\ (i.e. close to the \lya\ line wavelength).
Further, it is interesting to note that the estimate for the overall widths within the \lya\ structure are comparable to the velocity widths observed in
absorption in the CGM surrounding $z\sim2$ quasars ($\Delta v\approx 500$~km~s$^{-1}$; \citealt{QPQ3, QPQ8}). 
Both the emission and absorption kinematics are
comparable to the virial velocity $\sim300$~km~s$^{-1}$ of the massive dark matter haloes hosting quasars (M$_{\rm DM}\sim
10^{12.5}$~M$_{\odot}$; \citealt{white12}).

All these evidences suggest that resonant scattering of \lya\ photons do {\it not} play an important role in this system as opposed to intrinsic motions. 
Even though resonant scattering effects could be in place on small scales ($10$~kpc) 
and be hidden at the current spectral resolution (\citealt{Verhamme2012}) (e.g., ``double-peaked'' profiles, 
and/or faint absorption from low column density gas), we argue that on the much larger scales spanned by the Ly$\alpha$ nebula, 
resonant scattering seems to be negligible. Indeed, it has been demonstrated that the resonant scattering process results in a very efficient 
diffusion in velocity space, such that the vast majority of resonantly scattered photons produced at a certain location should escape the system 
after propagating for only small distances ($\lesssim10$~kpc; \citealt{Dijkstra_2006,Verhamme2006,cantalupo14}).  
Interestingly, to our knowledge, in all 
the currently known large-scale radio-quiet Ly$\alpha$ nebulosities where both Ly$\alpha$ emission and the 
non-resonant HeII emission are detected on $>50$~kpc scales (\citealt{Prescott2015, Cai2016}), the two lines show the 
same shapes, suggesting that the Ly$\alpha$ line on such large scales (hundreds of kpc) might trace the kinematics of the gas just as well.
For these reasons we thus claim that the extended \lya\ emission around the quasar \qso\ can be used to trace the cool gas motions on halo scales. 
Further, for the same reasons, we argue that resonant scattering effects are not able to produce the $\approx300$~km~s$^{-1}$ coherent velocity shear observed 
within the ELAN on hundreds of kpc.
However, future observations through additional diagnostics (e.g. polarization, higher resolution spectroscopy) are needed to firmly confirm the low importance of scattering within this system.

\subsubsection{Two Independent ``Blobs'' at Two Different Redshifts within a Large-scale Structure in Projection}
\label{sec:TwoBlobs}

\begin{figure}
\begin{center}
	\includegraphics[width=1.0\columnwidth, clip]{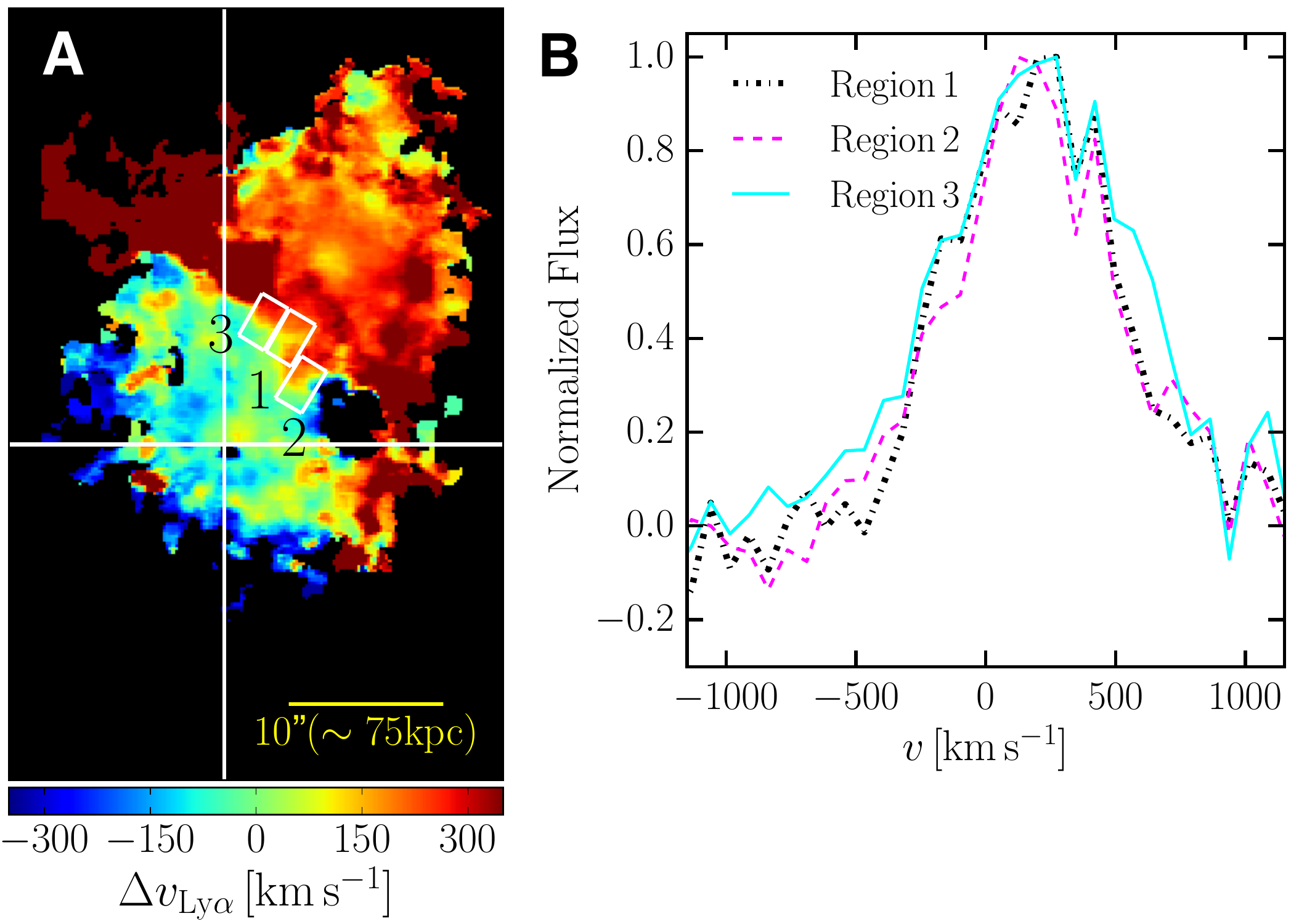}
\end{center}
\caption{{\bf \lya\ line shape at the ``boundary''.} {\bf (A)} The 
three $2\arcsec 
\times 3\arcsec$ boxes within which the \lya\ spectra have been extracted are 
overlaid 
on the flux-weighted centroid map for the \lya\ emission of the giant 
nebula (panel B of Fig.~\ref{figlya}). The large white cross indicates the position of the 
quasar \qso\ before PSF subtraction.
{\bf (B)} \lya\ emission-line within the three regions shown in panel A.
In each region, the \lya\ line do not show two distinctive peaks at two different redshifts (at $0$ and $300$~km~s$^{-1}$) 
expected in the case of two independent structures in projection. Note however, that there is a 
fainter and very narrow peak at $\approx420$~km~s$^{-1}$ in all of the spectra. Higher resolution data are needed to study the nature of this feature.}
\label{shapeLyaBorder}
\end{figure}

A first look at the velocity map in Fig.~\ref{figlya} could give the impression of 
two independent nebulae with different redshifts, i.e. one at the redshift of the quasar \qso\ 
and the other in the NW direction at $\approx300$~km~s$^{-1}$ (or $\Delta 
z\approx0.004$), separated at the ``boundary''. 
The velocity shift between the two nebulae could be interpreted as a large-scale structure in projection with a maximum extent of at least $600$~kpc (or a comoving distance of 
$\approx2.53\, h^{-1}$~Mpc), when converting the velocity difference into distance assuming the structure to be in the Hubble flow.
If this is indeed the case, to our knowledge, this structure could then represent the largest cosmic-web patch traced in \lya\ to date.
Even though we consider such a scenario of great interest, as we explicitly conduct our surveys (\citealt{fab+16} and QSO MUSEUM, Arrigoni Battaia et al., in prep.) 
to search for large-scale structures with the hope of directly detect the IGM, we argue that several lines of evidence 
contradict this interpretation. We list them in what follows.\\
1) The peak of the \lya\ emission continuously shifts in velocity with distance from the quasar \qso\ (Fig.~\ref{figchords}, Fig.~\ref{TwoDspectrum})  
and does {\it not} show any evidence for double-peaked \lya\ line profiles  at the ``boundary''. In other words, the line profiles do not show 
two distinct peaks separated by $300$~km~s$^{-1}$ (and thus at two different redshfits),  which would be clearly distinguishable at the MUSE 
spectral resolution and because of the small width of the \lya\ line within the ELAN.
This can be once again appreciated in Fig.~\ref{shapeLyaBorder} where we show  the \lya\ profiles in three boxes of $2\arcsec \times
3\arcsec$\footnote{Different sizes (e.g. boxes of $0.66\arcsec \times 5 \arcsec$) give similar results.} at this location (the ``boundary''), 
which one might assume to be the region where the two nebulae overlap. 
In all the apertures, the \lya\ line is characterised by a single peak slightly redshifted from the quasar systemic, as expected from our overall analysis. 
As also demonstrated by the Gaussian fit analysis (appendix~\ref{sec:kubeviz}), the same exercise do not reveal any double-peaked \lya\ profile in any region within the ELAN.
If what we see are two independent nebulosities, the bright emission from these two structures thus do not overlap, but has to stop exactly where they touch in projection. \\
2)  In the framework of two independent nebulosities, it would be difficult to explain the symmetrical gradual shift in velocities at both positive and negative projected 
distances ($-50<d<50$~kpc) from the quasar \qso\ (Fig.~\ref{figchords}, Fig.~\ref{TwoDspectrum}), and justify at the same time a separation of these signatures with 
respect to the coherent redshifted NW portion at higher positive distances without involving any kinematics (e.g. rotation-like and accretion).  \\
3)  In a two nebulae scenario, projection effects due to our vantage point conspire to perfectly mimic the expected kinematics and sizes of a dark matter halo hosting quasars. 
Indeed, it is intriguing that the \lya\ emission is detected only within  $\approx170$~projected kpc and with velocity shifts $<400$~km~s$^{-1}$, when the expected virial 
radius for a quasar is $\approx160$~kpc (\citealt{QPQ7}) and the expected virial velocity is $v_{\rm vir}=293$~km~s$^{-1}$ ($M_{\rm DM}=10^{12.5}$~M$_{\odot}$ \citealt{white12}). 
Such mimicking seems highly improbable for at least two reasons.   First, the observed velocities along the line-of-sight within a large-scale structure in projection 
would be greatly affected by the flow velocities of the gas within the structure itself. Indeed the gas within filaments in haloes is expected to flow with velocities 
$\sim200$~km~s$^{-1}$ (\citealt{Dekel2009,Goerdt2015}).  On top of this, resonant scattering effects of \lya\ 
photons may greatly affect any such structure aligned along the line-of-sight, as \lya\ photons have to pass through the structure itself to reach the observer.  
Distance information would therefore be highly distorted, especially for 
such a large 
reservoir of neutral hydrogen.
 \\
4)  If the structure is along the line of sight, the hard ionizing radiation of the quasar \qso\ has to impinge on the gas. Indeed, in accord with 
unified models of AGN (e.g. \citealt{Anton93}) quasars should emit in roughly symmetrical ionization cones. Being un-obscured along our line-of-sight, one 
should expect the quasar to be un-obscured also in the opposite direction.
If this is the case, given the high luminosity of \qso, photoionization would be undoubtedly the main powering mechanism for the whole extended 
emission (\citealt{cantalupo14,hennawi+15,fab+15b}).
In this scenario, the observed high \lya\ surface brightness and the absence of extended \heii require implausibly high densities ($n_{\rm H}>100$~cm$^{-3}$) 
within at least $350$~kpc spanned by the ELAN in the Hubble flow.
Such high density values would be in stark contrast with current  cosmological simulations which predict $n_{\rm H}\sim10^{-2}-10^{-5}$~cm$^{-3}$ in the CGM 
and IGM (e.g. \citealt{Rosdahl12}).
On top of this, a photoionization scenario would predict a smooth transition between optically thin and thick gas (to ionizing radiation) while moving from small to large 
distances from the quasar.  As the \lya\ surface brightness is proportional to $n_{\rm H}$, and column density $N_{\rm H}$ in the optically thin regime 
(i.e.  SB$_{\rm Lya}^{\rm thin}\propto n_{\rm H}N_{\rm H}$), while only to the ionizing luminosity in the optically thick regime 
(i.e.  SB$_{\rm Lya}^{\rm thick}\propto L_{\rm \nu}$) (\citealt{qpq4}), 
this scenario would imply a tuning between the different values so that the transition happens while roughly preserving 
the same level of observed \lya\ surface brightness in the whole extent of the ELAN (Fig.~\ref{figlya}). 
We discuss this analysis in detail in appendix~\ref{sec:unlikeCase}.

Notwithstanding these arguments, additional observations are needed to completely rule out this scenario, 
i.e. polarization study of the \lya\ emission (determination of the powering mechanism), 
observations in other gas tracers to confirm the \lya\ line shape (e.g. deeper MUSE observations, deep observations targeting the H$\alpha$ line), 
or observations with a higher spectral resolution (importance of scattering).

\subsubsection{A Large Rotating Disc}
\label{sec:LargeDisk}

A large rotating disc (in a dark-matter halo with $R=225$~kpc, $M_{\rm halo}=10^{13.1}M_{\odot}$, circular 
velocity $v_{\rm c} = 500$~km~s$^{-1}$) has been proposed to interpret the ELAN around the radio-quiet quasar UM~287 (\citealt{martin+15}). 
We thus discuss the same scenario for our observations, even though we consider the 
presence of an ordered disc 
extremely implausible given (i) the presence of \qso\ 
and two strong LAEs at the same redshift and embedded within the ELAN, 
and (ii) the morphology of the nebula, in which there are clear substructures.
Indeed a disc -- unlike accreting substructures from different directions -- is a coherent structure flattened on one plane and supported by rotation in the same plane. 
Hence, in the case of a disc, one would expect the largest and smallest velocity shifts along
the major axis (the pseudo-slit direction in our case), while the velocity should drop to zero as we go from the major to minor axis.

In such a scenario, the disc centre must 
be placed at the ``boundary'' (see Fig.~\ref{figlya}).
If one assumes the disc is sitting in a massive halo like in \citet{martin+15}, 
circular velocities as high as 400~km~s$^{-1}$ are expected at about 60~kpc. Given that the observed maximum velocity shift 
from the boundary is $\approx150$~km~s$^{-1}$,  an almost face-on disc would be required with an inclination $i<20^{\circ}$.
Indeed, the observed line-of-sight velocity within the \lya\ ``disc'' would depend on the inclination of the disc itself with a $\sin\,i$ term. 
The morphology of the \lya\ emission and the presence of the embedded sources run counter to this interpretation.
On the other hand, if we assume an almost edge-on \lya\ disc, the observed velocities would imply the presence of a less massive 
system $M\approx10^{11.32} M_{\odot}$, in strong contrast with the predicted average mass for quasar haloes (\citealt{white12}).
In addition, once again, it seems implausible that the \lya\ morphology is associated to a stable edge-on disc (see also Fig.~\ref{TwoDspectrum}).

\subsubsection{A Large-scale Outflow Driven by Quasar or Star-formation Activity}
\label{sec:outflowQSO}

Cosmological simulations of galaxy formation usually invoke the presence of AGN 
feedback to reproduce the observed properties of massive systems, in particular to 
not over-predict the stellar mass, (e.g., \citealt{Sijacki2007,Schaye2015}). 
This is because in principle  supermassive black-holes (SMBHs) have enough 
energy to be coupled  with the surrounding gas in a strong wind with velocities 
grater then the escape velocity of even the most massive galaxies (\citealt{SilkRees98}).  
SMBHs should thus quench star-formation by disrupting the gas reservoir within galaxies.
In addition, star-formation driven winds are expected to shape the galaxy properties especially at 
the low mass end of the galaxy population, where the injected supernovae energy is large enough to overcome 
the galaxy potential (e.g., \citealt{DekelSilk1986,Scannapieco2008}).
Both these feedback mechanisms are predicted to 
heavily affect the gas distribution and properties on hundred kpc scales, especially when 
implementations with a strong coupling are used (e.g., \citealt{Vogelsberger2014}). 
Such outflows would result in high velocity shifts and velocity dispersions clearly 
observable with current instrumentations. We here compare our data with such a scenario. 
However, given the large uncertainties on the current feedback implementations and modelling of outflows, 
we prefer to compare our observations with current data in the literature. 

The great effort in searching for AGN or star-formation driven outflows (or winds) 
has so far resulted in evidence for their presence on maximum ten kpc scales for radio-quiet 
objects at both low and intermediate redshifts (e.g., \citealt{Nesvadba2011,Harrison2014,Kakkad2016,Steidel2010, Rubin2014}). 
Hundreds kpc outflows have only been reported around HzRGs, where a strong radio-jet is able to displace the surrounding gas (e.g., \citealt{Swinbank2015}). 
In general, on kpc scales star-formation driven winds show lower peak velocities and velocity dispersions ($\sim300$~km~s$^{-1}$)  than AGN powered outflows ($\sim2000$~km~s$^{-1}$).
Importantly, even some of the most luminous galaxies at high redshift, such as the Ultraluminous Infrared Galaxies (ULIRG), 
show outflows in the ionized phase only out to few kpc ($\lesssim 15$~kpc) and with velocity up to $\sim1000$~km~s$^{-1}$ (e.g. \citealt{Harrison2012}), 
with faster outflows likely powered in the presence of an AGN.
On the other hand, the large scale (hundreds of kpc) outflows around HzRGs present velocity shifts and FWHM of $>1000$~km~s$^{-1}$. 
These outflows are often aligned with the radio axis and represent the highest surface brightness part of the extended \lya\ emission around HzRGs (\citealt{VillarMartin2003}).

This large body of observations are clearly at odds with our dataset. 
Indeed, given that all the embedded sources lack evidence for large displacements 
of the \lya\ line peak, the relatively quiescent and continuously rising kinematics 
traced by the \lya\ line around \qso\  can {\it not} be easily reconciled with an AGN wind scenario. 
The system around \qso\ is thus substantially different from the ELAN studied in \citet{Cai2016}, 
and currently interpreted as powered by an obscured QSO. 
Indeed, in that system, two velocity components are observed with a velocity shift of $\approx700$~km~s$^{-1}$ on $\gtrsim30$~kpc 
scales in both Ly$\alpha$, HeII and CIV, all presenting FWHM~$\approx 700 - 1000$~km~s$^{-1}$. The presence of these additional 
emission lines besides \lya\ clearly invoke the presence of a fast shock or an embedded hard ionizing source.

Further, the large size of the ELAN around \qso\ together with the small velocity dispersion 
throughout its extent with a continuous peak displacement (an ordered flow pattern), seems to disfavour 
a star-formation-driven wind. Indeed such a scenario would require a coherent high energy input from several 
coeval supernovae to sustain a massive wind for hundreds of kpc within a massive system. Even though supernova 
driven superwinds (with velocities up to 1000~km~s$^{-1}$) have been theorized to explain radio-quiet giant \lya\ nebulae 
(e.g. \citealt{TanShi2000}), a growing body of observations are in stark contrast with such a scenario (e.g. \citealt{Yang2014}). 
In addition, even a shock with low velocities ($\sim 100$~km~s$^{-1}$) may result in the production of hard ionizing photons 
and thus in detectable extended emission in the \heii and \civ lines within our sensitivity limits (\citealt{Allen2008,fab+15a}). 
Our upper limits on \ion{C}{iv}/\lya\ and \ion{He}{ii}/\lya\  (section~\ref{sec:heiiciv}) appear to be in contrast with the ratios 
expected (Figure~13 in \citealt{fab+15a}), and thus disfavour such a scenario.  
However, while we argue against a wind scenario to reproduce the velocity shear on hundreds of kpc, 
we cannot exclude the presence of both feedback effects on small scales (as usually reported in the literature) 
in close proximity to the compact sources embedded within the ELAN.

Finally, we stress the similarity of the kinematics traced within this ELAN around \qso\ with the  quiescent 
haloes observed around HzRGs (e.g., \citealt{VillarMartin2003}). 
Indeed, HzRGs not only present highly disturbed kinematics, but also the presence of low surface brightness haloes with 
velocity shifts and FWHM of few hundreds of km~s$^{-1}$. This emission is usually interpreted to be associated with the 
cool reservoir within a quiescent host halo not  perturbed by the radio activity of the HzRG (\citealt{VillarMartin2003}).

\subsection{Powering Mechanisms and Physical Properties of the Emitting Gas}
\label{sec:PowMech}

Our observations reveal emission from only the \lya\ transition from the ELAN associated with the quasar \qso.
Because several mechanisms could in principle act together to produce \lya\ emission on hundreds of kpc scales (section~\ref{sec:heiiciv}), 
characterizing the physical properties of the emitting gas is a problem of complexity, and only {\it ad-hoc} 
comprehensive simulations of massive haloes would shed light on the nature of bright and giant \lya\ nebulae, once the current 
computational problems would be bypassed (e.g. \citealt{cantalupo14, fab+15a, hennawi+15, McCourt2016} and appendix~\ref{sec:simulation}). 
Here we briefly discuss which mechanisms could be in play, and what physical properties are then expected.

The dominant powering mechanism has to reproduce the high level of the observed surface brightness of the \lya\ line 
SB${_{\rm Ly\alpha}\sim10^{-17}\, \cgssb}$ on scales of $\sim100$~kpc, together with its kinematics. Specifically, 
the ELAN around \qso\ shows overall quiescent kinematics (FWHM$\lesssim600$~km~s$^{-1}$), 
with more disturbed and active ones (larger velocity dispersion) in proximity of the three AGN, and a rotation-like pattern.

Such quiescent kinematics has been seen in several extended \lya\ 
nebulae discovered so far around radio-quiet quasars for which we have spectroscopic information (\citealt{Martin2014a, hennawi+15, Borisova2016}).
All these nebulosities have been overall interpreted as powered by fluorescence emission, i.e. recombination radiation boosted by the quasar ionizing 
radiation (e.g., \citealt{Cantalupo2005,kollmeier10}). 
However, if the luminous quasar does not shine directly on the surrounding gas (e.g. depending on its opening angle and orientation), then alternative powering mechanisms, 
i.e. photoionization from star-formation, shocks from superwinds (e.g., \citealt{TanShi2000, Mori2004}), scattering of \lya\ photons (e.g.,\citealt{Dijkstra2008}), 
and ``cooling radiation'' (e.g., \citealt{Haiman2000, Rosdahl12}), could still be relevant. 
In the case of \qso, we have already discussed the evidence against the 
presence of superwinds or a large contribution from resonant scattering, concluding that these processes could only be effective in altering the ELAN properties on small scales 
($\lesssim10$~kpc) (section~\ref{sec:scenShear}). In the remainder of this section we briefly consider ``cooling radiation'' and photoionization in the case of \qso.

\subsubsection{Cooling radiation}

``Cooling radiation'', i.e. collisional excitation driven \lya\ emission from gravitational 
accretion, is often invoked to explain extended \lya\ emission (e.g., \citealt{Fardal2001,Rosdahl12}).
The strength of such emission is largely ``controlled'' by the collisional excitation coefficient given by 
${C_{\rm Ly\alpha}=3.7\times10^{-17}}$ ${{\rm exp}(-h\nu/k_{\rm B}T)/T^{1/2}}$~erg~s$^{-1}$~cm$^3$ (\citealt{Osterbrock1989}), where $h$ is the planck constant and $k_{\rm
B}$ is the Boltzmann constant. 
Given the exponential dependence on temperature of $C_{\rm Ly\alpha}$ and (being a collisional process) on gas density squared (in the ionized case), this mechanism requires a 
``fine tuning'' between the temperature and density of the gas to reproduce 
the observed bright SB$_{\rm Ly\alpha}$, while taking into account the detail balance of heating and cooling 
within the gas itself. 
\citet{Dijkstra2009_hd} argue, while modelling LABs in an analytical way, that if $\gtrsim10$\% of the change in gravitational binding energy in 
ISM-like dense ($n_{\rm H}\gtrsim1$~cm$^{-3}$) cold ($T\sim10^4$~K) flows goes into heating of the gas, then cold flows in massive haloes ($M_{\rm halo}~10^{12}-10^{13}$~M$_{\odot}$)
would be detectable as LABs.  
The same picture, but with lower densities, has been reproduced with cosmological simulations (\citealt{Furlanetto05, FG2010, Rosdahl12}).  
These works show how gas with $n_{\rm H}\gtrsim0.3$~cm$^{-3}$ (the CGM for those simulations), whose emission is dominated by collisional excitation, 
accounts for 40\% of the total \lya\ luminosity (\citealt{Rosdahl12}), and could be detected as a LAB.
The high gas densities in the cold flows are caused by the confinement due to the presence of hot gas within the virial radius of the system
(\citealt{Dijkstra2009_hd}). For this reason, it is expected that outside the virial radius the emission powered by ``cooling radiation'' declines considerably as 
no hot gas is present  
and the cold flows would then be much more
rarefied (\citealt{Dijkstra2009_hd, Dekel2009, keres09a}). 

One could speculate that cooling radiation is indeed what we see around \qso. Indeed, in the favoured scenario (i.e., inspiraling cool gas), we probe the quasar halo, and  
therefore the \lya\ emission does not extend much beyond the expected virial radius. 
In addition, the infall of the gas is expected to occur with coherent velocities of the order of 
$200$~km~s$^{-1}$ (e.g., \citealt{Dekel2009,Goerdt2015}),
which are consistent with the observed velocity dispersion.
Further, in this scenario, our non detections on large scales for the \heii and \civ lines would reflect the low 
signal expected in \heii 
in massive haloes (SB$_{\rm HeII}\sim10^{-20}\, \cgssb$; \citealt{Yang2006}), 
and the low metallicity expected for inflowing gas (but note \citealt{QPQ8}), respectively.

\subsubsection{Photoionization}
\label{sec:photo}

Even though the emission around \qso\ seems to be interpretable in light of a ``cooling scenario'' in the aforementioned
qualitative way, the high \lya\ surface brightnesses on hundreds of kpc scales (SB$_{\rm Ly\alpha}\sim10^{-17}\, 
\cgssb$) are not easily reproduced in simulations with current expected densities and temperatures (e.g., \citealt{Rosdahl12,FG2010}), and it is thus difficult to firmly assess if 
this is the main powering mechanism. 
To reproduce the observed high SBs, lower temperatures or higher densities in the framework of a ``pure'' cooling 
radiation scenario are thus needed. 
In this regard, note that cosmological simulations could probably still miss high densities in the CGM due to 
computational issues or current subgrid prescriptions. 
Indeed, the above-mentioned works show that to reproduce 
the LABs, the \lya\ emission should come from gas dense enough
to be almost star-forming in their simulations (e.g., \citealt{Furlanetto05, FG2010, Rosdahl12}).

In addition, the sources embedded within the \lya\ emission could photoionize the surrounding gas and thus boost the 
\lya\ signal through recombination radiation
(fluorescence). This effect has not been accounted for when dealing with the ``cooling radiation'' 
scenario (e.g., \citealt{Dijkstra2009_hd,Rosdahl12}), 
as studies were focused on a conservative ``cooling flow'' framework. 
Given the very mild dependence on temperature and ionisation state for the recombination radiation, 
this process seems more plausible for extended \lya\ nebulae around quasars (e.g., \citealt{cantalupo14,Borisova2016}).
The far higher luminosity of \qso\ in comparison to its companions (see section~\ref{sec:compact}), likely makes it the dominant source 
of radiation if it illuminates the ELAN.
However, we don't exclude the possibility that the type-2 AGN1 may be brighter as seen from the nebula than from 
our perspective. Nevertheless, it is unlikely that its emission overcame 
the ionizing radiation from \qso, given the intensity of the observed narrow lines, e.g. \ion{C}{iv}.%

If we then assume the ELAN to span the CGM around \qso,
and the bright quasar to illuminate it, the emitting gas has to be highly ionized, and thus 
optically thin to the ionizing radiation (\citealt{fab+15b}). In this regime, the \lya\ emission would 
follow SB$_{\rm Ly\alpha}^{\rm thin}\propto n_{\rm  H}N_{\rm H}$, and thus
would not depend on the luminosity of the 
central source (as long as the quasar is able to keep the gas ionized) (\citealt{qpq4}). If the physical properties of the emitting 
gas ($n_{\rm  H}$, $N_{\rm H}$) are roughly the same, the aforementioned relation thus naturally explain the roughly constant high surface brightness in the whole extent of the ELAN. 

\begin{figure*}
\begin{center}
	\includegraphics[width=0.8\textwidth, clip]{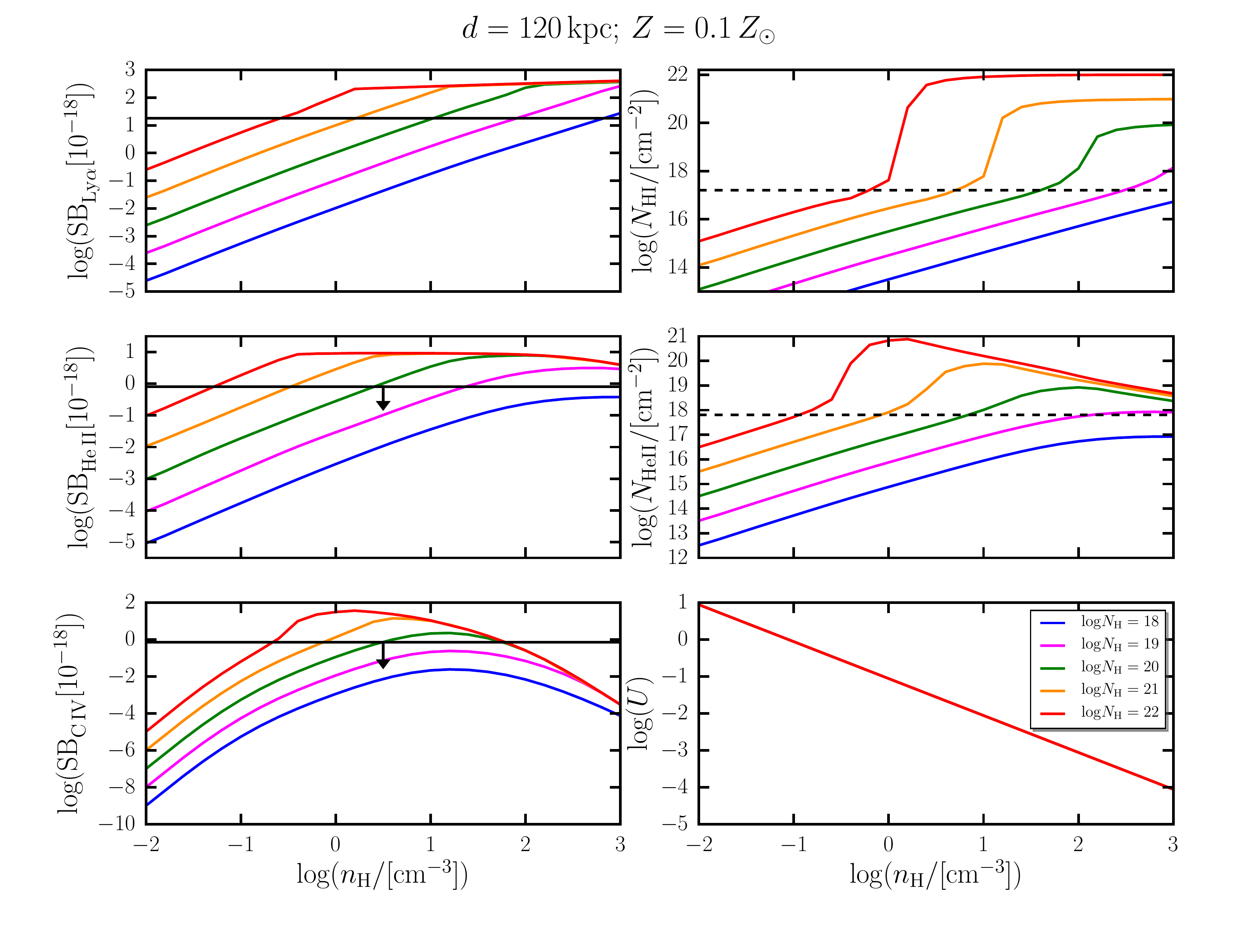}
\end{center}
\caption{{\bf Photoionization modeling of the maximum within the NW redshifted emission assumed to be at $120$~kpc, $Z=0.1$~Z$_{\odot}$.} 
The calculation has been performed using the photoionization code Cloudy (\citealt{ferland13}). We show, as a function of $n_{\rm H}$, 
the predicted SB$_{\rm Ly\alpha}$ in units of $10^{-18} \cgssb$, the predicted column density of neutral hydrogen $N_{\rm HI}$, 
the predicted SB$_{\rm He\, II}$ in units of $10^{-18} \cgssb$, the predicted column density of singly ionized helium $N_{\rm He\, II}$, 
the predicted SB$_{\rm C\, IV}$ in units of $10^{-18} \cgssb$, and the ionization parameter $U$, respectively in {\bf (A)}, {\bf (B)},  
{\bf (C)}, {\bf (D)}, {\bf (E)}, and {\bf (F)}. 
The solid horizontal lines show our measurement for  SB$_{\rm Ly\alpha}$ and the upper limits for SB$_{\rm He\, II}$ and SB$_{\rm C\, IV}$.
The horizontal dashed lines indicate the theoretical threshold dividing the optically thin regime from the optically thick case for that element. 
For neutral hydrogen is at $N_{\rm HI}=10^{17.2}$~cm$^{-2}$, while for helium is at  $N_{\rm He\, II}=10^{17.8}$~cm$^{-2}$. If the quasar \qso\ 
is illuminating the emitting gas, only models with  $n_{\rm H}\gtrsim 690$~cm$^{-3}$ and  ${\rm log}[N_{\rm H}/{\rm cm}^{-2}]\lesssim18$ would match 
our observational constraints at this position.}
\label{CloudyFirst}
\end{figure*}

As shown by \citet{fab+15b},  strong constraints on the level of the \heii line can break the degeneracy between $n_{\rm H}$ and $N_{\rm H}$ (and thus the cool gas mass) 
inherent in the observation of the SB$_{\rm Ly\alpha}$ alone in this optically thin regime.
We thus use the Cloudy photoionization code (v10.01), last described by \citet{ferland13}, to constrain the physical properties of the 
emitting gas around the quasar \qso, focusing in particular on the NW emission at $\approx120$~kpc (the high S/N portion, solid red contour in Fig.~\ref{civheii}).
In particular, we perform a similar analysis as in \citet{fab+15b}, using \\
\begin{enumerate}
\item their same prescriptions for the input quasar SED, using the MUSE spectrum of \qso\  (Fig.~\ref{figsources}) and masking the \lya\ and \civ lines 
to avoid contribution to both emissions from scattering of photons from the quasar. As in \citet{fab+15b}, for the extreme ultraviolet (UV) we adopt a 
slope of $\alpha_{\rm UV}=-1.7$ (\citealt{lusso+15}), and determine the luminosity at the Lyman limit of \qso\ to be $L_{\nu_{\rm LL}}=2.8\times 10^{31}$~erg~s$^{-1}$;\\
\item  a standard plane-parallel geometry for the emitting  clouds illuminated by the quasar at a distance of $120$~kpc; \\
\item a grid of models with this wide range of parameters given the dependence of SB$_{\rm Ly\alpha}$ on $n_{\rm  H}$, and $N_{\rm H}$, 
and the dependence on metallicity ($Z$) of the \civ and \heii lines, and of the collisional excitation:
\begin{enumerate}
\item  $n_{\rm H}=10^{-2}-10^{3}$~cm$^{-3}$ (steps of 0.2 dex),
\item  $N_{\rm H}=10^{18}-10^{22}$~cm$^{-2}$ (steps of 1 dex),
\item $Z=10^{-3}-1$~Z$_{\rm solar}$ (steps of 1 dex).
\end{enumerate}
\end{enumerate}
Note that photoionization models are self-similar in the ionization parameter $U\equiv \frac{\phi_{\rm LL}}{c n_{\rm H}}$, which is the ratio 
of the number density of ionizing photons to hydrogen atoms.
As the luminosity of the quasar is known, the variation of 
$U$ results from the variations of $n_{\rm H}$, as can be seen in panel F of Fig.~\ref{CloudyFirst}, where we show the predictions of this 
calculation for $Z=0.1$~Z$_{\rm solar}$ (expected for CGM gas; \citealt{QPQ8}).
As we discuss in section~\ref{sec:heiiciv}, our MUSE observations constrain the \lya\ emission of the NW 
bright clump at $120$~kpc to be SB$_{\rm Ly\alpha}=(1.81\pm0.08)\times10^{-17} \cgssb$, and 
yields $5\sigma$ upper limits of SB$_{\rm HeII}<8.1\times10^{-19} \cgssb$ and SB$_{\rm CIV}<7.2\times10^{-19} \cgssb$. 
On the other hand, each photoionization model in our grid predicts the intensity of these emission lines. Fig.~\ref{CloudyFirst} report the trajectory 
of these models of different $N_{\rm H}$ as a function of $n_{\rm H}$ and compare those values to our measurements. Our calculation confirms that optically thin gas 
(log$[N_{\rm HI}/{\rm cm}^{-2}]\ll 17.2$; see panel B) 
can reproduce the SB$_{\rm Ly\alpha}$ at the NW clump (horizontal line in panel A), while optically thick gas would result in $\approx20\times$ higher 
levels of emission. However, it is clear that the models that reproduce the observed SB$_{\rm Ly\alpha}$ present levels of \heii too high to be 
in agreement with our observations (see panels C), unless the emitting gas is characterized by very high densities and low column densities. 
Specifically, only models with $n_{\rm H}\gtrsim 690$~cm$^{-3}$ and ${\rm log}[N_{\rm H}/{\rm cm}^{-2}]\lesssim18$ are able to match our observations (${\rm SB}_{\rm Ly\alpha}$ and \heii upper limit).   
Given the stringent constraint \ion{He}{ii}/\lya$<0.045$, this result is also 
valid for the other metallicities here studied (see Figure 7 in \citealt{fab+15b}), and would change by roughly a factor of 3 ($n_{\rm H}>250$~cm$^{-3}$, \citealt{fab+15b}) 
if we adopt the softest ionizing slope ($\alpha_{\rm UV}$=-2.3) in agreement (within $1\sigma$) with the quasar average spectrum in \citet{lusso+15}.  
The high values for $n_{\rm H}$ and low values for  $N_{\rm H}$ are in strong contrast with the average values expected within the quasar 
CGM, i.e. $n_{\rm H}\approx10^{-2}-10^{-3}$~cm$^{-3}$ and log$N_{\rm H}=20.5$ (e.g., \citealt{Rosdahl12,QPQ8,fab+16}). This result exacerbates similar 
conclusions found when studying the ELAN around the quasar UM~287, i.e.  $n_{\rm H}\gtrsim3$~cm$^{-3}$ and log$N_{\rm H}\lesssim20$ (\citealt{fab+15b}), 
and would thus increase the tension with current simulations of cosmological structure formation (see discussion in 
\citealt{cantalupo14,crighton+15,fab+15b,hennawi+15,McCourt2016}), 
which are currently not able to 
follow such high densities within the diffuse CGM/IGM.
However, as the predicted values in the case of \qso\ seem implausible in comparison to the expectations,  
we conclude that the quasar is probably not directly shining on the ELAN, or/and it should have a spectral energy distribution much softer than the average quasar.

In the case of a photoionization scenario, the fainter embedded sources together with eventual obscured sources not detected in our 
observations (i.e. obscured starbursts, sub-millimeter galaxies) could thus probably largely contribute to the powering of the \lya\ emission. 
In particular, given the expected softer UV radiation from star-forming galaxies with respect to a quasar, photoionization from these objects
could explain the observed level of SB$_{\rm Ly\alpha}$, together with the non-detections in \heii and \civ with more reasonable densities ($\sim1$~cm$^{-3}$). 
In addition, part of the cool CGM in quasar haloes and emitting in \lya\ could be gas stripped from infalling satellites (e.g., \citealt{fab2012}).  
Such gas has been predicted to emit preferentially through collisional  excitation in low-redshift systems, though definite conclusions on this 
have not been reached (e.g., \citealt{Tonnesen2011}). If this is the case also at high redshift,  part of the \lya\ emission could be due  to this process, 
and would result in larger and brighter ELANe depending on the phase of the interaction we witness.
Further, powering mechanisms linked to companion objects can explain (i) the detections of the rotation-like pattern in a transverse direction 
with respect to the bright quasar without requiring ad-hoc orientation of the ionizing cone together with an ad-hoc wide opening angle for 
the AGN emission ($>90^{\circ}$), and (ii) the asymmetry of the ELAN, or in other words the absence of bright extended emission 
detected at negative distances from the quasar without requiring lack of cool gas within the ionizing cone of the quasar. 
Indeed at these locations no bright companions are detected down to our sensitivity limit. 

Finally, extended star-formation may be occurring in this large-scale nebula, and thus power, at least partially, the Ly$\alpha$ emission. 
Indeed, the presence of widespread star-formation and molecular gas on hundreds of kiloparsecs has been unveiled in the case of the 
Spiderweb radio-galaxy at $z=2.161$, where a giant Ly$\alpha$ nebula encompass several galaxies (\citealt{Emonts2016}).

Summarizing, we can {\it not} firmly constrain the powering mechanism for the extended \lya\ emission. 
Most likely, the high level of SB$_{\rm Ly\alpha}$ that we discovered is due to a complex combination of all the aforementioned processes.
Additional observations are thus needed to disentangle the contributions from each mechanism. A step towards a better understanding of this system could originate 
from our follow-up campaign in the submillimeter regime (to search for the  presence of dust-obscured highly star-forming galaxies associated with the ELAN, e.g., \citealt{Geach2016}, 
extended star formation, and molecular gas), or by our planned observations to estimate the polarization of the  \lya\ emission (to verify the importance 
of scattering within the ELAN and to understand which are the powering sources, e.g. \citealt{Prescott2011}).  
Notwithstanding these uncertainties, the bright levels of \lya\ emission seems to imply high $n_{\rm H}$ values ($\gtrsim1$~cm$^{-3}$) irrespective of the powering mechanism invoked.

\section{Summary and Conclusions}
\label{sec:summ}

To characterise the frequency of detection of 
the enormous \lya\ nebulae (ELANe) (${\rm SB}_{\rm Ly\alpha}\gtrsim 10^{-17}$\unitcgssb\ at 100~kpc) around 
quasars (\citealt{cantalupo14, hennawi+15, Cai2016}), and characterise the physical properties of the CGM/IGM
in emission, we have initiated a survey of the population of $z\sim3$ quasars using MUSE on VLT, i.e. QSO MUSEUM (Arrigoni Battaia et al., in prep.). 
In this framework, we here report the discovery of an additional ELAN around the radio-quiet quasar \qso\ at $z=3.167$. This ELAN spans a maximum projected distance of 297~kpc, and 
show an average \lya\ surface brightness ${\rm SB}_{\rm Ly\alpha}\sim 6.04\times10^{-18}$\unitcgssb (within the $2\sigma$ isophote).
Notwithstanding the high ${\rm SB}_{\rm Ly\alpha}$, the ELAN does {\it not} show extended emission in \heii and \civ down to our deep SB limits, resulting in
stringent upper limits on the line ratios, i.e. \ion{He}{ii}/\lya$<0.020$ and \ion{C}{iv}/\lya$<0.018$ ($5\sigma$).

Further, this ELAN is associated with an additional four embedded sources besides \qso, i.e. two LAEs and two faint AGN 
\qso\ thus seems to reside in an overdense environment as the ELAN associated with the only quadruple AGN known at high redshift (\citealt{hennawi+15}).
Such an occurrence hints to a scenario in which ELANe are preferentially observed in overdensities (\citealt{Matsuda2005,Matsuda2009,Saito2006,Prescott2008, Yang2009, hennawi+15}).

Also, this ELAN shows coherent kinematics, i.e. small gradient (300~km~s$^{-1}$) and velocity dispersion ($<270$~km~s$^{-1}$), 
on very large scales (hundreds of kpc) at high significance. Specifically, the ELAN shows a velocity shear of $\sim300$~km~s$^{-1}$ between its
SE and NW edges. 
After considering several scenarios (e.g., resonant scattering, outflows, coherent disc) to explain this observed velocity shear, 
and by comparing the velocity field of the ELAN with a cosmological zoom-in simulation, we conclude that we are likely witnessing the accretion 
of sub-structures onto a central massive halo.
Our discovery is in agreement with current theory of structure 
formation for infalling sub-structures which predict the presence of a rotation-like 
pattern on halo scales.
Our interpretation thus remarkably differs from previous studies in the literature which explain the kinematics within extended \lya\
nebulae as coherent thin discs, on similar sizes (\citealt{martin+15}) or on smaller scales (radius of $\approx3$\arcsec or $\approx25$~kpc; \citealt{prescott2015a}).

Independent of the origin for the observed cool 
halo gas (e.g., hot or cold mode accretion, stripping of satellite galaxies, 
feedback), our observations reveal the potential for high precision 
spatial mapping of gas kinematics on circumgalactic and intergalactic scales around massive systems, 
thought to be the progenitors of present-day elliptical galaxies (\citealt{white12}). 
Such observations 
encode precious 
information about the formation of such systems, possibly unveiling clues on the intimate connection between baryons, dark matter, and the large-scale structures 
at these early epochs (e.g. \citealt{vandenBosch2002, Sharma2005, AragonCalvo2007, Zhang2009, Codis2012, Zjupa2016}).
Ultimately, a large sample of similar observations would enable a statistical characterisation of the geometrical uncertainties inherent to this study, and open up 
the possibility of comparing the angular momentum of the observed halo gas 
to the predicted spin of their 
dark matter haloes. 
This would allow us to better 
comprehend the angular momentum 
build-up that eventually leads to the formation of a central massive disc (\citealt{FE1980,stewart+16}), 
before further accretion, violent mergers and interactions reshape it into an elliptical galaxy (e.g., \citealt{Gott1975, Oser2010}).
Although there is a wealth of theoretical work addressing the formation of Milky Way 
analogs, 
we stress the lack of simulations targeting more massive 
systems ($M_{\rm DM}>10^{12}\, M_\odot$ at $z\approx3$). 
Further development on this mass scales, with 
particular emphasis on the challenges in reproducing ELANe (\citealt{cantalupo14,fab+15b,hennawi+15,McCourt2016} and appendix {\bf D}), are crucial
to examine the processes of galaxy formation and gas 
accretion at early times.  
With the ongoing and planned 
observations with new sensitive integral-field spectrographs 
(e.g., \citealt{Borisova2016}, Arrigoni Battaia et al., in prep.) this field is rapidly emerging.

\section*{Acknowledgements}

The authors thank the anonymous referee for the punctual comments and suggestions that help 
improving the clarity of our manuscript. 
We thank H. Rahmani, A. De Cia, A. Zanella, F. Lelli for helpful discussions.
We thank R. Ellis, R. Ivison, and E. Emsellem for providing comments on an early version of this work.
This paper is based on observations collected at the European Organisation for Astronomical Research in the 
Southern Hemisphere under ESO programmes 094.A-0585(A), 096.A-0937(A) (PI: FAB).  
Simulations have been performed on the High Performance 
Computing resources at New York University Abu Dhabi. 
JXP acknowledges support from the National Science Foundation through
grant AST-1412981.
AO has been funded by the Deutsche Forschungsgemeinschaft (DFG, German Research Foundation) -- MO 2979/1-1.
TB acknowledges support from the Sonderforschungsbereich SFB 881 ``The Milky Way System'' (subproject A2) 
of the German Research Foundation (DFG).
This work made use of Pynbody, an open-source analysis package for astrophysical N-body and Smooth Particle Hydrodynamics simulations (\citealt{pynbody}). 
This work made also use of the KUBEVIZ software which is publicly available at \url{http://www.mpe.mpg.de/~mfossati/kubeviz}.




\bibliographystyle{mnras}
\bibliography{allrefs} 

\begin{thebibliography}{}
\makeatletter
\relax
\def\mn@urlcharsother{\let\do\@makeother \do\$\do\&\do\#\do\^\do\_\do\%\do\~}
\def\mn@doi{\begingroup\mn@urlcharsother \@ifnextchar [ {\mn@doi@}
  {\mn@doi@[]}}
\def\mn@doi@[#1]#2{\def\@tempa{#1}\ifx\@tempa\@empty \href
  {http://dx.doi.org/#2} {doi:#2}\else \href {http://dx.doi.org/#2} {#1}\fi
  \endgroup}
\def\mn@eprint#1#2{\mn@eprint@#1:#2::\@nil}
\def\mn@eprint@arXiv#1{\href {http://arxiv.org/abs/#1} {{\tt arXiv:#1}}}
\def\mn@eprint@dblp#1{\href {http://dblp.uni-trier.de/rec/bibtex/#1.xml}
  {dblp:#1}}
\def\mn@eprint@#1:#2:#3:#4\@nil{\def\@tempa {#1}\def\@tempb {#2}\def\@tempc
  {#3}\ifx \@tempc \@empty \let \@tempc \@tempb \let \@tempb \@tempa \fi \ifx
  \@tempb \@empty \def\@tempb {arXiv}\fi \@ifundefined
  {mn@eprint@\@tempb}{\@tempb:\@tempc}{\expandafter \expandafter \csname
  mn@eprint@\@tempb\endcsname \expandafter{\@tempc}}}

\bibitem[\protect\citeauthoryear{{Adelman-McCarthy} \& {et
  al.}}{{Adelman-McCarthy} \& {et al.}}{2009}]{Adelman2009}
{Adelman-McCarthy} J.~K.,  {et al.} 2009, VizieR Online Data Catalog, \href
  {http://cdsads.u-strasbg.fr/abs/2009yCat.2294....0A} {2294}

\bibitem[\protect\citeauthoryear{{Allen}, {Groves}, {Dopita}, {Sutherland}  \&
  {Kewley}}{{Allen} et~al.}{2008}]{Allen2008}
{Allen} M.~G.,  {Groves} B.~A.,  {Dopita} M.~A.,  {Sutherland} R.~S.,
  {Kewley} L.~J.,  2008, \mn@doi [\apjs] {10.1086/589652}, \href
  {http://adsabs.harvard.edu/abs/2008ApJS..178...20A} {178, 20}

\bibitem[\protect\citeauthoryear{{Antonucci}}{{Antonucci}}{1993}]{Anton93}
{Antonucci} R.,  1993, \mn@doi [\araa] {10.1146/annurev.aa.31.090193.002353},
  \href {http://adsabs.harvard.edu/abs/1993ARA%26A..31..473A} {31, 473}

\bibitem[\protect\citeauthoryear{{Arag{\'o}n-Calvo}, {van de Weygaert}, {Jones}
   \& {van der Hulst}}{{Arag{\'o}n-Calvo} et~al.}{2007}]{AragonCalvo2007}
{Arag{\'o}n-Calvo} M.~A.,  {van de Weygaert} R.,  {Jones} B.~J.~T.,   {van der
  Hulst} J.~M.,  2007, \mn@doi [\apjl] {10.1086/511633}, \href
  {http://adsabs.harvard.edu/abs/2007ApJ...655L...5A} {655, L5}

\bibitem[\protect\citeauthoryear{{Arrigoni Battaia} et~al.,}{{Arrigoni Battaia}
  et~al.}{2012}]{fab2012}
{Arrigoni Battaia} F.,  et~al., 2012, \mn@doi [\aap]
  {10.1051/0004-6361/201218895}, \href
  {http://adsabs.harvard.edu/abs/2012A%26A...543A.112A} {543, A112}

\bibitem[\protect\citeauthoryear{{Arrigoni Battaia}, {Yang}, {Hennawi},
  {Prochaska}, {Matsuda}, {Yamada}  \& {Hayashino}}{{Arrigoni Battaia}
  et~al.}{2015a}]{fab+15a}
{Arrigoni Battaia} F.,  {Yang} Y.,  {Hennawi} J.~F.,  {Prochaska} J.~X.,
  {Matsuda} Y.,  {Yamada} T.,   {Hayashino} T.,  2015a, \mn@doi [\apj]
  {10.1088/0004-637X/804/1/26}, \href
  {http://adsabs.harvard.edu/abs/2015ApJ...804...26A} {804, 26}

\bibitem[\protect\citeauthoryear{{Arrigoni Battaia}, {Hennawi}, {Prochaska}  \&
  {Cantalupo}}{{Arrigoni Battaia} et~al.}{2015b}]{fab+15b}
{Arrigoni Battaia} F.,  {Hennawi} J.~F.,  {Prochaska} J.~X.,   {Cantalupo} S.,
  2015b, \mn@doi [\apj] {10.1088/0004-637X/809/2/163}, \href
  {http://adsabs.harvard.edu/abs/2015ApJ...809..163A} {809, 163}

\bibitem[\protect\citeauthoryear{{Arrigoni Battaia}, {Hennawi}, {Cantalupo}  \&
  {Prochaska}}{{Arrigoni Battaia} et~al.}{2016}]{fab+16}
{Arrigoni Battaia} F.,  {Hennawi} J.~F.,  {Cantalupo} S.,   {Prochaska} J.~X.,
  2016, \mn@doi [\apj] {10.3847/0004-637X/829/1/3}, \href
  {http://adsabs.harvard.edu/abs/2016ApJ...829....3A} {829, 3}

\bibitem[\protect\citeauthoryear{{Bacon} et~al.,}{{Bacon}
  et~al.}{2010}]{Bacon2010}
{Bacon} R.,  et~al., 2010, in Society of Photo-Optical Instrumentation
  Engineers (SPIE) Conference Series. , \mn@doi{10.1117/12.856027}

\bibitem[\protect\citeauthoryear{{Becker}, {White}  \& {Helfand}}{{Becker}
  et~al.}{1994}]{Becker1994}
{Becker} R.~H.,  {White} R.~L.,   {Helfand} D.~J.,  1994, in {Crabtree} D.~R.,
  {Hanisch} R.~J.,   {Barnes} J.,  eds,  Astronomical Society of the Pacific
  Conference Series Vol. 61, Astronomical Data Analysis Software and Systems
  III. p.~165

\bibitem[\protect\citeauthoryear{{Behroozi}, {Wechsler}  \&
  {Conroy}}{{Behroozi} et~al.}{2013}]{Behroozi2013}
{Behroozi} P.~S.,  {Wechsler} R.~H.,   {Conroy} C.,  2013, \mn@doi [\apj]
  {10.1088/0004-637X/770/1/57}, \href
  {http://adsabs.harvard.edu/abs/2013ApJ...770...57B} {770, 57}

\bibitem[\protect\citeauthoryear{{Bertin} \& {Arnouts}}{{Bertin} \&
  {Arnouts}}{1996}]{bertin96}
{Bertin} E.,  {Arnouts} S.,  1996, Astronomy and Astrophysics Supplement, \href
  {http://adsabs.harvard.edu/abs/1996A%26AS..117..393B} {117, 393}

\bibitem[\protect\citeauthoryear{{Bolton} et~al.,}{{Bolton}
  et~al.}{2012}]{Bolton2012}
{Bolton} A.~S.,  et~al., 2012, \mn@doi [\aj] {10.1088/0004-6256/144/5/144},
  \href {http://adsabs.harvard.edu/abs/2012AJ....144..144B} {144, 144}

\bibitem[\protect\citeauthoryear{{Borisova} et~al.,}{{Borisova}
  et~al.}{2016}]{Borisova2016}
{Borisova} E.,  et~al., 2016, \mn@doi [\apj] {10.3847/0004-637X/831/1/39},
  \href {http://adsabs.harvard.edu/abs/2016ApJ...831...39B} {831, 39}

\bibitem[\protect\citeauthoryear{{Bullock}, {Dekel}, {Kolatt}, {Kravtsov},
  {Klypin}, {Porciani}  \& {Primack}}{{Bullock} et~al.}{2001}]{bullock01}
{Bullock} J.~S.,  {Dekel} A.,  {Kolatt} T.~S.,  {Kravtsov} A.~V.,  {Klypin}
  A.~A.,  {Porciani} C.,   {Primack} J.~R.,  2001, \mn@doi [\apj]
  {10.1086/321477}, \href {http://adsabs.harvard.edu/abs/2001ApJ...555..240B}
  {555, 240}

\bibitem[\protect\citeauthoryear{{Cabot}, {Cen}  \& {Zheng}}{{Cabot}
  et~al.}{2016}]{Cabot2016}
{Cabot} S.~H.~C.,  {Cen} R.,   {Zheng} Z.,  2016, \mn@doi [\mnras]
  {10.1093/mnras/stw1727}, \href
  {http://adsabs.harvard.edu/abs/2016MNRAS.462.1076C} {462, 1076}

\bibitem[\protect\citeauthoryear{{Cai} et~al.,}{{Cai} et~al.}{2016}]{Cai2016a}
{Cai} Z.,  et~al., 2016, preprint, \href
  {http://adsabs.harvard.edu/abs/2016arXiv160902913C} {} (\mn@eprint {arXiv}
  {1609.02913})

\bibitem[\protect\citeauthoryear{{Cai} et~al.,}{{Cai} et~al.}{2017}]{Cai2016}
{Cai} Z.,  et~al., 2017, \mn@doi [\apj] {10.3847/1538-4357/aa5d14}, \href
  {http://adsabs.harvard.edu/abs/2017ApJ...837...71C} {837, 71}

\bibitem[\protect\citeauthoryear{{Caminha} et~al.,}{{Caminha}
  et~al.}{2016}]{Caminha2016}
{Caminha} G.~B.,  et~al., 2016, \mn@doi [\aap] {10.1051/0004-6361/201527995},
  \href {http://adsabs.harvard.edu/abs/2016A%26A...595A.100C} {595, A100}

\bibitem[\protect\citeauthoryear{{Cantalupo}, {Porciani}, {Lilly}  \&
  {Miniati}}{{Cantalupo} et~al.}{2005}]{Cantalupo2005}
{Cantalupo} S.,  {Porciani} C.,  {Lilly} S.~J.,   {Miniati} F.,  2005, \mn@doi
  [\apj] {10.1086/430758}, \href
  {http://adsabs.harvard.edu/abs/2005ApJ...628...61C} {628, 61}

\bibitem[\protect\citeauthoryear{{Cantalupo}, {Arrigoni-Battaia}, {Prochaska},
  {Hennawi}  \& {Madau}}{{Cantalupo} et~al.}{2014}]{cantalupo14}
{Cantalupo} S.,  {Arrigoni-Battaia} F.,  {Prochaska} J.~X.,  {Hennawi} J.~F.,
  {Madau} P.,  2014, \mn@doi [\nat] {10.1038/nature12898}, \href
  {http://adsabs.harvard.edu/abs/2014Natur.506...63C} {506, 63}

\bibitem[\protect\citeauthoryear{{Cen} \& {Zheng}}{{Cen} \&
  {Zheng}}{2013}]{Cen2013}
{Cen} R.,  {Zheng} Z.,  2013, \mn@doi [\apj] {10.1088/0004-637X/775/2/112},
  \href {http://adsabs.harvard.edu/abs/2013ApJ...775..112C} {775, 112}

\bibitem[\protect\citeauthoryear{{Christensen}, {Jahnke}, {Wisotzki}  \&
  {S{\'a}nchez}}{{Christensen} et~al.}{2006}]{Christensen2006}
{Christensen} L.,  {Jahnke} K.,  {Wisotzki} L.,   {S{\'a}nchez} S.~F.,  2006,
  \mn@doi [\aap] {10.1051/0004-6361:20065318}, \href
  {http://adsabs.harvard.edu/abs/2006A%26A...459..717C} {459, 717}

\bibitem[\protect\citeauthoryear{{Codis}, {Pichon}, {Devriendt}, {Slyz},
  {Pogosyan}, {Dubois}  \& {Sousbie}}{{Codis} et~al.}{2012}]{Codis2012}
{Codis} S.,  {Pichon} C.,  {Devriendt} J.,  {Slyz} A.,  {Pogosyan} D.,
  {Dubois} Y.,   {Sousbie} T.,  2012, \mn@doi [\mnras]
  {10.1111/j.1365-2966.2012.21636.x}, \href
  {http://adsabs.harvard.edu/abs/2012MNRAS.427.3320C} {427, 3320}

\bibitem[\protect\citeauthoryear{{Condon} \& {Broderick}}{{Condon} \&
  {Broderick}}{1985}]{CondonBroderick1985}
{Condon} J.~J.,  {Broderick} J.~J.,  1985, \mn@doi [\aj] {10.1086/113959},
  \href {http://adsabs.harvard.edu/abs/1985AJ.....90.2540C} {90, 2540}

\bibitem[\protect\citeauthoryear{{Condon} \& {Broderick}}{{Condon} \&
  {Broderick}}{1986}]{CondonBroderick1986}
{Condon} J.~J.,  {Broderick} J.~J.,  1986, \mn@doi [\aj] {10.1086/114081},
  \href {http://adsabs.harvard.edu/abs/1986AJ.....91.1051C} {91, 1051}

\bibitem[\protect\citeauthoryear{{Condon}, {Cotton}, {Greisen}, {Yin},
  {Perley}, {Taylor}  \& {Broderick}}{{Condon} et~al.}{1998}]{Condon1998}
{Condon} J.~J.,  {Cotton} W.~D.,  {Greisen} E.~W.,  {Yin} Q.~F.,  {Perley}
  R.~A.,  {Taylor} G.~B.,   {Broderick} J.~J.,  1998, \mn@doi [\aj]
  {10.1086/300337}, \href {http://adsabs.harvard.edu/abs/1998AJ....115.1693C}
  {115, 1693}

\bibitem[\protect\citeauthoryear{{Crighton}, {Hennawi}, {Simcoe}, {Cooksey},
  {Murphy}, {Fumagalli}, {Prochaska}  \& {Shanks}}{{Crighton}
  et~al.}{2015}]{crighton+15}
{Crighton} N.~H.~M.,  {Hennawi} J.~F.,  {Simcoe} R.~A.,  {Cooksey} K.~L.,
  {Murphy} M.~T.,  {Fumagalli} M.,  {Prochaska} J.~X.,   {Shanks} T.,  2015,
  \mn@doi [\mnras] {10.1093/mnras/stu2088}, \href
  {http://adsabs.harvard.edu/abs/2015MNRAS.446...18C} {446, 18}

\bibitem[\protect\citeauthoryear{{Curran}, {Murphy}, {Webb}, {Rantakyr{\"o}},
  {Johansson}  \& {Nikoli{\'c}}}{{Curran} et~al.}{2002}]{Curran2002}
{Curran} S.~J.,  {Murphy} M.~T.,  {Webb} J.~K.,  {Rantakyr{\"o}} F.,
  {Johansson} L.~E.~B.,   {Nikoli{\'c}} S.,  2002, \mn@doi [\aap]
  {10.1051/0004-6361:20021231}, \href
  {http://adsabs.harvard.edu/abs/2002A%26A...394..763C} {394, 763}

\bibitem[\protect\citeauthoryear{{Danovich}, {Dekel}, {Hahn}, {Ceverino}  \&
  {Primack}}{{Danovich} et~al.}{2015}]{danovich+15}
{Danovich} M.,  {Dekel} A.,  {Hahn} O.,  {Ceverino} D.,   {Primack} J.,  2015,
  \mn@doi [\mnras] {10.1093/mnras/stv270}, \href
  {http://adsabs.harvard.edu/abs/2015MNRAS.449.2087D} {449, 2087}

\bibitem[\protect\citeauthoryear{{De Breuck}, {R{\"o}ttgering}, {Miley}, {van
  Breugel}  \& {Best}}{{De Breuck} et~al.}{2000}]{DeBreuck2000}
{De Breuck} C.,  {R{\"o}ttgering} H.,  {Miley} G.,  {van Breugel} W.,   {Best}
  P.,  2000, \aap, \href {http://adsabs.harvard.edu/abs/2000A%26A...362..519D}
  {362, 519}

\bibitem[\protect\citeauthoryear{{Dekel} \& {Birnboim}}{{Dekel} \&
  {Birnboim}}{2006}]{db06}
{Dekel} A.,  {Birnboim} Y.,  2006, \mn@doi [\mnras]
  {10.1111/j.1365-2966.2006.10145.x}, \href
  {http://adsabs.harvard.edu/abs/2006MNRAS.368....2D} {368, 2}

\bibitem[\protect\citeauthoryear{{Dekel} \& {Silk}}{{Dekel} \&
  {Silk}}{1986}]{DekelSilk1986}
{Dekel} A.,  {Silk} J.,  1986, \mn@doi [\apj] {10.1086/164050}, \href
  {http://adsabs.harvard.edu/abs/1986ApJ...303...39D} {303, 39}

\bibitem[\protect\citeauthoryear{{Dekel} et~al.,}{{Dekel}
  et~al.}{2009}]{Dekel2009}
{Dekel} A.,  et~al., 2009, \mn@doi [\nat] {10.1038/nature07648}, \href
  {http://adsabs.harvard.edu/abs/2009Natur.457..451D} {457, 451}

\bibitem[\protect\citeauthoryear{{Dey} et~al.,}{{Dey} et~al.}{2005}]{Dey2005}
{Dey} A.,  et~al., 2005, \mn@doi [\apj] {10.1086/430775}, \href
  {http://adsabs.harvard.edu/abs/2005ApJ...629..654D} {629, 654}

\bibitem[\protect\citeauthoryear{{Dijkstra} \& {Loeb}}{{Dijkstra} \&
  {Loeb}}{2008}]{Dijkstra2008}
{Dijkstra} M.,  {Loeb} A.,  2008, \mn@doi [\mnras]
  {10.1111/j.1365-2966.2008.13066.x}, \href
  {http://adsabs.harvard.edu/abs/2008MNRAS.386..492D} {386, 492}

\bibitem[\protect\citeauthoryear{{Dijkstra} \& {Loeb}}{{Dijkstra} \&
  {Loeb}}{2009}]{Dijkstra2009_hd}
{Dijkstra} M.,  {Loeb} A.,  2009, \mn@doi [\mnras]
  {10.1111/j.1365-2966.2009.15533.x}, \href
  {http://adsabs.harvard.edu/abs/2009MNRAS.400.1109D} {400, 1109}

\bibitem[\protect\citeauthoryear{{Dijkstra}, {Haiman}  \& {Spaans}}{{Dijkstra}
  et~al.}{2006}]{Dijkstra_2006}
{Dijkstra} M.,  {Haiman} Z.,   {Spaans} M.,  2006, \mn@doi [\apj]
  {10.1086/506243}, \href {http://adsabs.harvard.edu/abs/2006ApJ...649...14D}
  {649, 14}

\bibitem[\protect\citeauthoryear{{Djorgovski}, {Courbin}, {Meylan}, {Sluse},
  {Thompson}, {Mahabal}  \& {Glikman}}{{Djorgovski}
  et~al.}{2007}]{Djorgovski2007}
{Djorgovski} S.~G.,  {Courbin} F.,  {Meylan} G.,  {Sluse} D.,  {Thompson} D.,
  {Mahabal} A.,   {Glikman} E.,  2007, \mn@doi [\apjl] {10.1086/519162}, \href
  {http://adsabs.harvard.edu/abs/2007ApJ...662L...1D} {662, L1}

\bibitem[\protect\citeauthoryear{{Douglas}, {Bash}, {Bozyan}, {Torrence}  \&
  {Wolfe}}{{Douglas} et~al.}{1996}]{Douglas1996}
{Douglas} J.~N.,  {Bash} F.~N.,  {Bozyan} F.~A.,  {Torrence} G.~W.,   {Wolfe}
  C.,  1996, \mn@doi [\aj] {10.1086/117932}, \href
  {http://adsabs.harvard.edu/abs/1996AJ....111.1945D} {111, 1945}

\bibitem[\protect\citeauthoryear{{Draine}}{{Draine}}{2011}]{Draine2011}
{Draine} B.~T.,  2011, {Physics of the Interstellar and Intergalactic Medium}

\bibitem[\protect\citeauthoryear{{Duval}, {Schaerer}, {{\"O}stlin}  \&
  {Laursen}}{{Duval} et~al.}{2014}]{Duval2014}
{Duval} F.,  {Schaerer} D.,  {{\"O}stlin} G.,   {Laursen} P.,  2014, \mn@doi
  [\aap] {10.1051/0004-6361/201220455}, \href
  {http://adsabs.harvard.edu/abs/2014A%26A...562A..52D} {562, A52}

\bibitem[\protect\citeauthoryear{{Ekers}}{{Ekers}}{1969}]{Ekers1969}
{Ekers} J.~A.,  1969, Australian Journal of Physics Astrophysical Supplement,
  \href {http://adsabs.harvard.edu/abs/1969AuJPA...7....3E} {7}

\bibitem[\protect\citeauthoryear{{Ellison}, {York}, {Pettini}  \&
  {Kanekar}}{{Ellison} et~al.}{2008}]{Ellison2008}
{Ellison} S.~L.,  {York} B.~A.,  {Pettini} M.,   {Kanekar} N.,  2008, \mn@doi
  [\mnras] {10.1111/j.1365-2966.2008.13482.x}, \href
  {http://adsabs.harvard.edu/abs/2008MNRAS.388.1349E} {388, 1349}

\bibitem[\protect\citeauthoryear{{Emonts} et~al.,}{{Emonts}
  et~al.}{2016}]{Emonts2016}
{Emonts} B.~H.~C.,  et~al., 2016, \mn@doi [Science] {10.1126/science.aag0512},
  \href {http://adsabs.harvard.edu/abs/2016Sci...354.1128E} {354, 1128}

\bibitem[\protect\citeauthoryear{{Fall} \& {Efstathiou}}{{Fall} \&
  {Efstathiou}}{1980}]{FE1980}
{Fall} S.~M.,  {Efstathiou} G.,  1980, \mn@doi [\mnras]
  {10.1093/mnras/193.2.189}, \href
  {http://adsabs.harvard.edu/abs/1980MNRAS.193..189F} {193, 189}

\bibitem[\protect\citeauthoryear{{Fanidakis}, {Macci{\`o}}, {Baugh}, {Lacey}
  \& {Frenk}}{{Fanidakis} et~al.}{2013}]{fanidakis13}
{Fanidakis} N.,  {Macci{\`o}} A.~V.,  {Baugh} C.~M.,  {Lacey} C.~G.,   {Frenk}
  C.~S.,  2013, \mn@doi [\mnras] {10.1093/mnras/stt1567}, \href
  {http://adsabs.harvard.edu/abs/2013MNRAS.436..315F} {436, 315}

\bibitem[\protect\citeauthoryear{{Fardal}, {Katz}, {Gardner}, {Hernquist},
  {Weinberg}  \& {Dav{\'e}}}{{Fardal} et~al.}{2001}]{Fardal2001}
{Fardal} M.~A.,  {Katz} N.,  {Gardner} J.~P.,  {Hernquist} L.,  {Weinberg}
  D.~H.,   {Dav{\'e}} R.,  2001, \mn@doi [\apj] {10.1086/323519}, \href
  {http://adsabs.harvard.edu/abs/2001ApJ...562..605F} {562, 605}

\bibitem[\protect\citeauthoryear{{Farina}, {Montuori}, {Decarli}  \&
  {Fumagalli}}{{Farina} et~al.}{2013}]{Farina2013}
{Farina} E.~P.,  {Montuori} C.,  {Decarli} R.,   {Fumagalli} M.,  2013, \mn@doi
  [\mnras] {10.1093/mnras/stt209}, \href
  {http://adsabs.harvard.edu/abs/2013MNRAS.431.1019F} {431, 1019}

\bibitem[\protect\citeauthoryear{{Faucher-Gigu{\`e}re} \& {Kere{\v
  s}}}{{Faucher-Gigu{\`e}re} \& {Kere{\v s}}}{2011}]{fg11}
{Faucher-Gigu{\`e}re} C.-A.,  {Kere{\v s}} D.,  2011, \mn@doi [\mnras]
  {10.1111/j.1745-3933.2011.01018.x}, \href
  {http://adsabs.harvard.edu/abs/2011MNRAS.412L.118F} {412, L118}

\bibitem[\protect\citeauthoryear{{Faucher-Gigu{\`e}re}, {Kere{\v s}},
  {Dijkstra}, {Hernquist}  \& {Zaldarriaga}}{{Faucher-Gigu{\`e}re}
  et~al.}{2010}]{FG2010}
{Faucher-Gigu{\`e}re} C.-A.,  {Kere{\v s}} D.,  {Dijkstra} M.,  {Hernquist} L.,
    {Zaldarriaga} M.,  2010, \mn@doi [\apj] {10.1088/0004-637X/725/1/633},
  \href {http://adsabs.harvard.edu/abs/2010ApJ...725..633F} {725, 633}

\bibitem[\protect\citeauthoryear{{Ferland} et~al.,}{{Ferland}
  et~al.}{2013}]{ferland13}
{Ferland} G.~J.,  et~al., 2013, \rmxaa, \href
  {http://adsabs.harvard.edu/abs/2013RMxAA..49..137F} {49, 137}

\bibitem[\protect\citeauthoryear{{Fossati}, {Fumagalli}, {Boselli}, {Gavazzi},
  {Sun}  \& {Wilman}}{{Fossati} et~al.}{2016}]{Fossati2016}
{Fossati} M.,  {Fumagalli} M.,  {Boselli} A.,  {Gavazzi} G.,  {Sun} M.,
  {Wilman} D.~J.,  2016, \mn@doi [\mnras] {10.1093/mnras/stv2400}, \href
  {http://adsabs.harvard.edu/abs/2016MNRAS.455.2028F} {455, 2028}

\bibitem[\protect\citeauthoryear{{Fumagalli}, {Prochaska}, {Kasen}, {Dekel},
  {Ceverino}  \& {Primack}}{{Fumagalli} et~al.}{2011}]{fumagalli11a}
{Fumagalli} M.,  {Prochaska} J.~X.,  {Kasen} D.,  {Dekel} A.,  {Ceverino} D.,
  {Primack} J.~R.,  2011, \mn@doi [\mnras] {10.1111/j.1365-2966.2011.19599.x},
  \href {http://adsabs.harvard.edu/abs/2011MNRAS.418.1796F} {418, 1796}

\bibitem[\protect\citeauthoryear{{Fumagalli}, {Cantalupo}, {Dekel}, {Morris},
  {O'Meara}, {Prochaska}  \& {Theuns}}{{Fumagalli}
  et~al.}{2016}]{Fumagalli2016}
{Fumagalli} M.,  {Cantalupo} S.,  {Dekel} A.,  {Morris} S.~L.,  {O'Meara}
  J.~M.,  {Prochaska} J.~X.,   {Theuns} T.,  2016, preprint, \href
  {http://adsabs.harvard.edu/abs/2016arXiv160703893F} {} (\mn@eprint {arXiv}
  {1607.03893})

\bibitem[\protect\citeauthoryear{{Furlanetto}, {Schaye}, {Springel}  \&
  {Hernquist}}{{Furlanetto} et~al.}{2005}]{Furlanetto05}
{Furlanetto} S.~R.,  {Schaye} J.,  {Springel} V.,   {Hernquist} L.,  2005,
  \mn@doi [\apj] {10.1086/426808}, \href
  {http://adsabs.harvard.edu/abs/2005ApJ...622....7F} {622, 7}

\bibitem[\protect\citeauthoryear{{Geach} et~al.,}{{Geach}
  et~al.}{2016}]{Geach2016}
{Geach} J.~E.,  et~al., 2016, preprint, \href
  {http://adsabs.harvard.edu/abs/2016arXiv160802941G} {} (\mn@eprint {arXiv}
  {1608.02941})

\bibitem[\protect\citeauthoryear{{Geller}, {Hwang}, {Diaferio}, {Kurtz}, {Coe}
  \& {Rines}}{{Geller} et~al.}{2014}]{Geller2014}
{Geller} M.~J.,  {Hwang} H.~S.,  {Diaferio} A.,  {Kurtz} M.~J.,  {Coe} D.,
  {Rines} K.~J.,  2014, \mn@doi [\apj] {10.1088/0004-637X/783/1/52}, \href
  {http://adsabs.harvard.edu/abs/2014ApJ...783...52G} {783, 52}

\bibitem[\protect\citeauthoryear{{Gibson} et~al.,}{{Gibson}
  et~al.}{2009}]{Gibson2009}
{Gibson} R.~R.,  et~al., 2009, \mn@doi [\apj] {10.1088/0004-637X/692/1/758},
  \href {http://adsabs.harvard.edu/abs/2009ApJ...692..758G} {692, 758}

\bibitem[\protect\citeauthoryear{{Goerdt} \& {Ceverino}}{{Goerdt} \&
  {Ceverino}}{2015}]{Goerdt2015}
{Goerdt} T.,  {Ceverino} D.,  2015, \mn@doi [\mnras] {10.1093/mnras/stv786},
  \href {http://adsabs.harvard.edu/abs/2015MNRAS.450.3359G} {450, 3359}

\bibitem[\protect\citeauthoryear{{Gott}}{{Gott}}{1975}]{Gott1975}
{Gott} III J.~R.,  1975, \mn@doi [\apj] {10.1086/153887}, \href
  {http://adsabs.harvard.edu/abs/1975ApJ...201..296G} {201, 296}

\bibitem[\protect\citeauthoryear{{Gould} \& {Weinberg}}{{Gould} \&
  {Weinberg}}{1996}]{GW96}
{Gould} A.,  {Weinberg} D.~H.,  1996, \mn@doi [\apj] {10.1086/177707}, \href
  {http://adsabs.harvard.edu/cgi-bin/nph-bib_query?bibcode=1996ApJ...468..462G&db_key=AST}
  {468, 462}

\bibitem[\protect\citeauthoryear{{Haardt} \& {Madau}}{{Haardt} \&
  {Madau}}{2012}]{hm12}
{Haardt} F.,  {Madau} P.,  2012, \mn@doi [\apj] {10.1088/0004-637X/746/2/125},
  \href {http://adsabs.harvard.edu/abs/2012ApJ...746..125H} {746, 125}

\bibitem[\protect\citeauthoryear{{Haiman}, {Spaans}  \& {Quataert}}{{Haiman}
  et~al.}{2000}]{Haiman2000}
{Haiman} Z.,  {Spaans} M.,   {Quataert} E.,  2000, \mn@doi [\apjl]
  {10.1086/312754}, \href {http://adsabs.harvard.edu/abs/2000ApJ...537L...5H}
  {537, L5}

\bibitem[\protect\citeauthoryear{{Hainline}, {Shapley}, {Greene}  \&
  {Steidel}}{{Hainline} et~al.}{2011}]{Hainline2011}
{Hainline} K.~N.,  {Shapley} A.~E.,  {Greene} J.~E.,   {Steidel} C.~C.,  2011,
  \mn@doi [\apj] {10.1088/0004-637X/733/1/31}, \href
  {http://adsabs.harvard.edu/abs/2011ApJ...733...31H} {733, 31}

\bibitem[\protect\citeauthoryear{{Harrison} et~al.,}{{Harrison}
  et~al.}{2012}]{Harrison2012}
{Harrison} C.~M.,  et~al., 2012, \mn@doi [\mnras]
  {10.1111/j.1365-2966.2012.21723.x}, \href
  {http://adsabs.harvard.edu/abs/2012MNRAS.426.1073H} {426, 1073}

\bibitem[\protect\citeauthoryear{{Harrison}, {Alexander}, {Mullaney}  \&
  {Swinbank}}{{Harrison} et~al.}{2014}]{Harrison2014}
{Harrison} C.~M.,  {Alexander} D.~M.,  {Mullaney} J.~R.,   {Swinbank} A.~M.,
  2014, \mn@doi [\mnras] {10.1093/mnras/stu515}, \href
  {http://adsabs.harvard.edu/abs/2014MNRAS.441.3306H} {441, 3306}

\bibitem[\protect\citeauthoryear{{Heckman}, {Miley}, {Lehnert}  \& {van
  Breugel}}{{Heckman} et~al.}{1991a}]{heckman91a}
{Heckman} T.~M.,  {Miley} G.~K.,  {Lehnert} M.~D.,   {van Breugel} W.,  1991a,
  \mn@doi [\apj] {10.1086/169794}, \href
  {http://adsabs.harvard.edu/cgi-bin/nph-bib_query?bibcode=1991ApJ...370...78H&db_key=AST}
  {370, 78}

\bibitem[\protect\citeauthoryear{{Heckman}, {Lehnert}, {Miley}  \& {van
  Breugel}}{{Heckman} et~al.}{1991b}]{heckman91b}
{Heckman} T.~M.,  {Lehnert} M.~D.,  {Miley} G.~K.,   {van Breugel} W.,  1991b,
  \mn@doi [\apj] {10.1086/170660}, \href
  {http://adsabs.harvard.edu/cgi-bin/nph-bib_query?bibcode=1991ApJ...381..373H&db_key=AST}
  {381, 373}

\bibitem[\protect\citeauthoryear{{Hennawi} \& {Prochaska}}{{Hennawi} \&
  {Prochaska}}{2013}]{qpq4}
{Hennawi} J.~F.,  {Prochaska} J.~X.,  2013, \mn@doi [\apj]
  {10.1088/0004-637X/766/1/58}, \href
  {http://adsabs.harvard.edu/abs/2013ApJ...766...58H} {766, 58 (QPQ4)}

\bibitem[\protect\citeauthoryear{{Hennawi} et~al.,}{{Hennawi}
  et~al.}{2006}]{Hennawi2006}
{Hennawi} J.~F.,  et~al., 2006, \mn@doi [\aj] {10.1086/498235}, \href
  {http://adsabs.harvard.edu/abs/2006AJ....131....1H} {131, 1}

\bibitem[\protect\citeauthoryear{{Hennawi}, {Prochaska}, {Cantalupo}  \&
  {Arrigoni-Battaia}}{{Hennawi} et~al.}{2015}]{hennawi+15}
{Hennawi} J.~F.,  {Prochaska} J.~X.,  {Cantalupo} S.,   {Arrigoni-Battaia} F.,
  2015, \mn@doi [Science] {10.1126/science.aaa5397}, \href
  {http://adsabs.harvard.edu/abs/2015Sci...348..779H} {348, 779}

\bibitem[\protect\citeauthoryear{{Herenz}, {Wisotzki}, {Roth}  \&
  {Anders}}{{Herenz} et~al.}{2015}]{Herenz2015}
{Herenz} E.~C.,  {Wisotzki} L.,  {Roth} M.,   {Anders} F.,  2015, \mn@doi
  [\aap] {10.1051/0004-6361/201425580}, \href
  {http://esoads.eso.org/abs/2015A%26A...576A.115H} {576, A115}

\bibitem[\protect\citeauthoryear{{Hewett} \& {Wild}}{{Hewett} \&
  {Wild}}{2010}]{Hewett2010}
{Hewett} P.~C.,  {Wild} V.,  2010, \mn@doi [\mnras]
  {10.1111/j.1365-2966.2010.16648.x}, \href
  {http://adsabs.harvard.edu/abs/2010MNRAS.405.2302H} {405, 2302}

\bibitem[\protect\citeauthoryear{{Hoyle}}{{Hoyle}}{1951}]{hoyle51}
{Hoyle} F.,  1951, in Problems of Cosmical Aerodynamics. p.~195

\bibitem[\protect\citeauthoryear{{Hu} \& {Cowie}}{{Hu} \&
  {Cowie}}{1987}]{HuCowie1987}
{Hu} E.~M.,  {Cowie} L.~L.,  1987, \mn@doi [\apjl] {10.1086/184902}, \href
  {http://adsabs.harvard.edu/abs/1987ApJ...317L...7H} {317, L7}

\bibitem[\protect\citeauthoryear{{Hubble}}{{Hubble}}{1929}]{hubble29}
{Hubble} E.,  1929, \mn@doi [Proceedings of the National Academy of Science]
  {10.1073/pnas.15.3.168}, \href
  {http://adsabs.harvard.edu/abs/1929PNAS...15..168H} {15, 168}

\bibitem[\protect\citeauthoryear{{Humphrey}, {Villar-Mart{\'{\i}}n}, {Vernet},
  {Fosbury}, {di Serego Alighieri}  \& {Binette}}{{Humphrey}
  et~al.}{2008}]{Humphrey2008}
{Humphrey} A.,  {Villar-Mart{\'{\i}}n} M.,  {Vernet} J.,  {Fosbury} R.,  {di
  Serego Alighieri} S.,   {Binette} L.,  2008, \mn@doi [\mnras]
  {10.1111/j.1365-2966.2007.12506.x}, \href
  {http://adsabs.harvard.edu/abs/2008MNRAS.383...11H} {383, 11}

\bibitem[\protect\citeauthoryear{{Husemann}, {Wisotzki}, {S{\'a}nchez}  \&
  {Jahnke}}{{Husemann} et~al.}{2013}]{Husemann2013}
{Husemann} B.,  {Wisotzki} L.,  {S{\'a}nchez} S.~F.,   {Jahnke} K.,  2013,
  \mn@doi [\aap] {10.1051/0004-6361/201220076}, \href
  {http://adsabs.harvard.edu/abs/2013A%26A...549A..43H} {549, A43}

\bibitem[\protect\citeauthoryear{{Husemann}, {Jahnke}, {S{\'a}nchez},
  {Wisotzki}, {Nugroho}, {Kupko}  \& {Schramm}}{{Husemann}
  et~al.}{2014}]{Husemann2014}
{Husemann} B.,  {Jahnke} K.,  {S{\'a}nchez} S.~F.,  {Wisotzki} L.,  {Nugroho}
  D.,  {Kupko} D.,   {Schramm} M.,  2014, \mn@doi [\mnras]
  {10.1093/mnras/stu1167}, \href
  {http://esoads.eso.org/abs/2014MNRAS.443..755H} {443, 755}

\bibitem[\protect\citeauthoryear{{Kakkad} et~al.,}{{Kakkad}
  et~al.}{2016}]{Kakkad2016}
{Kakkad} D.,  et~al., 2016, \mn@doi [\aap] {10.1051/0004-6361/201527968}, \href
  {http://adsabs.harvard.edu/abs/2016A%26A...592A.148K} {592, A148}

\bibitem[\protect\citeauthoryear{{Keller}, {Wadsley}, {Benincasa}  \&
  {Couchman}}{{Keller} et~al.}{2014}]{Keller2014}
{Keller} B.~W.,  {Wadsley} J.,  {Benincasa} S.~M.,   {Couchman} H.~M.~P.,
  2014, \mn@doi [\mnras] {10.1093/mnras/stu1058}, \href
  {http://adsabs.harvard.edu/abs/2014MNRAS.442.3013K} {442, 3013}

\bibitem[\protect\citeauthoryear{{Keller}, {Wadsley}  \& {Couchman}}{{Keller}
  et~al.}{2015}]{Keller2015}
{Keller} B.~W.,  {Wadsley} J.,   {Couchman} H.~M.~P.,  2015, \mn@doi [\mnras]
  {10.1093/mnras/stv1789}, \href
  {http://adsabs.harvard.edu/abs/2015MNRAS.453.3499K} {453, 3499}

\bibitem[\protect\citeauthoryear{{Kellermann}, {Sramek}, {Schmidt}, {Shaffer}
  \& {Green}}{{Kellermann} et~al.}{1989}]{Kellermann1989}
{Kellermann} K.~I.,  {Sramek} R.,  {Schmidt} M.,  {Shaffer} D.~B.,   {Green}
  R.,  1989, \mn@doi [\aj] {10.1086/115207}, \href
  {http://adsabs.harvard.edu/abs/1989AJ.....98.1195K} {98, 1195}

\bibitem[\protect\citeauthoryear{{Kent} \& {Gunn}}{{Kent} \&
  {Gunn}}{1982}]{KentGunn1982}
{Kent} S.~M.,  {Gunn} J.~E.,  1982, \mn@doi [\aj] {10.1086/113178}, \href
  {http://adsabs.harvard.edu/abs/1982AJ.....87..945K} {87, 945}

\bibitem[\protect\citeauthoryear{{Kere{\v s}}, {Katz}, {Weinberg}  \&
  {Dav{\'e}}}{{Kere{\v s}} et~al.}{2005}]{kkw+05}
{Kere{\v s}} D.,  {Katz} N.,  {Weinberg} D.~H.,   {Dav{\'e}} R.,  2005, \mn@doi
  [\mnras] {10.1111/j.1365-2966.2005.09451.x}, \href
  {http://adsabs.harvard.edu/abs/2005MNRAS.363....2K} {363, 2}

\bibitem[\protect\citeauthoryear{{Kere{\v s}}, {Katz}, {Fardal}, {Dav{\'e}}  \&
  {Weinberg}}{{Kere{\v s}} et~al.}{2009}]{keres09a}
{Kere{\v s}} D.,  {Katz} N.,  {Fardal} M.,  {Dav{\'e}} R.,   {Weinberg} D.~H.,
  2009, \mn@doi [\mnras] {10.1111/j.1365-2966.2009.14541.x}, \href
  {http://adsabs.harvard.edu/abs/2009MNRAS.395..160K} {395, 160}

\bibitem[\protect\citeauthoryear{{Kollmeier}, {Zheng}, {Dav{\'e}}, {Gould},
  {Katz}, {Miralda-Escud{\'e}}  \& {Weinberg}}{{Kollmeier}
  et~al.}{2010}]{kollmeier10}
{Kollmeier} J.~A.,  {Zheng} Z.,  {Dav{\'e}} R.,  {Gould} A.,  {Katz} N.,
  {Miralda-Escud{\'e}} J.,   {Weinberg} D.~H.,  2010, \mn@doi [\apj]
  {10.1088/0004-637X/708/2/1048}, \href
  {http://adsabs.harvard.edu/abs/2010ApJ...708.1048K} {708, 1048}

\bibitem[\protect\citeauthoryear{{Kreimeyer} \& {Veilleux}}{{Kreimeyer} \&
  {Veilleux}}{2013}]{Kreimeyer2013}
{Kreimeyer} K.,  {Veilleux} S.,  2013, \mn@doi [\apjl]
  {10.1088/2041-8205/772/1/L11}, \href
  {http://adsabs.harvard.edu/abs/2013ApJ...772L..11K} {772, L11}

\bibitem[\protect\citeauthoryear{{Lacy} et~al.,}{{Lacy}
  et~al.}{2013}]{Lacy2013}
{Lacy} M.,  et~al., 2013, \mn@doi [\apjs] {10.1088/0067-0049/208/2/24}, \href
  {http://adsabs.harvard.edu/abs/2013ApJS..208...24L} {208, 24}

\bibitem[\protect\citeauthoryear{{Lau}, {Prochaska}  \& {Hennawi}}{{Lau}
  et~al.}{2016}]{QPQ8}
{Lau} M.~W.,  {Prochaska} J.~X.,   {Hennawi} J.~F.,  2016, \mn@doi [\apjs]
  {10.3847/0067-0049/226/2/25}, \href
  {http://adsabs.harvard.edu/abs/2016ApJS..226...25L} {226, 25}

\bibitem[\protect\citeauthoryear{{Lawrence}}{{Lawrence}}{1991}]{Lawrence1991}
{Lawrence} A.,  1991, \mn@doi [\mnras] {10.1093/mnras/252.4.586}, \href
  {http://adsabs.harvard.edu/abs/1991MNRAS.252..586L} {252, 586}

\bibitem[\protect\citeauthoryear{{Lusso}, {Worseck}, {Hennawi}, {Prochaska},
  {Vignali}, {Stern}  \& {O'Meara}}{{Lusso} et~al.}{2015}]{lusso+15}
{Lusso} E.,  {Worseck} G.,  {Hennawi} J.~F.,  {Prochaska} J.~X.,  {Vignali} C.,
   {Stern} J.,   {O'Meara} J.~M.,  2015, \mn@doi [\mnras]
  {10.1093/mnras/stv516}, \href
  {http://adsabs.harvard.edu/abs/2015MNRAS.449.4204L} {449, 4204}

\bibitem[\protect\citeauthoryear{{Martin}, {Chang}, {Matuszewski}, {Morrissey},
  {Rahman}, {Moore}  \& {Steidel}}{{Martin} et~al.}{2014}]{Martin2014a}
{Martin} D.~C.,  {Chang} D.,  {Matuszewski} M.,  {Morrissey} P.,  {Rahman} S.,
  {Moore} A.,   {Steidel} C.~C.,  2014, \mn@doi [\apj]
  {10.1088/0004-637X/786/2/106}, \href
  {http://adsabs.harvard.edu/abs/2014ApJ...786..106M} {786, 106}

\bibitem[\protect\citeauthoryear{{Martin}, {Matuszewski}, {Morrissey}, {Neill},
  {Moore}, {Cantalupo}, {Prochaska}  \& {Chang}}{{Martin}
  et~al.}{2015}]{martin+15}
{Martin} D.~C.,  {Matuszewski} M.,  {Morrissey} P.,  {Neill} J.~D.,  {Moore}
  A.,  {Cantalupo} S.,  {Prochaska} J.~X.,   {Chang} D.,  2015, \mn@doi [\nat]
  {10.1038/nature14616}, \href
  {http://adsabs.harvard.edu/abs/2015Natur.524..192M} {524, 192}

\bibitem[\protect\citeauthoryear{{Matsuda} et~al.,}{{Matsuda}
  et~al.}{2005}]{Matsuda2005}
{Matsuda} Y.,  et~al., 2005, \mn@doi [\apjl] {10.1086/499071}, \href
  {http://adsabs.harvard.edu/abs/2005ApJ...634L.125M} {634, L125}

\bibitem[\protect\citeauthoryear{{Matsuda} et~al.,}{{Matsuda}
  et~al.}{2009}]{Matsuda2009}
{Matsuda} Y.,  et~al., 2009, \mn@doi [\mnras]
  {10.1111/j.1745-3933.2009.00764.x}, \href
  {http://adsabs.harvard.edu/abs/2009MNRAS.400L..66M} {400, L66}

\bibitem[\protect\citeauthoryear{{McCarthy}}{{McCarthy}}{1993}]{McCarthy1993}
{McCarthy} P.~J.,  1993, \mn@doi [\araa] {10.1146/annurev.aa.31.090193.003231},
  \href {http://adsabs.harvard.edu/abs/1993ARA%26A..31..639M} {31, 639}

\bibitem[\protect\citeauthoryear{{McCourt}, {Oh}, {O'Leary}  \&
  {Madigan}}{{McCourt} et~al.}{2016}]{McCourt2016}
{McCourt} M.,  {Oh} S.~P.,  {O'Leary} R.~M.,   {Madigan} A.-M.,  2016,
  preprint, \href {http://adsabs.harvard.edu/abs/2016arXiv161001164M} {}
  (\mn@eprint {arXiv} {1610.01164})

\bibitem[\protect\citeauthoryear{{Miley} et~al.,}{{Miley}
  et~al.}{2006}]{Miley2006}
{Miley} G.~K.,  et~al., 2006, \mn@doi [\apjl] {10.1086/508534}, \href
  {http://adsabs.harvard.edu/abs/2006ApJ...650L..29M} {650, L29}

\bibitem[\protect\citeauthoryear{{More}, {Diemer}  \& {Kravtsov}}{{More}
  et~al.}{2015}]{More2015}
{More} S.,  {Diemer} B.,   {Kravtsov} A.~V.,  2015, \mn@doi [\apj]
  {10.1088/0004-637X/810/1/36}, \href
  {http://adsabs.harvard.edu/abs/2015ApJ...810...36M} {810, 36}

\bibitem[\protect\citeauthoryear{{Mori}, {Umemura}  \& {Ferrara}}{{Mori}
  et~al.}{2004}]{Mori2004}
{Mori} M.,  {Umemura} M.,   {Ferrara} A.,  2004, \mn@doi [\apjl]
  {10.1086/425255}, \href {http://adsabs.harvard.edu/abs/2004ApJ...613L..97M}
  {613, L97}

\bibitem[\protect\citeauthoryear{{Moster}, {Naab}  \& {White}}{{Moster}
  et~al.}{2013}]{Moster2013}
{Moster} B.~P.,  {Naab} T.,   {White} S.~D.~M.,  2013, \mn@doi [\mnras]
  {10.1093/mnras/sts261}, \href
  {http://adsabs.harvard.edu/abs/2013MNRAS.428.3121M} {428, 3121}

\bibitem[\protect\citeauthoryear{{Nagao}, {Maiolino}  \& {Marconi}}{{Nagao}
  et~al.}{2006}]{Nagao2006}
{Nagao} T.,  {Maiolino} R.,   {Marconi} A.,  2006, \mn@doi [\aap]
  {10.1051/0004-6361:20065216}, \href
  {http://adsabs.harvard.edu/abs/2006A%26A...459...85N} {459, 85}

\bibitem[\protect\citeauthoryear{{Nelson}, {Genel}, {Pillepich},
  {Vogelsberger}, {Springel}  \& {Hernquist}}{{Nelson}
  et~al.}{2016}]{Nelson2016}
{Nelson} D.,  {Genel} S.,  {Pillepich} A.,  {Vogelsberger} M.,  {Springel} V.,
   {Hernquist} L.,  2016, \mn@doi [\mnras] {10.1093/mnras/stw1191}, \href
  {http://adsabs.harvard.edu/abs/2016MNRAS.tmp..884N} {}

\bibitem[\protect\citeauthoryear{{Nesvadba}, {Polletta}, {Lehnert}, {Bergeron},
  {De Breuck}, {Lagache}  \& {Omont}}{{Nesvadba} et~al.}{2011}]{Nesvadba2011}
{Nesvadba} N.~P.~H.,  {Polletta} M.,  {Lehnert} M.~D.,  {Bergeron} J.,  {De
  Breuck} C.,  {Lagache} G.,   {Omont} A.,  2011, \mn@doi [\mnras]
  {10.1111/j.1365-2966.2011.18862.x}, \href
  {http://adsabs.harvard.edu/abs/2011MNRAS.415.2359N} {415, 2359}

\bibitem[\protect\citeauthoryear{{Neufeld}}{{Neufeld}}{1990}]{Neufeld_1990}
{Neufeld} D.~A.,  1990, \mn@doi [\apj] {10.1086/168375}, \href
  {http://adsabs.harvard.edu/abs/1990ApJ...350..216N} {350, 216}

\bibitem[\protect\citeauthoryear{{North}, {Courbin}, {Eigenbrod}  \&
  {Chelouche}}{{North} et~al.}{2012}]{North2012}
{North} P.~L.,  {Courbin} F.,  {Eigenbrod} A.,   {Chelouche} D.,  2012, \mn@doi
  [\aap] {10.1051/0004-6361/201015153}, \href
  {http://adsabs.harvard.edu/abs/2012A%26A...542A..91N} {542, A91}

\bibitem[\protect\citeauthoryear{{O'Meara} et~al.,}{{O'Meara}
  et~al.}{2015}]{Omeara15}
{O'Meara} J.~M.,  et~al., 2015, \mn@doi [\aj] {10.1088/0004-6256/150/4/111},
  \href {http://adsabs.harvard.edu/abs/2015AJ....150..111O} {150, 111}

\bibitem[\protect\citeauthoryear{{Oke}}{{Oke}}{1974}]{Oke1974}
{Oke} J.~B.,  1974, \mn@doi [\apjs] {10.1086/190287}, \href
  {http://adsabs.harvard.edu/abs/1974ApJS...27...21O} {27, 21}

\bibitem[\protect\citeauthoryear{{Oser}, {Ostriker}, {Naab}, {Johansson}  \&
  {Burkert}}{{Oser} et~al.}{2010}]{Oser2010}
{Oser} L.,  {Ostriker} J.~P.,  {Naab} T.,  {Johansson} P.~H.,   {Burkert} A.,
  2010, \mn@doi [\apj] {10.1088/0004-637X/725/2/2312}, \href
  {http://adsabs.harvard.edu/abs/2010ApJ...725.2312O} {725, 2312}

\bibitem[\protect\citeauthoryear{{Osterbrock}}{{Osterbrock}}{1989}]{Osterbrock1989}
{Osterbrock} D.~E.,  1989, {Astrophysics of gaseous nebulae and active galactic
  nuclei}

\bibitem[\protect\citeauthoryear{{Ouchi} et~al.,}{{Ouchi}
  et~al.}{2010}]{Ouchi2010}
{Ouchi} M.,  et~al., 2010, \mn@doi [\apj] {10.1088/0004-637X/723/1/869}, \href
  {http://adsabs.harvard.edu/abs/2010ApJ...723..869O} {723, 869}

\bibitem[\protect\citeauthoryear{{P{\^a}ris} et~al.,}{{P{\^a}ris}
  et~al.}{2012}]{Paris2012}
{P{\^a}ris} I.,  et~al., 2012, \mn@doi [\aap] {10.1051/0004-6361/201220142},
  \href {http://adsabs.harvard.edu/abs/2012A%26A...548A..66P} {548, A66}

\bibitem[\protect\citeauthoryear{{Peebles}}{{Peebles}}{1969}]{peebles69}
{Peebles} P.~J.~E.,  1969, \mn@doi [\apj] {10.1086/149876}, \href
  {http://adsabs.harvard.edu/abs/1969ApJ...155..393P} {155, 393}

\bibitem[\protect\citeauthoryear{{Persic}, {Salucci}  \& {Stel}}{{Persic}
  et~al.}{1996}]{Persic1996}
{Persic} M.,  {Salucci} P.,   {Stel} F.,  1996, \mn@doi [\mnras]
  {10.1093/mnras/281.1.27}, \href
  {http://adsabs.harvard.edu/abs/1996MNRAS.281...27P} {281, 27}

\bibitem[\protect\citeauthoryear{{Pilkington} \& {Scott}}{{Pilkington} \&
  {Scott}}{1965}]{Pilkington1965}
{Pilkington} J.~D.~H.,  {Scott} J.~F.,  1965, \memras, \href
  {http://cdsads.u-strasbg.fr/abs/1965MmRAS..69..183P} {69, 183}

\bibitem[\protect\citeauthoryear{{Planck Collaboration} et~al.,}{{Planck
  Collaboration} et~al.}{2014}]{Planck2014}
{Planck Collaboration} et~al., 2014, \mn@doi [\aap]
  {10.1051/0004-6361/201321591}, \href
  {http://adsabs.harvard.edu/abs/2014A%26A...571A..16P} {571, A16}

\bibitem[\protect\citeauthoryear{{Pontzen}, {Ro{\v s}kar}, {Stinson}, {Woods},
  {Reed}, {Coles}  \& {Quinn}}{{Pontzen} et~al.}{2013}]{pynbody}
{Pontzen} A.,  {Ro{\v s}kar} R.,  {Stinson} G.~S.,  {Woods} R.,  {Reed} D.~M.,
  {Coles} J.,   {Quinn} T.~R.,  2013, {pynbody: Astrophysics Simulation
  Analysis for Python}

\bibitem[\protect\citeauthoryear{{Porciani}, {Dekel}  \& {Hoffman}}{{Porciani}
  et~al.}{2002}]{porciani02}
{Porciani} C.,  {Dekel} A.,   {Hoffman} Y.,  2002, \mn@doi [\mnras]
  {10.1046/j.1365-8711.2002.05306.x}, \href
  {http://adsabs.harvard.edu/abs/2002MNRAS.332..339P} {332, 339}

\bibitem[\protect\citeauthoryear{{Prescott}, {Kashikawa}, {Dey}  \&
  {Matsuda}}{{Prescott} et~al.}{2008}]{Prescott2008}
{Prescott} M.~K.~M.,  {Kashikawa} N.,  {Dey} A.,   {Matsuda} Y.,  2008, \mn@doi
  [\apjl] {10.1086/588606}, \href
  {http://adsabs.harvard.edu/abs/2008ApJ...678L..77P} {678, L77}

\bibitem[\protect\citeauthoryear{{Prescott}, {Dey}  \& {Jannuzi}}{{Prescott}
  et~al.}{2009}]{Prescott2009}
{Prescott} M.~K.~M.,  {Dey} A.,   {Jannuzi} B.~T.,  2009, \mn@doi [\apj]
  {10.1088/0004-637X/702/1/554}, \href
  {http://adsabs.harvard.edu/abs/2009ApJ...702..554P} {702, 554}

\bibitem[\protect\citeauthoryear{{Prescott}, {Smith}, {Schmidt}  \&
  {Dey}}{{Prescott} et~al.}{2011}]{Prescott2011}
{Prescott} M.~K.~M.,  {Smith} P.~S.,  {Schmidt} G.~D.,   {Dey} A.,  2011,
  \mn@doi [\apjl] {10.1088/2041-8205/730/2/L25}, \href
  {http://adsabs.harvard.edu/abs/2011ApJ...730L..25P} {730, L25}

\bibitem[\protect\citeauthoryear{{Prescott}, {Dey}  \& {Jannuzi}}{{Prescott}
  et~al.}{2013}]{Prescott2013}
{Prescott} M.~K.~M.,  {Dey} A.,   {Jannuzi} B.~T.,  2013, \mn@doi [\apj]
  {10.1088/0004-637X/762/1/38}, \href
  {http://adsabs.harvard.edu/abs/2013ApJ...762...38P} {762, 38}

\bibitem[\protect\citeauthoryear{{Prescott}, {Martin}  \& {Dey}}{{Prescott}
  et~al.}{2015a}]{Prescott2015}
{Prescott} M.~K.~M.,  {Martin} C.~L.,   {Dey} A.,  2015a, \mn@doi [\apj]
  {10.1088/0004-637X/799/1/62}, \href
  {http://adsabs.harvard.edu/abs/2015ApJ...799...62P} {799, 62}

\bibitem[\protect\citeauthoryear{{Prescott}, {Martin}  \& {Dey}}{{Prescott}
  et~al.}{2015b}]{prescott2015a}
{Prescott} M.~K.~M.,  {Martin} C.~L.,   {Dey} A.,  2015b, \mn@doi [\apj]
  {10.1088/0004-637X/799/1/62}, \href
  {http://adsabs.harvard.edu/abs/2015ApJ...799...62P} {799, 62}

\bibitem[\protect\citeauthoryear{{Prochaska} \& {Hennawi}}{{Prochaska} \&
  {Hennawi}}{2009}]{QPQ3}
{Prochaska} J.~X.,  {Hennawi} J.~F.,  2009, \mn@doi [\apj]
  {10.1088/0004-637X/690/2/1558}, \href
  {http://adsabs.harvard.edu/abs/2009ApJ...690.1558P} {690, 1558}

\bibitem[\protect\citeauthoryear{{Prochaska}, {Weiner}, {Chen}, {Mulchaey}  \&
  {Cooksey}}{{Prochaska} et~al.}{2011}]{pwc+11}
{Prochaska} J.~X.,  {Weiner} B.,  {Chen} H.-W.,  {Mulchaey} J.,   {Cooksey} K.,
   2011, \mn@doi [\apj] {10.1088/0004-637X/740/2/91}, \href
  {http://adsabs.harvard.edu/abs/2011ApJ...740...91P} {740, 91}

\bibitem[\protect\citeauthoryear{{Prochaska}, {Hennawi}  \&
  {Simcoe}}{{Prochaska} et~al.}{2013a}]{QPQ5}
{Prochaska} J.~X.,  {Hennawi} J.~F.,   {Simcoe} R.~A.,  2013a, \mn@doi [\apjl]
  {10.1088/2041-8205/762/2/L19}, \href
  {http://adsabs.harvard.edu/abs/2013ApJ...762L..19P} {762, L19 (QPQ5)}

\bibitem[\protect\citeauthoryear{{Prochaska} et~al.,}{{Prochaska}
  et~al.}{2013b}]{qpq6}
{Prochaska} J.~X.,  et~al., 2013b, \mn@doi [\apj]
  {10.1088/0004-637X/776/2/136}, \href
  {http://adsabs.harvard.edu/abs/2013ApJ...776..136P} {776, 136}

\bibitem[\protect\citeauthoryear{{Prochaska}, {Lau}  \& {Hennawi}}{{Prochaska}
  et~al.}{2014}]{QPQ7}
{Prochaska} J.~X.,  {Lau} M.~W.,   {Hennawi} J.~F.,  2014, \mn@doi [\apj]
  {10.1088/0004-637X/796/2/140}, \href
  {http://adsabs.harvard.edu/abs/2014ApJ...796..140P} {796, 140}

\bibitem[\protect\citeauthoryear{{Richards} et~al.,}{{Richards}
  et~al.}{2011}]{Richards2011}
{Richards} G.~T.,  et~al., 2011, \mn@doi [\aj] {10.1088/0004-6256/141/5/167},
  \href {http://adsabs.harvard.edu/abs/2011AJ....141..167R} {141, 167}

\bibitem[\protect\citeauthoryear{{Richardson}, {Zheng}, {Chatterjee}, {Nagai}
  \& {Shen}}{{Richardson} et~al.}{2012}]{richardson12}
{Richardson} J.,  {Zheng} Z.,  {Chatterjee} S.,  {Nagai} D.,   {Shen} Y.,
  2012, \mn@doi [\apj] {10.1088/0004-637X/755/1/30}, \href
  {http://adsabs.harvard.edu/abs/2012ApJ...755...30R} {755, 30}

\bibitem[\protect\citeauthoryear{{Roche}, {Humphrey}  \& {Binette}}{{Roche}
  et~al.}{2014}]{Roche2014}
{Roche} N.,  {Humphrey} A.,   {Binette} L.,  2014, \mn@doi [\mnras]
  {10.1093/mnras/stu1430}, \href
  {http://adsabs.harvard.edu/abs/2014MNRAS.443.3795R} {443, 3795}

\bibitem[\protect\citeauthoryear{{Rosdahl} \& {Blaizot}}{{Rosdahl} \&
  {Blaizot}}{2012}]{Rosdahl12}
{Rosdahl} J.,  {Blaizot} J.,  2012, \mn@doi [\mnras]
  {10.1111/j.1365-2966.2012.20883.x}, \href
  {http://adsabs.harvard.edu/abs/2012MNRAS.423..344R} {423, 344}

\bibitem[\protect\citeauthoryear{{Rubin}, {Prochaska}, {Koo}, {Phillips},
  {Martin}  \& {Winstrom}}{{Rubin} et~al.}{2014}]{Rubin2014}
{Rubin} K.~H.~R.,  {Prochaska} J.~X.,  {Koo} D.~C.,  {Phillips} A.~C.,
  {Martin} C.~L.,   {Winstrom} L.~O.,  2014, \mn@doi [\apj]
  {10.1088/0004-637X/794/2/156}, \href
  {http://adsabs.harvard.edu/abs/2014ApJ...794..156R} {794, 156}

\bibitem[\protect\citeauthoryear{{Rudie} et~al.,}{{Rudie}
  et~al.}{2012}]{rudie12}
{Rudie} G.~C.,  et~al., 2012, \mn@doi [\apj] {10.1088/0004-637X/750/1/67},
  \href {http://adsabs.harvard.edu/abs/2012ApJ...750...67R} {750, 67}

\bibitem[\protect\citeauthoryear{{Saito}, {Shimasaku}, {Okamura}, {Ouchi},
  {Akiyama}  \& {Yoshida}}{{Saito} et~al.}{2006}]{Saito2006}
{Saito} T.,  {Shimasaku} K.,  {Okamura} S.,  {Ouchi} M.,  {Akiyama} M.,
  {Yoshida} M.,  2006, \mn@doi [\apj] {10.1086/505678}, \href
  {http://adsabs.harvard.edu/abs/2006ApJ...648...54S} {648, 54}

\bibitem[\protect\citeauthoryear{{Scannapieco}, {Tissera}, {White}  \&
  {Springel}}{{Scannapieco} et~al.}{2008}]{Scannapieco2008}
{Scannapieco} C.,  {Tissera} P.~B.,  {White} S.~D.~M.,   {Springel} V.,  2008,
  \mn@doi [\mnras] {10.1111/j.1365-2966.2008.13678.x}, \href
  {http://adsabs.harvard.edu/abs/2008MNRAS.389.1137S} {389, 1137}

\bibitem[\protect\citeauthoryear{{Schaye} et~al.,}{{Schaye}
  et~al.}{2015}]{Schaye2015}
{Schaye} J.,  et~al., 2015, \mn@doi [\mnras] {10.1093/mnras/stu2058}, \href
  {http://adsabs.harvard.edu/abs/2015MNRAS.446..521S} {446, 521}

\bibitem[\protect\citeauthoryear{{Shapley}, {Steidel}, {Pettini}  \&
  {Adelberger}}{{Shapley} et~al.}{2003}]{shapley03}
{Shapley} A.~E.,  {Steidel} C.~C.,  {Pettini} M.,   {Adelberger} K.~L.,  2003,
  \mn@doi [\apj] {10.1086/373922}, \href
  {http://adsabs.harvard.edu/cgi-bin/nph-bib_query?bibcode=2003ApJ...588...65S&db_key=AST}
  {588, 65}

\bibitem[\protect\citeauthoryear{{Sharma} \& {Steinmetz}}{{Sharma} \&
  {Steinmetz}}{2005}]{Sharma2005}
{Sharma} S.,  {Steinmetz} M.,  2005, \mn@doi [\apj] {10.1086/430660}, \href
  {http://adsabs.harvard.edu/abs/2005ApJ...628...21S} {628, 21}

\bibitem[\protect\citeauthoryear{{Shen}, {Wadsley}  \& {Stinson}}{{Shen}
  et~al.}{2010}]{Shen2010}
{Shen} S.,  {Wadsley} J.,   {Stinson} G.,  2010, \mn@doi [\mnras]
  {10.1111/j.1365-2966.2010.17047.x}, \href
  {http://adsabs.harvard.edu/abs/2010MNRAS.407.1581S} {407, 1581}

\bibitem[\protect\citeauthoryear{{Shen} et~al.,}{{Shen}
  et~al.}{2016}]{Shen2016}
{Shen} Y.,  et~al., 2016, \mn@doi [\apj] {10.3847/0004-637X/831/1/7}, \href
  {http://adsabs.harvard.edu/abs/2016ApJ...831....7S} {831, 7}

\bibitem[\protect\citeauthoryear{{Sijacki}, {Springel}, {Di Matteo}  \&
  {Hernquist}}{{Sijacki} et~al.}{2007}]{Sijacki2007}
{Sijacki} D.,  {Springel} V.,  {Di Matteo} T.,   {Hernquist} L.,  2007, \mn@doi
  [\mnras] {10.1111/j.1365-2966.2007.12153.x}, \href
  {http://adsabs.harvard.edu/abs/2007MNRAS.380..877S} {380, 877}

\bibitem[\protect\citeauthoryear{{Silk} \& {Rees}}{{Silk} \&
  {Rees}}{1998}]{SilkRees98}
{Silk} J.,  {Rees} M.~J.,  1998, \aap, \href
  {http://adsabs.harvard.edu/abs/1998A%26A...331L...1S} {331, L1}

\bibitem[\protect\citeauthoryear{{Skrutskie} et~al.,}{{Skrutskie}
  et~al.}{2006}]{Skrutskie2006}
{Skrutskie} M.~F.,  et~al., 2006, \mn@doi [\aj] {10.1086/498708}, \href
  {http://adsabs.harvard.edu/abs/2006AJ....131.1163S} {131, 1163}

\bibitem[\protect\citeauthoryear{{Steidel}, {Erb}, {Shapley}, {Pettini},
  {Reddy}, {Bogosavljevi{\'c}}, {Rudie}  \& {Rakic}}{{Steidel}
  et~al.}{2010}]{Steidel2010}
{Steidel} C.~C.,  {Erb} D.~K.,  {Shapley} A.~E.,  {Pettini} M.,  {Reddy} N.,
  {Bogosavljevi{\'c}} M.,  {Rudie} G.~C.,   {Rakic} O.,  2010, \mn@doi [\apj]
  {10.1088/0004-637X/717/1/289}, \href
  {http://adsabs.harvard.edu/abs/2010ApJ...717..289S} {717, 289}

\bibitem[\protect\citeauthoryear{{Stern}, {Hennawi}, {Prochaska}  \&
  {Werk}}{{Stern} et~al.}{2016}]{Stern2016}
{Stern} J.,  {Hennawi} J.~F.,  {Prochaska} J.~X.,   {Werk} J.~K.,  2016,
  preprint, \href {http://adsabs.harvard.edu/abs/2016arXiv160402168S} {}
  (\mn@eprint {arXiv} {1604.02168})

\bibitem[\protect\citeauthoryear{{Stewart}, {Kaufmann}, {Bullock}, {Barton},
  {Maller}, {Diemand}  \& {Wadsley}}{{Stewart} et~al.}{2011}]{stewart11}
{Stewart} K.~R.,  {Kaufmann} T.,  {Bullock} J.~S.,  {Barton} E.~J.,  {Maller}
  A.~H.,  {Diemand} J.,   {Wadsley} J.,  2011, \mn@doi [\apj]
  {10.1088/0004-637X/738/1/39}, \href
  {http://adsabs.harvard.edu/abs/2011ApJ...738...39S} {738, 39}

\bibitem[\protect\citeauthoryear{{Stewart} et~al.,}{{Stewart}
  et~al.}{2016}]{stewart+16}
{Stewart} K.,  et~al., 2016, preprint, \href
  {http://adsabs.harvard.edu/abs/2016arXiv160608542S} {} (\mn@eprint {arXiv}
  {1606.08542})

\bibitem[\protect\citeauthoryear{{Stinson}, {Seth}, {Katz}, {Wadsley},
  {Governato}  \& {Quinn}}{{Stinson} et~al.}{2006}]{Stinson2006}
{Stinson} G.,  {Seth} A.,  {Katz} N.,  {Wadsley} J.,  {Governato} F.,   {Quinn}
  T.,  2006, \mn@doi [\mnras] {10.1111/j.1365-2966.2006.11097.x}, \href
  {http://adsabs.harvard.edu/abs/2006MNRAS.373.1074S} {373, 1074}

\bibitem[\protect\citeauthoryear{{Stinson}, {Brook}, {Macci{\`o}}, {Wadsley},
  {Quinn}  \& {Couchman}}{{Stinson} et~al.}{2013}]{Stinson2013}
{Stinson} G.~S.,  {Brook} C.,  {Macci{\`o}} A.~V.,  {Wadsley} J.,  {Quinn}
  T.~R.,   {Couchman} H.~M.~P.,  2013, \mn@doi [\mnras] {10.1093/mnras/sts028},
  \href {http://adsabs.harvard.edu/abs/2013MNRAS.428..129S} {428, 129}

\bibitem[\protect\citeauthoryear{{Stockton}, {Fu}  \& {Canalizo}}{{Stockton}
  et~al.}{2006}]{Stockton2006}
{Stockton} A.,  {Fu} H.,   {Canalizo} G.,  2006, \mn@doi [\nar]
  {10.1016/j.newar.2006.06.024}, \href
  {http://adsabs.harvard.edu/abs/2006NewAR..50..694S} {50, 694}

\bibitem[\protect\citeauthoryear{{Stuik}, {Bacon}, {Conzelmann}, {Delabre},
  {Fedrigo}, {Hubin}, {Le Louarn}  \& {Str{\"o}bele}}{{Stuik}
  et~al.}{2006}]{Stuik2006}
{Stuik} R.,  {Bacon} R.,  {Conzelmann} R.,  {Delabre} B.,  {Fedrigo} E.,
  {Hubin} N.,  {Le Louarn} M.,   {Str{\"o}bele} S.,  2006, \mn@doi [\nar]
  {10.1016/j.newar.2005.10.015}, \href
  {http://adsabs.harvard.edu/abs/2006NewAR..49..618S} {49, 618}

\bibitem[\protect\citeauthoryear{{Swinbank} et~al.,}{{Swinbank}
  et~al.}{2015}]{Swinbank2015}
{Swinbank} A.~M.,  et~al., 2015, \mn@doi [\mnras] {10.1093/mnras/stv366}, \href
  {http://adsabs.harvard.edu/abs/2015MNRAS.449.1298S} {449, 1298}

\bibitem[\protect\citeauthoryear{{Taniguchi} \& {Shioya}}{{Taniguchi} \&
  {Shioya}}{2000}]{TanShi2000}
{Taniguchi} Y.,  {Shioya} Y.,  2000, \mn@doi [\apjl] {10.1086/312557}, \href
  {http://adsabs.harvard.edu/abs/2000ApJ...532L..13T} {532, L13}

\bibitem[\protect\citeauthoryear{{Tonnesen} \& {Bryan}}{{Tonnesen} \&
  {Bryan}}{2011}]{Tonnesen2011}
{Tonnesen} S.,  {Bryan} G.~L.,  2011, \mn@doi [\apj]
  {10.1088/0004-637X/727/2/133}, \href
  {http://adsabs.harvard.edu/abs/2011ApJ...727..133T} {727, 133}

\bibitem[\protect\citeauthoryear{{Trainor} \& {Steidel}}{{Trainor} \&
  {Steidel}}{2012}]{Trainor2012}
{Trainor} R.~F.,  {Steidel} C.~C.,  2012, \mn@doi [\apj]
  {10.1088/0004-637X/752/1/39}, \href
  {http://adsabs.harvard.edu/abs/2012ApJ...752...39T} {752, 39}

\bibitem[\protect\citeauthoryear{{Tumlinson} et~al.,}{{Tumlinson}
  et~al.}{2013}]{tumlinson+13}
{Tumlinson} J.,  et~al., 2013, \mn@doi [\apj] {10.1088/0004-637X/777/1/59},
  \href {http://adsabs.harvard.edu/abs/2013ApJ...777...59T} {777, 59}

\bibitem[\protect\citeauthoryear{{Venemans} et~al.,}{{Venemans}
  et~al.}{2007}]{Venemans2007}
{Venemans} B.~P.,  et~al., 2007, \mn@doi [\aap] {10.1051/0004-6361:20053941},
  \href {http://adsabs.harvard.edu/abs/2007A%26A...461..823V} {461, 823}

\bibitem[\protect\citeauthoryear{{Verhamme}, {Schaerer}  \&
  {Maselli}}{{Verhamme} et~al.}{2006}]{Verhamme2006}
{Verhamme} A.,  {Schaerer} D.,   {Maselli} A.,  2006, \mn@doi [\aap]
  {10.1051/0004-6361:20065554}, \href
  {http://adsabs.harvard.edu/abs/2006A%26A...460..397V} {460, 397}

\bibitem[\protect\citeauthoryear{{Verhamme}, {Dubois}, {Blaizot}, {Garel},
  {Bacon}, {Devriendt}, {Guiderdoni}  \& {Slyz}}{{Verhamme}
  et~al.}{2012}]{Verhamme2012}
{Verhamme} A.,  {Dubois} Y.,  {Blaizot} J.,  {Garel} T.,  {Bacon} R.,
  {Devriendt} J.,  {Guiderdoni} B.,   {Slyz} A.,  2012, \mn@doi [\aap]
  {10.1051/0004-6361/201218783}, \href
  {http://adsabs.harvard.edu/abs/2012A%26A...546A.111V} {546, A111}

\bibitem[\protect\citeauthoryear{{V{\'e}ron-Cetty} \&
  {V{\'e}ron}}{{V{\'e}ron-Cetty} \& {V{\'e}ron}}{2010}]{VeronCetty2010}
{V{\'e}ron-Cetty} M.-P.,  {V{\'e}ron} P.,  2010, \mn@doi [\aap]
  {10.1051/0004-6361/201014188}, \href
  {http://adsabs.harvard.edu/abs/2010A%26A...518A..10V} {518, A10}

\bibitem[\protect\citeauthoryear{{Villar-Mart{\'{\i}}n}, {Vernet}, {di Serego
  Alighieri}, {Fosbury}, {Humphrey}  \& {Pentericci}}{{Villar-Mart{\'{\i}}n}
  et~al.}{2003}]{VillarMartin2003}
{Villar-Mart{\'{\i}}n} M.,  {Vernet} J.,  {di Serego Alighieri} S.,  {Fosbury}
  R.,  {Humphrey} A.,   {Pentericci} L.,  2003, \mn@doi [\mnras]
  {10.1046/j.1365-2966.2003.07090.x}, \href
  {http://adsabs.harvard.edu/abs/2003MNRAS.346..273V} {346, 273}

\bibitem[\protect\citeauthoryear{{Vogelsberger} et~al.,}{{Vogelsberger}
  et~al.}{2014}]{Vogelsberger2014}
{Vogelsberger} M.,  et~al., 2014, \mn@doi [\mnras] {10.1093/mnras/stu1536},
  \href {http://adsabs.harvard.edu/abs/2014MNRAS.444.1518V} {444, 1518}

\bibitem[\protect\citeauthoryear{{Vogt} et~al.,}{{Vogt} et~al.}{1994}]{vogt94}
{Vogt} S.~S.,  et~al., 1994, in Proc. SPIE Instrumentation in Astronomy VIII,
  David L. Crawford; Eric R. Craine; Eds., Volume 2198, p. 362. pp 362--+

\bibitem[\protect\citeauthoryear{{Wadsley}, {Stadel}  \& {Quinn}}{{Wadsley}
  et~al.}{2004}]{Wadsley2004}
{Wadsley} J.~W.,  {Stadel} J.,   {Quinn} T.,  2004, \mn@doi [\na]
  {10.1016/j.newast.2003.08.004}, \href
  {http://adsabs.harvard.edu/abs/2004NewA....9..137W} {9, 137}

\bibitem[\protect\citeauthoryear{{Wadsley}, {Veeravalli}  \&
  {Couchman}}{{Wadsley} et~al.}{2008}]{Wadsley2008}
{Wadsley} J.~W.,  {Veeravalli} G.,   {Couchman} H.~M.~P.,  2008, \mn@doi
  [\mnras] {10.1111/j.1365-2966.2008.13260.x}, \href
  {http://adsabs.harvard.edu/abs/2008MNRAS.387..427W} {387, 427}

\bibitem[\protect\citeauthoryear{{Wang}, {Dutton}, {Stinson}, {Macci{\`o}},
  {Penzo}, {Kang}, {Keller}  \& {Wadsley}}{{Wang} et~al.}{2015}]{Nihao2015}
{Wang} L.,  {Dutton} A.~A.,  {Stinson} G.~S.,  {Macci{\`o}} A.~V.,  {Penzo} C.,
   {Kang} X.,  {Keller} B.~W.,   {Wadsley} J.,  2015, \mn@doi [\mnras]
  {10.1093/mnras/stv1937}, \href
  {http://adsabs.harvard.edu/abs/2015MNRAS.454...83W} {454, 83}

\bibitem[\protect\citeauthoryear{{Weilbacher}, {Streicher}, {Urrutia},
  {P{\'e}contal-Rousset}, {Jarno}  \& {Bacon}}{{Weilbacher}
  et~al.}{2014}]{Weilbacher2014}
{Weilbacher} P.~M.,  {Streicher} O.,  {Urrutia} T.,  {P{\'e}contal-Rousset} A.,
   {Jarno} A.,   {Bacon} R.,  2014, in {Manset} N.,  {Forshay} P.,  eds,
  Astronomical Society of the Pacific Conference Series Vol. 485, Astronomical
  Data Analysis Software and Systems XXIII. p.~451 (\mn@eprint {arXiv}
  {1507.00034})

\bibitem[\protect\citeauthoryear{{White} \& {Becker}}{{White} \&
  {Becker}}{1992}]{WhiteBecker1992}
{White} R.~L.,  {Becker} R.~H.,  1992, \mn@doi [\apjs] {10.1086/191656}, \href
  {http://adsabs.harvard.edu/abs/1992ApJS...79..331W} {79, 331}

\bibitem[\protect\citeauthoryear{{White} \& {Rees}}{{White} \&
  {Rees}}{1978}]{WR1978}
{White} S.~D.~M.,  {Rees} M.~J.,  1978, \mn@doi [\mnras]
  {10.1093/mnras/183.3.341}, \href
  {http://adsabs.harvard.edu/abs/1978MNRAS.183..341W} {183, 341}

\bibitem[\protect\citeauthoryear{{White} et~al.,}{{White}
  et~al.}{2012}]{white12}
{White} M.,  et~al., 2012, \mn@doi [\mnras] {10.1111/j.1365-2966.2012.21251.x},
  \href {http://adsabs.harvard.edu/abs/2012MNRAS.424..933W} {424, 933}

\bibitem[\protect\citeauthoryear{{Yang}, {Zabludoff}, {Dav{\'e}}, {Eisenstein},
  {Pinto}, {Katz}, {Weinberg}  \& {Barton}}{{Yang} et~al.}{2006}]{Yang2006}
{Yang} Y.,  {Zabludoff} A.~I.,  {Dav{\'e}} R.,  {Eisenstein} D.~J.,  {Pinto}
  P.~A.,  {Katz} N.,  {Weinberg} D.~H.,   {Barton} E.~J.,  2006, \mn@doi [\apj]
  {10.1086/497898}, \href {http://adsabs.harvard.edu/abs/2006ApJ...640..539Y}
  {640, 539}

\bibitem[\protect\citeauthoryear{{Yang}, {Zabludoff}, {Tremonti}, {Eisenstein}
  \& {Dav{\'e}}}{{Yang} et~al.}{2009}]{Yang2009}
{Yang} Y.,  {Zabludoff} A.,  {Tremonti} C.,  {Eisenstein} D.,   {Dav{\'e}} R.,
  2009, \mn@doi [\apj] {10.1088/0004-637X/693/2/1579}, \href
  {http://adsabs.harvard.edu/abs/2009ApJ...693.1579Y} {693, 1579}

\bibitem[\protect\citeauthoryear{{Yang}, {Zabludoff}, {Jahnke}  \&
  {Dav{\'e}}}{{Yang} et~al.}{2014}]{Yang2014}
{Yang} Y.,  {Zabludoff} A.,  {Jahnke} K.,   {Dav{\'e}} R.,  2014, \mn@doi
  [\apj] {10.1088/0004-637X/793/2/114}, \href
  {http://adsabs.harvard.edu/abs/2014ApJ...793..114Y} {793, 114}

\bibitem[\protect\citeauthoryear{{Zhang}, {Yang}, {Faltenbacher}, {Springel},
  {Lin}  \& {Wang}}{{Zhang} et~al.}{2009}]{Zhang2009}
{Zhang} Y.,  {Yang} X.,  {Faltenbacher} A.,  {Springel} V.,  {Lin} W.,   {Wang}
  H.,  2009, \mn@doi [\apj] {10.1088/0004-637X/706/1/747}, \href
  {http://adsabs.harvard.edu/abs/2009ApJ...706..747Z} {706, 747}

\bibitem[\protect\citeauthoryear{{Zjupa} \& {Springel}}{{Zjupa} \&
  {Springel}}{2017}]{Zjupa2016}
{Zjupa} J.,  {Springel} V.,  2017, \mn@doi [\mnras] {10.1093/mnras/stw2945},
  \href {http://adsabs.harvard.edu/abs/2017MNRAS.466.1625Z} {466, 1625}

\bibitem[\protect\citeauthoryear{{van den Bosch}, {Abel}, {Croft}, {Hernquist}
  \& {White}}{{van den Bosch} et~al.}{2002}]{vandenBosch2002}
{van den Bosch} F.~C.,  {Abel} T.,  {Croft} R.~A.~C.,  {Hernquist} L.,
  {White} S.~D.~M.,  2002, \mn@doi [\apj] {10.1086/341619}, \href
  {http://adsabs.harvard.edu/abs/2002ApJ...576...21V} {576, 21}

\makeatother
\end{thebibliography}




\appendix

\section{The Systemic Redshift of the Quasar \qso}
\label{appZ}

Because of shifts between broad lines and the presence/shape of the broad lines themselves, it is notoriously difficult to obtain precise estimates
of the systemic redshifts of quasars (e.g., \citealt{Hewett2010, Bolton2012, Paris2012, Shen2016}).
The redshift reported in the literature for \qso\ is $z=3.1589\pm0.0003$ (SDSS-DR12 Quasar 
catalogue\footnote{\url{http://www.sdss.org/dr12/algorithms/boss-dr12-quasar-catalog/}}), 
or previously $z=3.1678\pm0.0008$ (\citealt{Adelman2009}).
However, the uncertainty on the redshift seems underestimated in both works, given the absence of data covering narrow emission lines. 
For this reason, in this study, we use as systemic redshift the estimate from the \civ line after correcting for the known luminosity-dependent blueshift of this 
line with respect to systemic (\citealt{Richards2011, Shen2016}), i.e. $z_{\rm SDSS\,J1020+1040}=3.164\pm0.006$. Indeed, 
\citet{Shen2016} has shown that, according to the quasar luminosity, the redshift determined from \civ is shifted by 
${v=14-122\times({\rm log}(\lambda L_{1700\angstrom})-45)}$~km~s$^{-1}$ 
(where $L_{1700\angstrom}$ is the monochromatic luminosity at rest-frame 1700\AA). In this formula, 
negative velocities indicate blueshifts from systemic, and thus a positive correction needed to get the correct
systemic redshift.
From the MUSE spectrum of \qso\ extracted within a circular region ($R=1.5\arcsec$; see Fig.~\ref{figsources}), we obtain 
$f_{1700\angstrom}=(19.4\pm0.5)\times10^{-17}$\unitfluxdcgs\  \footnote{Note that this value is in agreement with the SDSS flux of 
$f_{1700\angstrom}=(17.2\pm1.3)\times10^{-17}$\unitfluxdcgs. Note also that the $i$ magnitudes (fiber magnitude within circular 
aperture with radius of $1.5\arcsec$) from our MUSE data and from SDSS agree, i.e. $17.94 \pm 0.03$ and $17.98 \pm 0.02$, respectively. 
In this regard, note that the quasar \qso\ has been cataloged as broad absorption quasar (BAL, \citealt{Gibson2009}) and its spectrum shows 
an increase of a factor $1.8$ (at rest-frame $1700\angstrom$) between the SDSS (UT December 26th 2004), and BOSS (UT January 31st 2012) 
observations. On the other hand, our MUSE observations are in agreement with SDSS. The shape of the broad-lines (e.g. \lya, \civ) slightly 
changes between the different observations. We checked our redshift estimate against these changes, but they do not affect our calculation, 
mainly because of the logarithmic dependence on $L_{1700\angstrom}$. }, 
and thus $\lambda L_{1700\angstrom} = (2.87\pm0.07)\times10^{46}$~erg~s$^{-1}$. 
Using this value, we estimate an expected blueshift of -164~km~s$^{-1}$.
We then estimate the peak of the \civ line, as specified in \citet{Shen2016}, and obtained $z=3.162$. 
Adding the previously determined shift gives then $z_{\rm SDSS\,J1020+1040}=3.164\pm0.006$.
The error on the redshift reflects the 415~km~s$^{-1}$ intrinsic uncertainty of the empirical calibration in \citet{Shen2016}.
To have a firm interpretation of the system surrounding \qso, a better determination of the redshift is certainly needed. 
However, we note that the narrow and bright emission evident after the PSF subtraction in close proximity to the bright quasar, 
which we interpret as the extended emission line regions (EELR; \citealt{Husemann2014}) associated with \qso, 
has a redshift in agreement with the above calculation, i.e. $z=3.167\pm0.001$ (see section~\ref{sec:gb_lya_neb}).
In this work, we thus assume that the quasar and the ELAN are at the same redshift, and thus physically associated.

\section{The Quasar \qso\  is {\it Not} Radio-Loud}
\label{appB}

As discussed in section~\ref{sec:obs}, while building our survey QSO MUSEUM, we have carefully 
selected our targeted quasars to try to cover any dependence on the presence of an ELAN with radio activity, 
companion sources, luminosity, etc. Thus, 
we collect from the literature these information, if present, and check for any 
inconsistency.

In the case of \qso, we found that it has been so far considered as the counterpart of a strong radio emission 
($F=515$~mJy at 1.4~GHz; \citealt{Condon1998}), and thus as a radio-loud object.  However, the available radio data 
together with our MUSE data clearly rule out an association of the quasar \qso\ with the strong radio emission, 
which instead appear related to an unknown HzRG at $\sim30\arcsec$ from the quasar \qso, but at much lower redshift. 

Specifically, \qso\ was initially catalogued as counterpart of the radio emission within the Fourth Cambridge Survey (\citealt{Pilkington1965}), 
with the name 4C~10.29, and subsequently in the Parkes survey (\citealt{Ekers1969}), with the name PKS~1017+109.
Also,  \citet{CondonBroderick1985, CondonBroderick1986} and \citet{WhiteBecker1992} reported \qso\ as the counterpart of an emission with  $F=538$~mJy at 1.4~GHz.
None the less, the spatial resolution of these datasets were not optimal for source matching at arcsec precision, 
e.g. \citet{CondonBroderick1985, CondonBroderick1986} has a resolution of $700\arcsec$.

Also the Texas Survey of Radio Sources at 365~MHz (TXS, \citealt{Douglas1996}) and the NRAO VLA Sky Survey at 1.4~GHz 
(NVSS, \citealt{Condon1998}),  with a resolution of $5\arcsec$ and $45\arcsec$, reported the quasar as counterpart of the radio emission.
This happened even though both these surveys, with a position precision estimated to be $<1\arcsec$, were able to detect the position shift of $\sim 33\arcsec$
between their reported position for the radio emission and the quasar \qso\  (see Table~\ref{tabHzRG}). 
Probably the absence of detected optical sources at the location of the radio emission made the quasar the best alternative given the available optical catalogs at that time.
Recently, data at 1.4~GHz with higher spatial resolution undoubtedly show that there is a shift of $\approx33\arcsec$ between \qso\ and 
the strong radio emission (\citealt{Condon1998,Becker1994}).

Notwithstanding the work by \citet{Condon1998} and \citet{Becker1994}, due to the former erroneous association, 
the quasar \qso\ has been included till now in radio-loud samples (e.g., \citealt{Curran2002,Ellison2008}).
In particular, it is worth noting that 
if one follows the last edition (13th) of the catalogue by \citet{VeronCetty2010}, one would end up classifying \qso\ as definitely radio-loud. 
Indeed, the most used definition of radio-loudness requires that the ratio $R=f_{\nu,\, {\rm 5GHz}}/f_{\nu,\, {\rm 4400\angstrom}}>10$ (\citealt{Kellermann1989}), 
where $f_{\nu,\,{\rm 5GHz}}$ is the 5~GHz radio
rest-frame flux density, and $f_{\nu,\, {\rm 4400\angstrom}}$ is the 4400\AA\ optical rest-frame flux density.
At the redshift of \qso, one can use the information at 1.4~GHz (observed) for the radio emission, 
and interpolate the data for the $H$ and $K$ bands to get the optical rest-frame flux at $4400$\AA. 
In this regard, \citet{VeronCetty2010} list $F=538$~mJy at 1.4~GHz (\citealt{CondonBroderick1985, CondonBroderick1986, WhiteBecker1992}), 
while the infrared information come from the Two 
Micron All Sky Survey (2MASS; \citealt{Skrutskie2006}), with a magnitude of $H=15.49$, and $K=15.32$. 
It is thus evident that \qso\ would be classified as radio-loud. 

\begin{figure}
\centering	
\includegraphics[width=0.5\textwidth, clip]{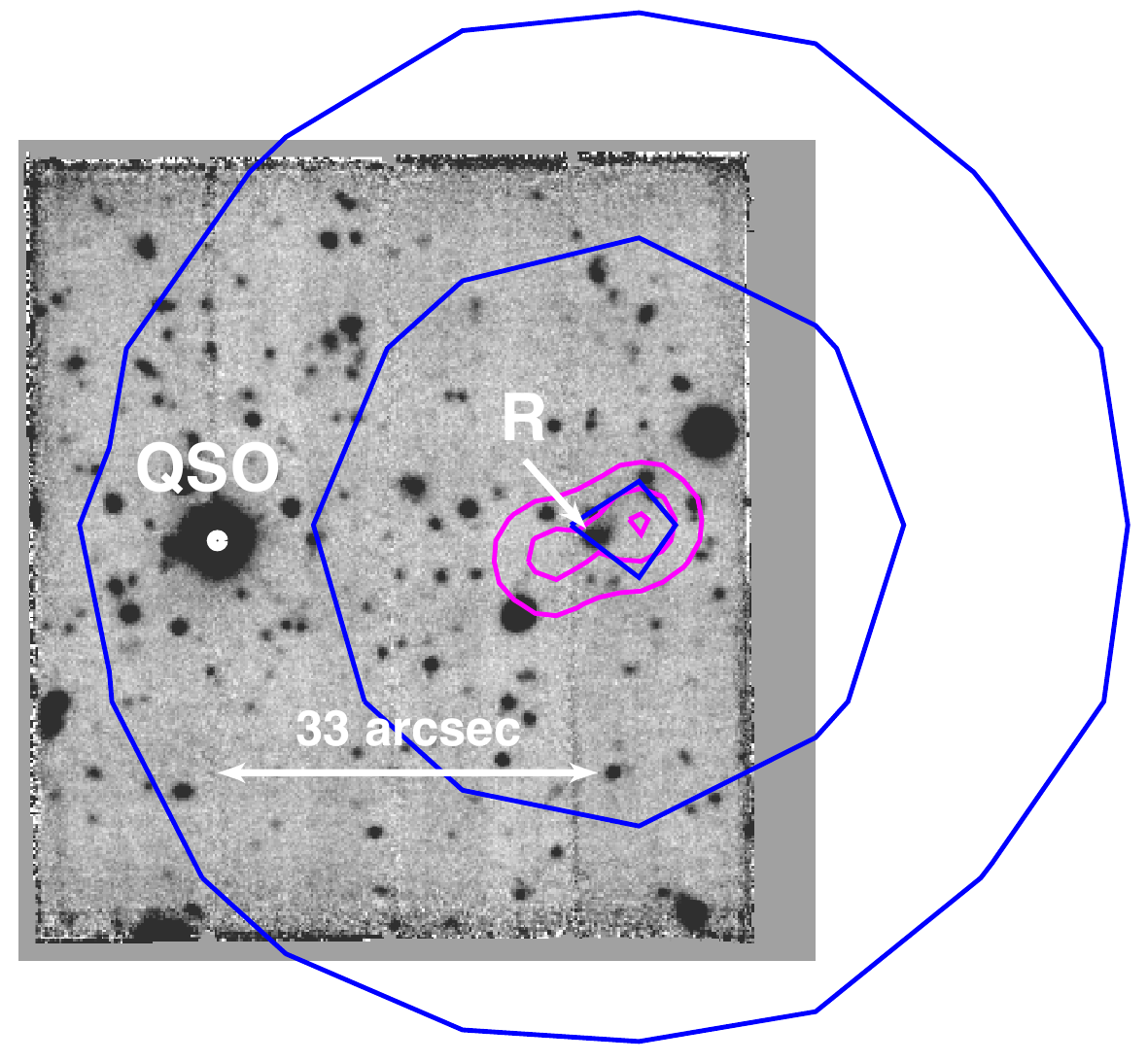}
\caption{{\bf \qso\ is radio-quiet: discovery of an HzRG.}
White-light image of the combined exposures of our final MUSE datacube (as in panel A of Fig.~\ref{Fig1}) with superimposed the radio contours from the 
NVSS (blue; \citealt{Condon1998}) and VLA FIRST Survey (magenta; \citealt{Becker1994}) for the radio source at $\approx30\arcsec$ from \qso. 
Both sets of contours are shown at ${\rm S/N}=2$, $10$, and $20$. Owing to their higher spatial resolution ($5\arcsec$) 
with respect to the NVSS ($45\arcsec$), the data from the FIRST survey show two resolved structures, which we interpret as the radio lobes of a
previously unknown radio galaxy. The MUSE spectrum for the optical counterpart (host galaxy) ``R'' of the radio 
emission is shown in Fig.~\ref{Fig2appC}, confirming this interpretation.}
\label{Fig1appC}    
\end{figure}

In Fig.~\ref{Fig1appC} we show the ``white-light'' image from our final MUSE datacube with 
radio contours superimposed. Specifically, the $S/N=2$, $10$, and $20$ contours from the NVSS 
(blue; \citealt{Condon1998}), and from the VLA Faint Images of the Radio Sky at Twenty-Centimeters Survey (magenta; FIRST, resolution of $5\arcsec$, \citealt{Becker1994}) 
are indicated in blue and magenta, respectively.  
The shift between the quasar and the radio emission is evident, with values of 
$33.1\arcsec$ and $32.35\arcsec$ (average), from the NVSS and FIRST survey positions, respectively. 
While the superior resolution of the FIRST data reveals two bright structures separated by $7.6\arcsec$, 
the total integrated flux is in agreement between the two surveys (i.e. the sum of the 
two sources in FIRST is equal to the flux observed in the NVSS; see Table~\ref{tabHzRG}). 

\begin{table*}
\caption{Summary of the radio data able to resolve the distance between \qso\ and the discovered HzRG.}
\centering
\begin{tabular}{lccccc}
\hline
\hline
Survey	& RA	& Dec	& Separation	& Flux	& Reference \\	
	& (J2000) & (J2000)	& (arcsec)	& (Jy)	& \\
\hline
TXS  		       &  10:20:07.79	&  +10:40:02.8   &  32.6 &  1.642 (at 365~MHz)	& \cite{Douglas1996} \\
NVSS 		       &  10:20:07.76	&  +10:40:03.5   &  33.1 &  0.515 (at 1.4~GHz)	& \cite{Condon1998}  \\
FIRST 		       &  10:20:08.05	&  +10:40:01.5	 &  28.7 &  0.218 (at 1.4~GHz)	& \cite{Becker1994}  \\
 		       &  10:20:07.56	&  +10:40:04.1	 &  36.0 &  0.309 (at 1.4~GHz)	&	              \\
\hline
\end{tabular}
\flushleft{\small For each survey we report the coordinates of the newly discovered HzRG (or lobes when resolved), 
its projected distance from \qso, the integrated flux, and the references.}
\label{tabHzRG}
\end{table*}

\begin{figure}
\centering	
\includegraphics[width=1.0\columnwidth, clip]{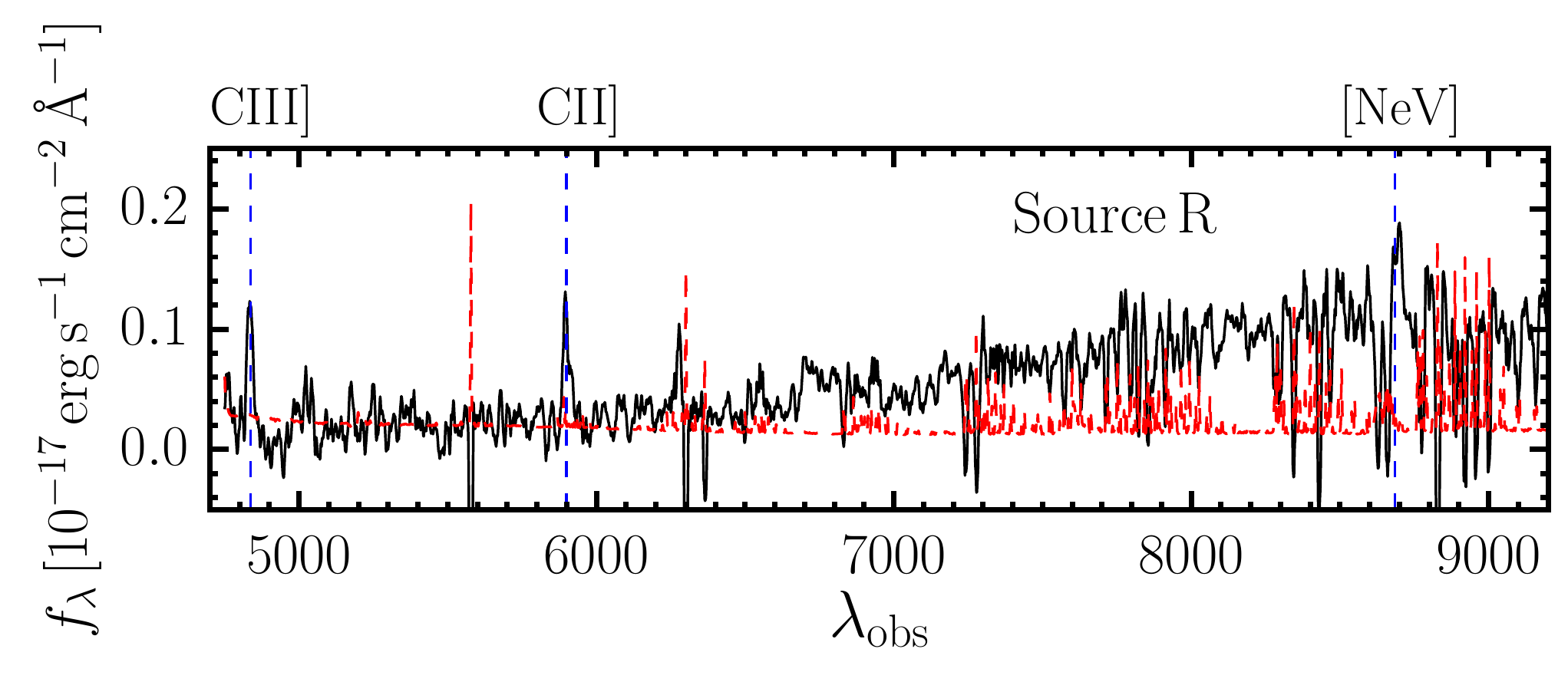}
\caption{{\bf 1D spectrum for the optical counterpart ``R'' of the radio 
emission shown in Fig.~\ref{Fig1appC}.} 
The spectrum has been extracted within the isophotal aperture defined by the continuum 
detection in the MUSE white-light image.  The source ``R'' is at RA 10:20:07.8 and Dec 10:40:02.93.
The red dashed line represents the noise spectra extracted within the same aperture. The location of 
the detected \ciii, \cii, and [\ion{Ne}{V}] emission lines is indicated with blue dashed lines. 
Note the red upturn of the continuum, expected in HzRGs due to a combination of old stellar 
population of the host and reddened AGN (e.g., \citealt{McCarthy1993}).}
\label{Fig2appC}
\end{figure}

Furthermore, we are able to classify 
this radio source. As discussed in section~\ref{sec:obs}, 
between the program 094.A-0585(A) and 096.A-0937(A) we offset the position of \qso\ by about $15\arcsec$ to 
the East to try to obtain a spectrum of any optical counterpart to the radio emission. Inspection of the final MUSE 
datacube at the location of the two bright radio spots detected in FIRST, 
reveals that there are no optical counterparts down to our continuum sensitivity level (in the ``white-light'' image, 
$f_{\lambda,1\sigma}=1.1\times10^{-21}$~erg~s$^{-1}$~cm$^{-2}$~\AA$^{-1}$~pixel$^{-1}$), but there is a faint source  
in between ($i=23.94 \pm 0.02$), source ``R'' (Fig.~\ref{Fig1appC}).
Given this geometric configuration we realised that this source is likely a radio galaxy, with the two radio sources being its bright radio lobes.
This association is confirmed by the optical spectrum of ``R'' extracted from the MUSE datacube (see Fig.~\ref{Fig2appC}), which 
shows the emission lines expected in this wavelength range from a radio galaxy, i.e. \ciii$\lambda$1909, \cii$\lambda$2327, and [\ion{Ne}{v}]$\lambda$3426, 
and the characteristic red upturn of the continuum due to the sum of the contribution from the old stellar population in the host 
galaxy and the reddened AGN at the centre of the system (e.g., \citealt{McCarthy1993}). 
Further confirmation is given by the agreement between 
the ratios of the rest-frame emission line equivalent widths for our newly discovered HzRG, EW$_{\rm rest}^{\rm C\,III]}$/EW$_{\rm rest}^{\rm C\,II]}=1.81 \pm 0.41$ and 
EW$_{\rm rest}^{\rm [Ne\,V]}$/EW$_{\rm rest}^{\rm C\,II]}=1.51 \pm 0.31$ (see Table~\ref{tabTwoHzRG}), and the values reported for a 
composite of HzRGs presented in \citet{McCarthy1993},  
EW$_{\rm rest}^{\rm C\,III]}$/EW$_{\rm rest}^{\rm C\,II]}=1.68$ and EW$_{\rm rest}^{\rm [Ne\,V]}$/EW$_{\rm rest}^{\rm C\,II]}=1.16$.
In addition, the analysis of the emission-lines indicates that this newly discovered radio galaxy is
at a redshift 
of $z=1.536\pm0.001$ (see Table~\ref{tabTwoHzRG}), and thus is not physically related to \qso. 

\begin{table*}
\caption{Information on the emission lines detected in the spectrum of source ``R'' (see Figure~\ref{Fig1appC}).}
\begin{tabular}{lcccccc}
\hline
\hline
Line	    	           & Line Center      & Redshift 	 & Line Flux 			       & Continuum Flux			       & EW$_{\rm rest}$ & Line Width	\\
		           & (\AA)	      & 	 	 & (10$^{-17}$ erg s$^{-1}$ cm$^{-2}$) & (10$^{-19}$ erg s$^{-1}$ cm$^{-2}$ Hz$^{-1}$)   & (\AA)  	 & (km s$^{-1}$)\\
\hline
\ciii$\lambda$1909   &  4840.4$\pm$0.8  &  1.535$\pm$0.002 & 1.91$\pm$0.10                       & 1.63$\pm$0.87	                       & 46.2$\pm$12.6   & 244$\pm$76   \\
\cii$\lambda$2327 	   &  5896.6$\pm$0.7  &  1.534$\pm$0.002 & 1.62$\pm$0.09	               & 2.51$\pm$0.75	    		       & 25.5$\pm$7.8    & 250$\pm$58   \\
{[}\ion{Ne}{v}{]}$\lambda$3426   &  8700.4$\pm$0.6  &  1.539$\pm$0.002 & 3.71$\pm$0.29		       & 9.67$\pm$0.44	                       & 38.4$\pm$3.5    & 272$\pm$38   \\
\hline
\end{tabular}
\flushleft{For each detected emission-line, we report the line center, correspective redshift, line flux, continuum flux, rest-frame equivalent width, and the line width as the 
$\sigma_v$ of a Gaussian fit to the line. In addition, note that \ciii\ is actually a doublet, a combination of [\ion{C}{iii}]$\lambda$1907, 
a forbidden magnetic quadrupole transition, and \ion{C}{iii}]$\lambda$1909, a semi-forbidden 
electrodipole transition. Here, we report the data only for \ion{C}{iii}]$\lambda$1909, which is clearly detected. 
There could be a hint for [\ion{C}{iii}]$\lambda$1907, but higher 
resolution and deeper data are needed to clearly separate and detect the two emission lines.}
\label{tabTwoHzRG}
\end{table*}

\begin{figure*}
\centering	
\includegraphics[width=0.90\textwidth, clip]{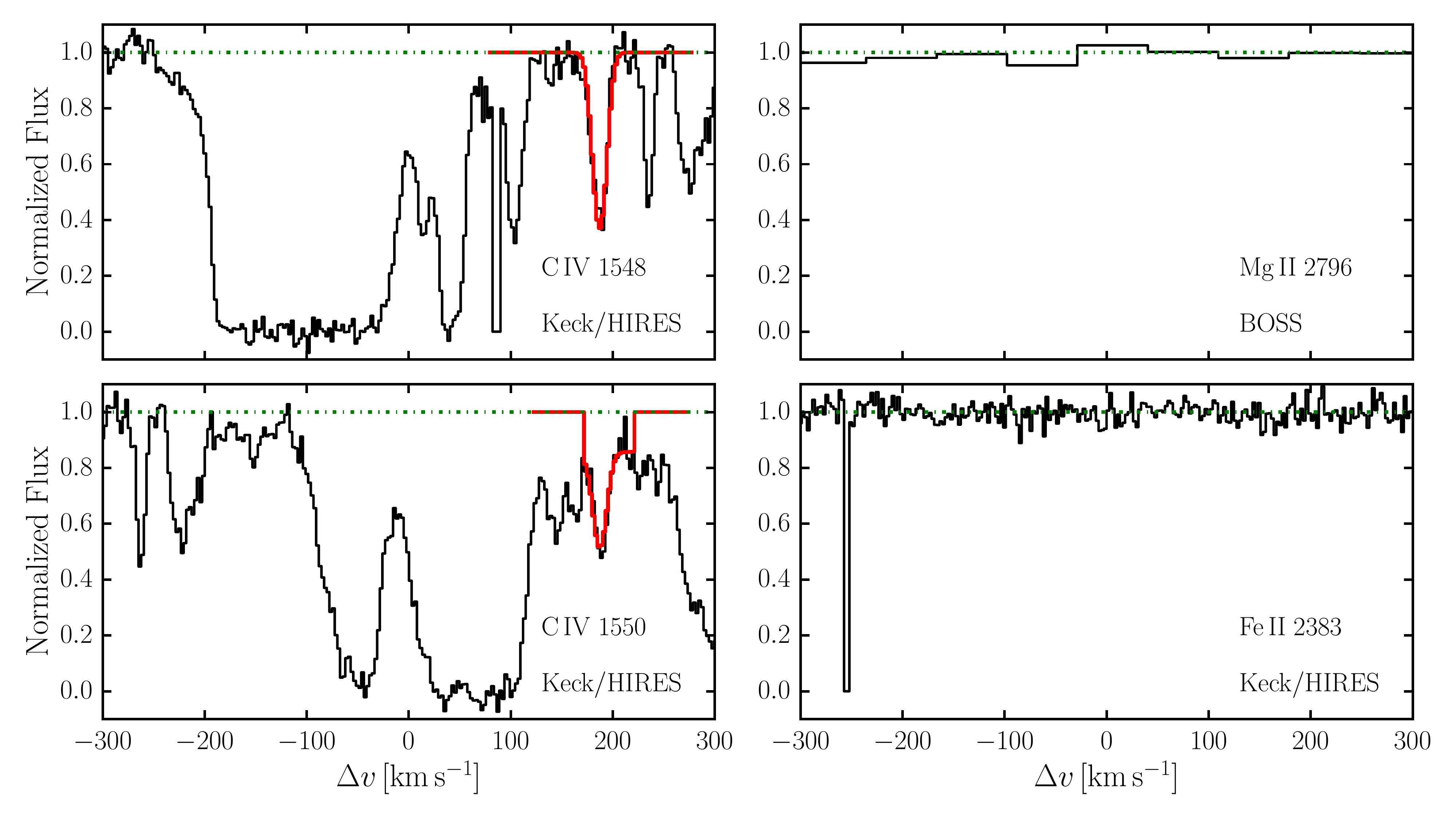}
\caption{{\bf Absorption line spectrum of gas at $\sim275$~kpc from the HzRG.} 
The spectrum of the absorbing gas detected in the sightline of the quasar \qso\ at an impact parameter of $\sim275$~kpc 
from the HzRG is shown.  We found absorption from the high-ionization line \ion{C}{iv}, offset by $188$~km~s$^{-1}$ from the HzRG's 
systemic redshift ($z=1.536$). No low-ionization-state metal absorption is detected; 
we show two representative non-detection, i.e. the first line of the \ion{Mg}{ii} doublet, and \ion{Fe}{ii}.  
The absorbing gas thus seems highly ionized. We fitted the \civ\ absorption (red) and estimated a column density ${\rm log}[N_{\rm C\, IV}/{\rm cm}^{2}]=13.32\pm0.02$. 
This absorber  could be associated with the HzRG itself or with a companion object.}
\label{Fig3appC}
\end{figure*}

Finally, we stress that this system is interesting because one could use the bright background quasar \qso\ to study any absorption feature arising 
from the environment around the newly discovered HzRG, at an impact parameter of $\sim275$~kpc (at $z=1.536$).  
Indeed, the quasar \qso\ has been observed with the High Resolution Echelle Spectrometer (HIRES, \citealt{vogt94}) on the Keck telescope (e.g., \citealt{Omeara15}).  
To search for absorption signatures from low (e.g., \ion{Mg}{ii}, \ion{Fe}{II}) and high (e.g., \ion{C}{iv}) ionization lines, 
we have inspected both the Keck/HIRES and the BOSS/SDSS spectra in a 2000~km~s$^{-1}$ window around the systemic redshift of the newly discovered HzRG.
This search resulted in the identification of \civ\ absorption at $\sim 200$~km~s$^{-1}$ from the HzRG systemic, 
but no other metal absorption lines are found. Using VPFIT\footnote{\url{http://www.ast.cam.ac.uk/~rfc/vpfit.html}}, 
we have constrained the \civ\ doublet to be  at $z=1.53759\pm0.000006$ (or uncertainty of $0.7$~km~s$^{-1}$), 
and characterised by a parameter $b=7.5\pm0.5$~km~s$^{-1}$, and a column density of log~$N_{\rm C\,IV}=13.32\pm0.02$. Fig.~\ref{Fig3appC}
shows the fit for this \civ absorber and also the spectra at the expected correspective locations for low-ionization transitions (\ion{Mg}{ii}, \ion{Fe}{II}).
The absence of such low-ionization-state metal absorptions suggests that this absorber is highly ionized. 
This absorption could be thus associated to the HzRG itself, or to a companion object.
The further characterisation of this HzRG and its environment is left to future studies.

\clearpage
\section{Testing our Analysis Tools: PSF Subtraction and Lyman-Alpha Analysis}
\label{appC}

As described in section~\ref{sec:cubex}, we have presented results based on the {\tt CubExtractor} package (Cantalupo in prep., \citealt{Borisova2016}), 
and showed the overall surface brightness, velocity, and dispersion maps as sum (along wavelength), first moment, and second moment of the 
flux distribution within the extracted three-dimensional mask (Fig.~\ref{figlya}), respectively.
Here we test our results by using a different method for the PSF subtraction and the analysis of the extended nebula. Specifically, 
we first compare the previously obtained maps with the maps obtained by using {\tt QDeblend$^{\rm 3D}$} (\citealt{Husemann2013}) 
for the PSF subtraction. Secondly, we compare the moments analysis to a Gaussian fitting approach. Finding complete agreement 
between different techniques, both tests clearly show that our results are not affected by our method.

\subsection{PSF Subtraction with {\tt QDeblend$^{\rm 3D}$}} 
\label{sec:QDeb}

To study extended emission around AGN without contamination from the central source, \citet{Husemann2013} have 
developed the software tool {\tt QDeblend$^{\rm 3D}$}, which differs from the method described in \citet{Borisova2016} and used in our analysis. 
We have thus used this tool to test our PSF subtraction.
For completeness, we briefly summarise the iterative subtraction algorithm used in {\tt QDeblend$^{\rm 3D}$}, but for further 
details we refer the reader to \citet{Husemann2013, Husemann2014} or to the manual of the 
software\footnote{\url{http://www.bhusemann-astro.org/?q=qdeblend3d}}.

Whereas {\tt CubExtractor} treats the IFU datacube
as a sequence of layers for the PSF subtraction (section~\ref{sec:cubex}), {\tt QDeblend$^{\rm 3D}$} 
considers each spaxel (spatial pixel of the cube) as an independent spectrum. In this framework, the spectrum of a point source 
differs in different spaxels only by a scale factor $s(x,y)$ defined by the PSF during observations. 
The relative strength of broad emission lines and the adjacent continuum can be used to constrain this scale 
factor for a quasar.
More specifically, in the case of \qso, {\tt QDeblend$^{\rm 3D}$} extracts a high S/N quasar spectrum, from which we select the red 
portion of the \lya\ line (to avoid the contamination from the \lya\ forest) and the whole \civ\ line emission as broad emission lines, 
and the interval $5300-5500$~\AA\ as continuum. 
For each spaxel, these wavelength ranges are then used to determine $s(x,y)$ with respect to the high S/N quasar spectrum used 
as quasar template in the first iteration, i.e. $f_{\rm QSO, temp}(\lambda)$.
{\tt QDeblend$^{\rm 3D}$} then subtracts $s(x,y)f_{\rm QSO, temp}(\lambda)$ from the initial datacube. 
To avoid over-subtraction due to eventual host galaxy light on small scales, {\tt QDeblend$^{\rm 3D}$} 
estimates this contribution $f_{\rm host}(\lambda)$ in a user-defined annulus from the residual datacube, subtracts it from $f_{\rm QSO, temp}(\lambda)$, 
and iteratively converges to a stable solution. This iterative procedure usually results in a PSF-subtracted datacube in 3-4 iterations. 
However, here we turn off the estimation of the host contribution to have a fair comparison with {\tt CubExtractor}, 
which assumes the host to be much fainter than the quasar, and thus negligible. Indeed this should be the case for the bright quasar \qso.

\begin{figure*}
\begin{center}
\includegraphics[width=0.9\textwidth]{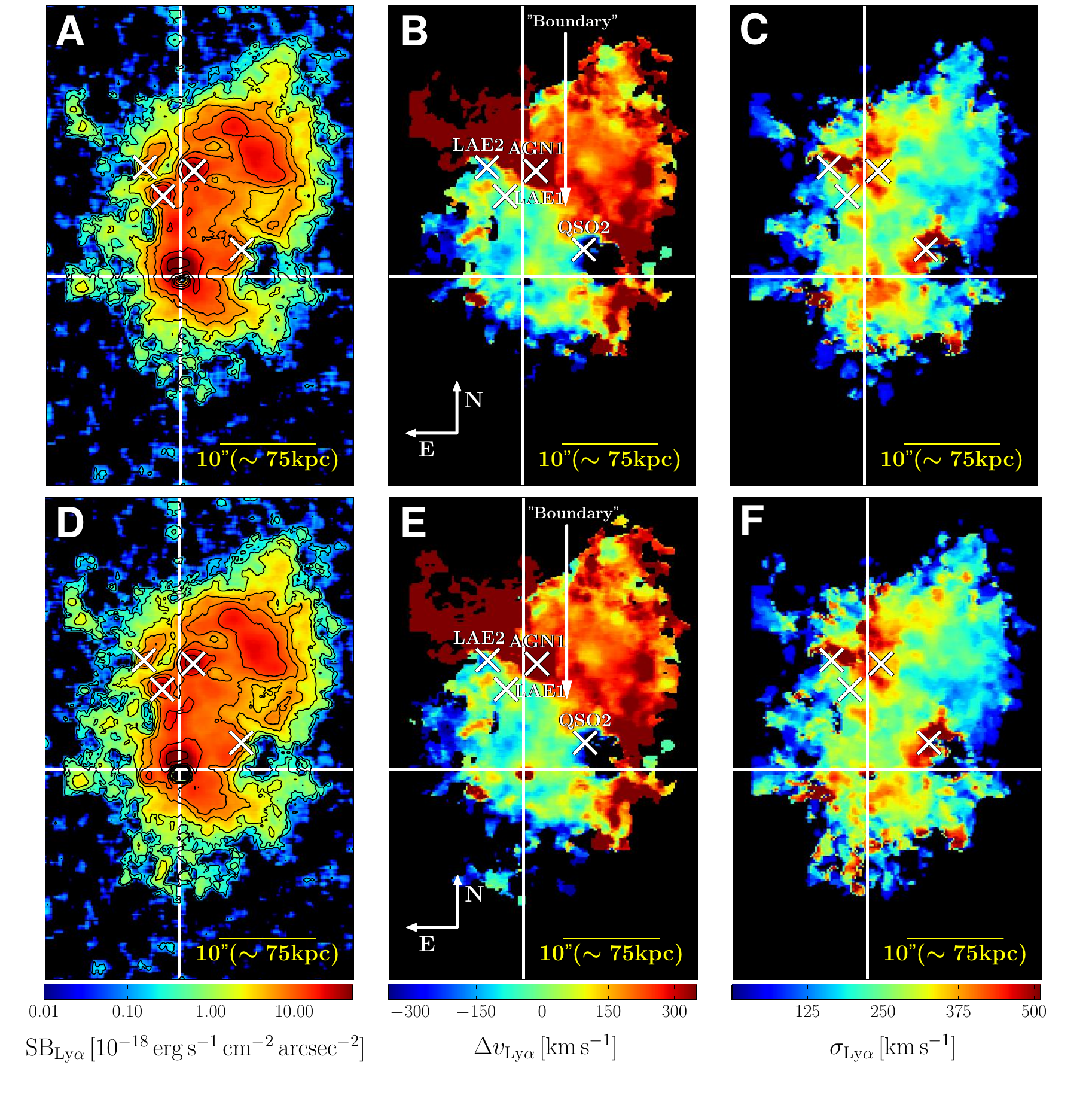}
\end{center}
\caption{{\bf Testing the PSF subtraction: {\tt CubExtractor} versus {\tt QDeblend$^{\rm 3D}$}.} {\bf (A), (B), and (C)} as in Fig.~\ref{figlya}.
{\bf (D), (E), and (F)} as in Fig.~\ref{figlya} but 
using {\tt QDeblend$^{\rm 3D}$} (\citealt{Husemann2013}) for the PSF subtraction, instead of the {\tt CubExtractor} package (\citealt{Borisova2016}). 
It is evident that our results do not depend  on the PSF-subtraction algorithm used.}
\label{QDthreeD}
\end{figure*}

Having obtained the new PSF subtracted datacube using 
{\tt QDeblend$^{\rm 3D}$}, we then extracted a new three dimensional mask for the ELAN, and computed the ``optimally extracted'' NB image, and
the first and the second moments of the flux distribution as previously done with {\tt CubExtractor}.
Fig.~\ref{QDthreeD} shows the comparison between the maps shown in Fig.~\ref{figlya} (i.e. PSF subtraction done with the {\tt CubExtractor} package) and the new ones.
For the purposes of this work (i.e. study of the structure on large scales), the maps are in complete agreement. However, we note that the residuals on 
very small scales ($\sim1\arcsec$) greatly depends on the PSF normalization used and thus differs between the two methods.
Notwithstanding these differences, the peak of the ELAN remains at the same position in close proximity to the location of \qso, 
confirming the robustness of the detection of such emission at close separation.

\subsection{Analysis of the \lya\ Nebula with {\tt KUBEVIZ}: Gaussian Fit}
\label{sec:kubeviz}

\begin{figure*}
\begin{center}
\includegraphics[width=0.9\textwidth]{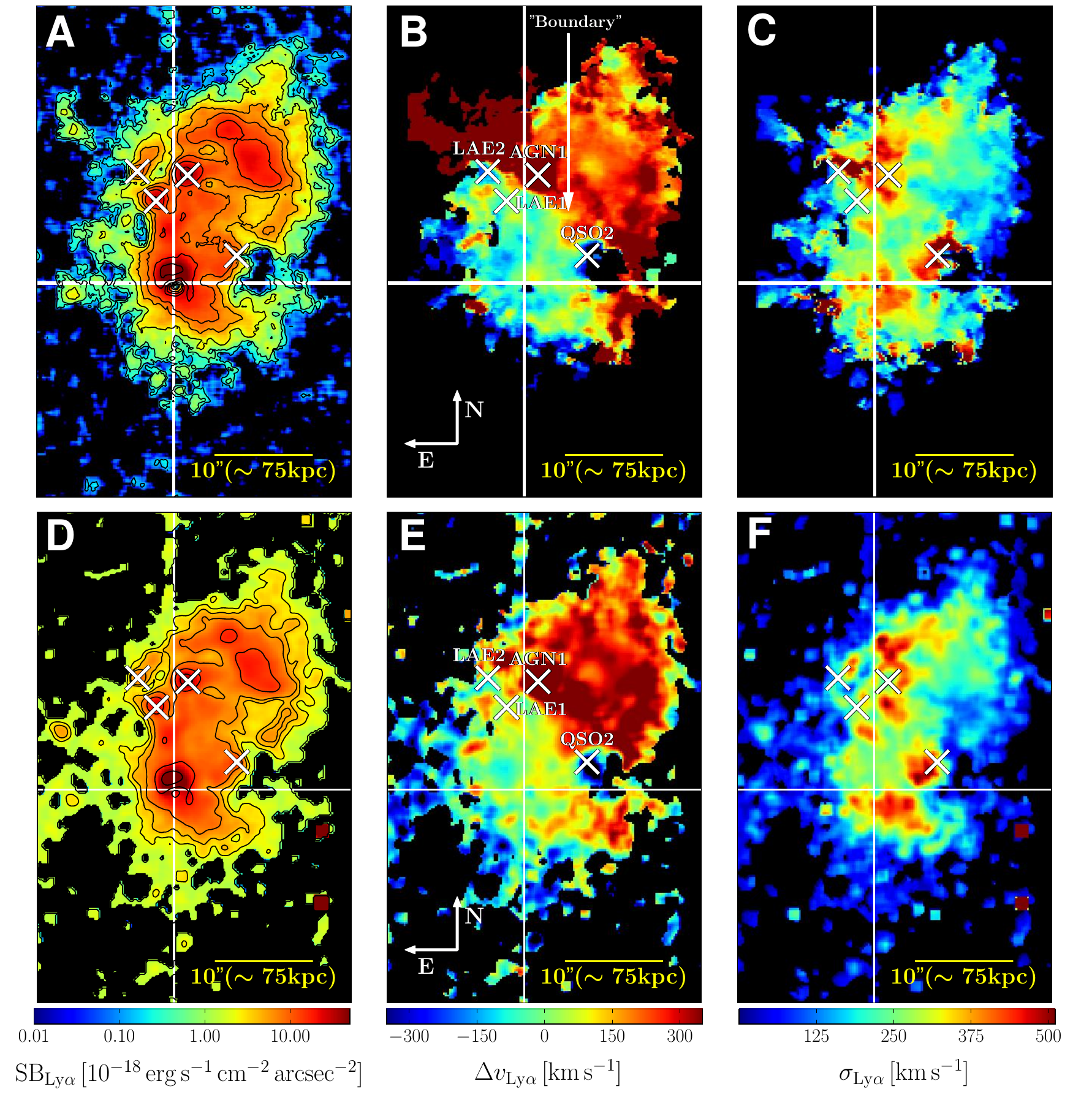}
\end{center}
\caption{{\bf Testing the kinematic analysis of the ELAN: moments of the flux distribution versus Gaussian fit.} 
{\bf (A), (B), and (C)} as in Fig.~\ref{figlya}. {\bf (D), (E), and (F)} \lya\ surface brightness, velocity shift from 
the systemic redshift of \qso, and velocity dispersion within the ELAN obtained from a Gaussian fit with {\tt KUBEVIZ} (\citealt{Fossati2016}). 
In both methods, the maps are smoothed with a $3$ pixels ($0.6\arcsec$) boxcar kernel.
The similarity of the analysis with the two different methods (top and bottom row) is striking, probably revealing that scattering 
of \lya\ photons does not play an important role on the large scales spanned by the ELAN, and down to the current spectral resolution.} 
\label{QdebKub}
\end{figure*}

In this section we show that a Gaussian can model in first approximation the kinematics of the \lya\ emission within the whole nebula, 
resulting in  complete agreement with the previous non-parametric approach, yielding consistent 
surface brightness, velocity shift and dispersion maps.
To perform this additional analysis, we have used the custom IDL software {\tt KUBEVIZ} (\citealt{Fossati2016}) to fit the \lya\ 
emission within the ELAN. This code uses ``linesets'', defined as groups of lines that are fitted simultaneously (e.g., H$\alpha$ and \ion{N}{II}). 
However, here, in the case of the Ly$\alpha$ line, the ``lineset'' is just a 1D Gaussian. Two symmetric windows free 
of contamination around each ``lineset'' are used to evaluate the continuum level\footnote{Note that, in our case, the continuum level is the background level, 
which is consistent with 0 (see e.g. Fig.~\ref{QSONebComparison}).}.  Before the fit, the datacube is smoothed in the spatial direction with a kernel of 
$3\times3$ pixels (or $0.6\arcsec \times 0.6\arcsec$) to increase the S/N per pixel 
without compromising the spatial resolution of the data. No spectral smoothing is 
performed.
During the fit, {\tt KUBEVIZ} takes into account the noise from the variance datacube. However, 
the adopted variance underestimates the real error, most notably because it does not account for correlated 
noise introduced by resampling and smoothing (both applied while using {\tt KUBEVIZ}). We therefore renormalize 
the final errors on the line fluxes assuming a $\chi^{2} = 1$ per degree of freedom. In the end, we mask spaxels 
where the S/N in the \lya\ line is $<2$ (see also \citealt{Fossati2016} for more details).
The results are shown in Fig.~\ref{QdebKub}, and are in striking agreement with the ``moments'' analysis (section~\ref{sec:cubex}). 
Specifically, all the maps are consistent, showing (i) the same high surface brightness 
level (SB$_{\rm Ly\alpha}\sim 10^{-17} \cgssb$), (ii) the same rotation-like pattern, and (iii) the same low velocity 
dispersion on large scales ($\sigma_v\approx200$~km~s$^{-1}$), while higher near compact objects.

\begin{figure}
\begin{center}
\includegraphics[width=0.9\columnwidth]{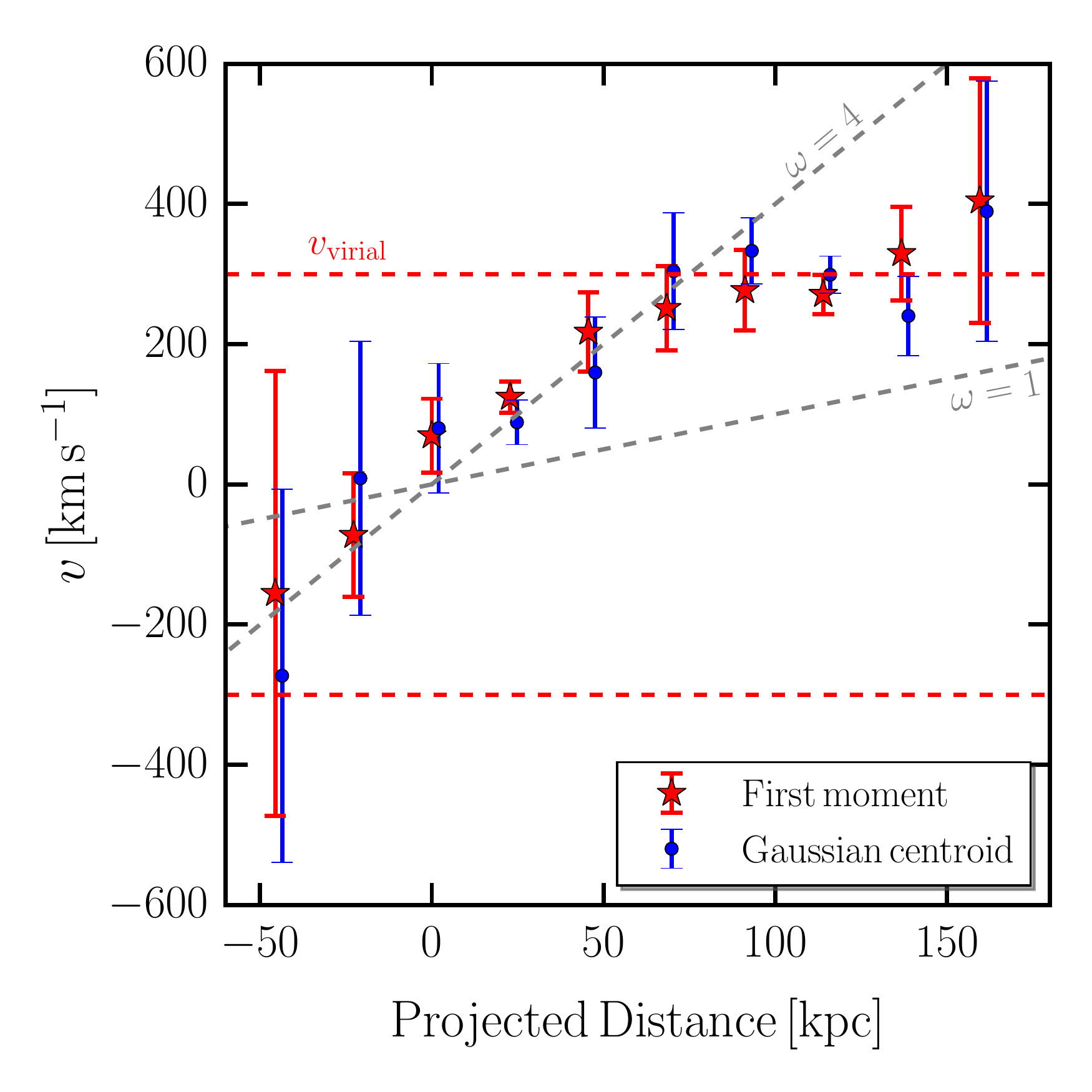}
\end{center}
\caption{{\bf \lya\ velocity centroids along ``Pseudo-slit 1'' through the ELAN: comparison between the first moment and a Gaussian fit.} 
Velocity shift computed along 
pseudo-slit 1 using the first moment of the flux distribution (red, as in Fig.~\ref{figchords}) or a Gaussian fit (blue). 
The data-points are estimated with respect to the position and systemic redshift of the quasar \qso. 
In both cases, the errorbars show the 1$\sigma$ uncertainties. 
The red dashed line indicates the 
virial velocity $v_{\rm vir}=293$~km~s$^{-1}$ for a halo of $M_{\rm 
DM}=10^{12.5}$~M$_{\odot}$.
While the grey dashed lines represent the velocity curve for 
solid-body rotation with $\omega=1$, or $4$ in units of 
$3.1\times10^{13}$~Hz.
Once again we find good agreement between the moment analysis and the Gaussian fit.}
\label{figchordsSup}
\end{figure}

To further stress the good agreement between the moments analysis and the Gaussian fit,  we have computed the velocity 
shift within ``Pseudo-slit 1'' using a Gaussian fit, and compare it to the velocity curves shown in Fig.~\ref{figchords}.
Fig.~\ref{figchordsSup} shows this test, reflecting once again the consistency between the two approaches  for the system 
here studied. The similarity between the two different methods probably reveals that scattering of \lya\ photons 
does not play an important role on the large scales of interest here ($\sim100$~kpc), down to the current resolution.

\section{Our Cosmological Zoom-in Simulation}
\label{sec:simulation}

A powerful tool to interpret our observations comes from cosmological simulations of structure formation. 
The huge dynamical range 
involved in these types of simulations is the main limiting factor on the level of detail in modelling the gas physics.

Indeed, a complete simulation that can cover the full \lya\ radiative transfer dynamics 
requires parsec resolution with a proper treatment of the mediums clumpiness (\citealt{Verhamme2012}). 
These high resolutions are only beginning to be achieved in cosmological simulations, and even so only in 
the inner regions of dark matter haloes where the galaxies reside and not in the CGM. On top of that coupling 
radiative transfer with gas hydrodynamics at these scales is still prohibitive in terms of CPU time.

To avoid the introduction of complex uncertainties and biases in our 
analysis, we thus decided 
to directly compare the observed velocity shear traced by the \lya\ 
emission with the velocity patterns of the cool gas in a cosmological zoom-in simulation.
Indeed, notwithstanding the challenges in play on small scales (e.g., \citealt{McCourt2016}), 
current cosmological simulations are able to follow in detail the 
overall distribution of the cool gas and its kinematics 
on the cosmological scales of ELANe (\citealt{Goerdt2015,Nelson2016,Dekel2009,stewart+16}). 

\subsection{The Simulated Halo}

We have selected a system with the dark-matter halo mass at $z\approx3$ similar 
to the expected value for DM haloes hosting quasars inferred 
from clustering studies at $z=2-3$, i.e. $M_{\rm DM}=10^{12.5}\, M_{\odot}$ (\citealt{white12, fanidakis13}) 
($M_{\rm DM}=10^{12.3\pm0.5}$ for bright quasars; \citealt{Trainor2012}).  
Such a halo has been modelled within the test-runs of 
the high-resolution, ``zoom-in'' cosmological simulation suite NIHAO 
(\citealt{Nihao2015}), and was not included in the original sample, which is limited to less massive systems 
(i.e. $M_{\rm star} < 2\times10^{11}$~M$_\odot$ at $z=0$). It is important to keep in mind that this selected halo has not been run with the aim
of modelling our observations\footnote{We performed the same analysis on the most massive halo in \citet{Nihao2015}, which hosts a Milky Way-like galaxy at z=0, 
and found similar velocity patterns for the infalling gas at z=3, albeit with a lower maximum velocity.}. Specifically, the massive halo 
has been simulated with the new version of the $N$-body SPH code, {\tt GASOLINE} (\citealt{Wadsley2004}), which incorporates fundamental 
improvements for the hydro-solver as explained in \citet{Keller2014,Keller2015}. The resulting halo has
$M_{200}^{\rm DM}=10^{12.29}\, M_{\odot}$, $R_{200}=101$~kpc, and $M_{\rm star}=10^{10.77}\, M_{\odot}$ at $z=3$ 
(the snapshot closest to our observations). The system has
been selected 
from a cosmological box of $90$~Mpc~$h^{-1}$, which has been run 
with force softenings 
$\epsilon_{\rm DM}=0.7$~kpc~$h^{-1}$, $\epsilon_{\rm gas}=0.3$~kpc~$h^{-1}$, 
and particle masses $M_{\rm DM}=1.4 \times 10^7\, M_{\odot}$, ${M_{\rm 
gas}=5\times10^6\, M_{\odot}}$ for the highest refinement level for 
dark matter and gas, respectively.   
The simulation is based on the latest Planck cosmology (\citealt{Planck2014}).

The code allows for metal diffusion as reported in \citet{Wadsley2008}.  The heating function takes into account photoionization and 
photoheating by a redshift-dependent UV background (\citealt{hm12}), while the cooling function includes the effects of metal lines and 
Compton scattering (\citealt{Shen2010}). Metal line cooling allows the gas to cool more efficiently and to 
temperatures below $10^4$~K. Specifically, the gas could reach temperatures as low as $\sim100$~K in the 
densest regions with $n_{\rm H}\sim10^{3.5}$~cm$^{-3}$ (\citealt{Shen2010}). 
A Kennicutt-Schmidt relation regulates star formation, with gas cooler than $15000$~K and denser than $10.3$~cm$^{-3}$ able to form stars. 
Stellar feedback includes two mechanisms: photoionizing radiation from pre-supernova massive stars (\citealt{Stinson2013}), and blast-waves from supernovae (\citealt{Stinson2006}). 
The version of the {\tt GASOLINE} code used here does not include AGN feedback, which is often invoked to prevent the formation of galaxies with too high a 
stellar mass content in massive haloes (e.g., \citealt{Schaye2015}). 
However, at the redshift considered this feedback process should have marginal effects on the properties of the host galaxy of a quasar 
as the peak of the halo mass-stellar mass relation is moved at higher halo masses at $z=3$ with respect to $z=2$ (\citealt{Moster2013}). 
Indeed, the simulated halo agrees well with halo abundance matching measurements at $z=3$ (\citealt{Moster2013,Behroozi2013}), but starts to 
considerably overpredict the stellar mass for redshifts $z<1.5$.

\subsubsection{A View of the Simulated Halo with Velocity Shear similar to Observations}
\label{sec:HaloView}

To analyse the simulated halo and to select the orientation used in section~\ref{sec:favInterp}, we proceed as follows.
We focus on the snapshot at $z=3$, selecting a box of $4\times$ the virial radius of the halo, $L_{\rm box}=404$~kpc, centred on the massive halo. 
This box size has been chosen to encompass the whole size of the observed ELAN (if the centre of the halo is assumed to be \qso). 
Next, we select the cool gas ($T<10^{4.5}$~K)\footnote{Only $8.7\%$ of the gas with $T<10^{4.5}$~K within the selected box has a temperature below $5000$~K, and thus 
should be less efficient in emitting \lya. Given this small fraction and the fact that this low temperature component is in compact object, 
a selection with a lower cut would not change our conclusions.} within the selected region as this phase should be responsible for the \lya\ emission (e.g., \citealt{GW96}).
Indeed, as stated in the main text and in section~\ref{sec:PowMech}, the most favoured powering 
mechanism for the observed \lya\ emission is photoionization from embedded sources in an  optically thin 
regime. In this framework, the \lya\ emission depends on the amount of cool gas and on its density, SB$_{\rm Ly\alpha}\propto n_{\rm H}N_{\rm H}$ (e.g., \citealt{qpq4}).  
As current simulations seem to be unable to reproduce the high densities  ($n_{\rm H}\gtrsim 1$~cm$^{-3}$)
required by the ELANe on CGM scales (\citealt{fab+15b,hennawi+15}),  the most reliable tracer of the \lya\ emission in simulations is the amount of cool gas, 
which simulations tend to correctly reproduce (\citealt{cantalupo14}).
For this reason, we compute mass-weighted line-of-sight velocities and velocity dispersions for the cool gas to 
recover patterns visible in the observed \lya\ maps (Fig.~\ref{figlya}).

To find a view similar to our observations, we align our line-of-sight to the direction perpendicular to the angular momentum 
of the cool gas within the selected box. This simple requirement allow us to find the rotation-like pattern similar to the observations 
presented in Fig.~\ref{figmodel}. However, it is important to stress that many other directions would show a similar rotation pattern ($\sim20$\%; see section~\ref{sec:favInterp}). 
To match the seeing of our dataset, we apply a Gaussian smoothing to the mass-weighted line-of-sight velocities and velocity dispersion maps 
before comparison (Fig.~\ref{figmodel}). The pseudo-slit orientation has been chosen to best cover  
the positive and negative shift of the line-of-sight velocity within the simulated halo, $140$~deg, 
which by chance happens to be close to the angle used for the slit in the observations, $149$~deg (E from N). 

\begin{figure*}
\begin{center}	
\includegraphics[width=0.95\textwidth, clip]{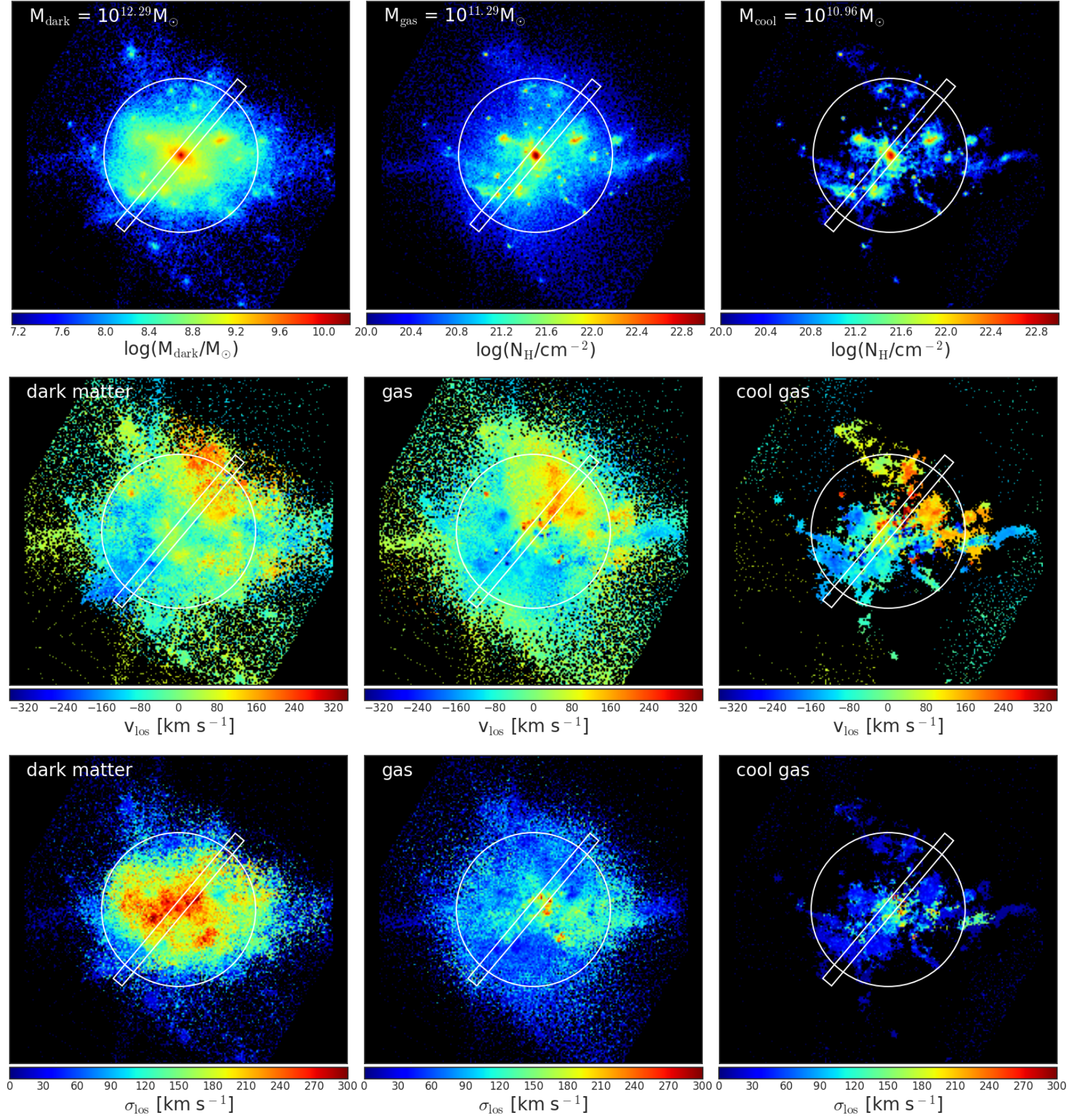}
\end{center}
\caption{{\bf The simulated halo at $z=3$.} 
We show the mass distribution (first row), mass-weighted line-of-sight velocities (second row), and velocity dispersion (third row) 
for dark matter (first column), gas (second column), and cool gas ($T < 10^{4.5}$~K, third column) in a box of $\approx400$~kpc 
centred on the simulated dark-matter halo and oriented as explained in the main text and in section~\ref{sec:HaloView}. 
The white circle gives the halo virial radius, $R_{200} = 101$~kpc, while the white box represents the chord used for the comparison with the observations in Fig.~\ref{figmodel}.
Note that the black corners are due to our analysis technique, i.e. we first select the volume and then analyse it.} 
\label{AtlasSim}
\end{figure*}

In Fig.~\ref{AtlasSim} we present the  mass distribution (first row), mass-weighted line-of-sight velocity (second row), and mass-weighted velocity dispersion (third row) for the
view of the simulated halo perpendicular to the angular momentum of the cool gas and discussed in section~\ref{sec:favInterp}. 
In order to better highlight the presence of substructures, we show the raw (un-smoothed) images.  
Specifically, the first, second, and third column of Fig.~\ref{AtlasSim} show the dark matter, gas, and cool gas, respectively. 
For each component (dark matter, gas, cool gas), the mass used for the weighting is the mass of the component itself at each particular location.
The line-of-sight velocity maps once again reveal the rotation-like pattern discussed in the main text, while the velocity dispersion 
maps show very low values ($\sigma_v\sim90$~km~s$^{-1}$). The predicted velocity dispersions are thus smaller than the observed \lya\ kinematics 
($\sigma_v < 270$~km~s$^{-1}$). However we expect our simulation to show lower velocity dispersions than observations because of the combined effect 
of (i) the spectral resolution of MUSE at these wavelengths ($\sigma= 72$~km~s$^{-1}$), (ii) uncertainties on the mass for the quasar halo 
(i.e. the quasar \qso\ could reside in a more massive halo than here considered), and (iii) scattering of \lya\ photons on small scales ($\lesssim 10$~kpc). 
Note that the (upper left and bottom right) black corners in all the simulated images are due to our strategy for the analysis, i.e. we first select the box and then rotate.

\begin{figure*}
\begin{center}	
\includegraphics[width=0.70\textwidth, clip]{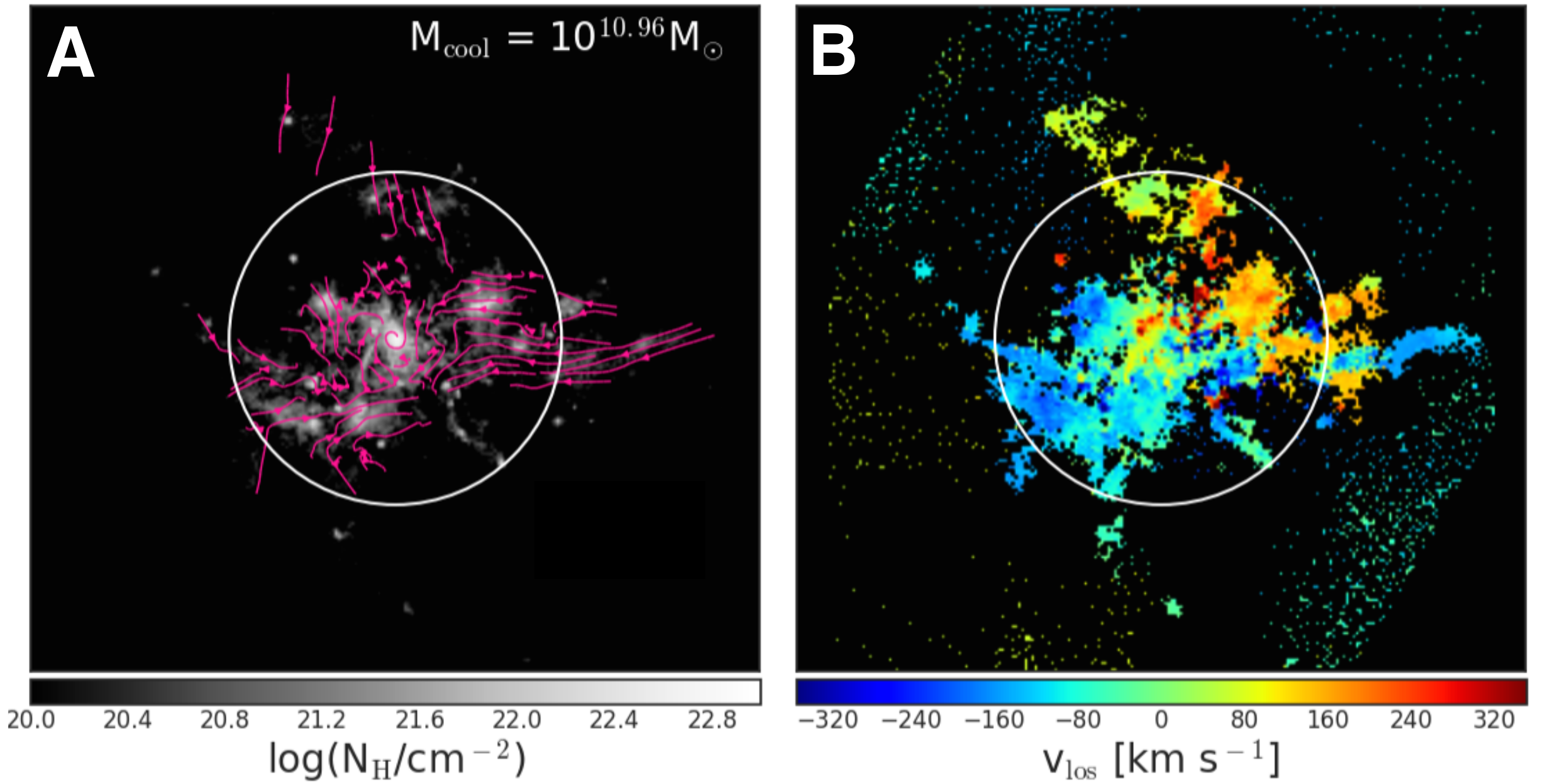}
\end{center}
\caption{{\bf The rotation-like signal is due to inspiraling accreting cool gas.} 
{\bf (A)} cool gas ($T<10^{4.5}$~K) mass distribution (as total hydrogen column density) 
viewed through the same direction as in Fig.~\ref{figmodel} and Fig.~\ref{AtlasSim}.  
The magenta lines indicate  the velocity vector field in the plane of the image. 
It is clear the presence of an overall accreting pattern. {\bf (B)} cool gas mass-weighted line-of-sight velocity. 
The rotation-like pattern is driven by inspiraling motions, whose accretion component is best visible in the in-plane velocity vector field in the left panel. 
The white circles indicate the halo virial radius, $R_{200} = 101$~kpc.}
\label{AccretingMotions}
\end{figure*}

Finally, by analysing the  velocity vector field within the simulated halo, we confirm that the rotation-like pattern 
observed in our simulation is indeed due to inspiraling 
cool gaseous structures as expected by current theories (\citealt{FE1980,stewart+16}). 
In Fig.~\ref{AccretingMotions} we present this analysis. In particular, we overlay on the cool gas mass distribution 
the velocity vector field perpendicular to the line-of-sight (panel A). The substructures are clearly accreting 
towards the centre of the halo. This motion together with the line-of-sight motion (panel B), describes 
the inspiraling kinematics which we have argued to lead to the velocity shear visible in our observations.

\section{Photoionization: the Unlikely Case of a Large-scale Structure along the Line-of-sight}
\label{sec:unlikeCase}

As a larger distance between the quasar \qso\ and the bright NW emission (i.e. $\gg 120$~kpc) could in principle explain the absence of 
\heii emission (because it would simply reduce the incidence of ionizing photons), and thus alleviate the tension in a photoionization 
scenario powered by the bright quasar (see section~\ref{sec:photo}),  we report here a calculation which
assumes the ELAN is within the Hubble flow and along our line-of-sight. 
However, as anticipated in section~\ref{sec:scenShear}, the photoionization modelling of such a projected structure requires a tuning 
of the physical properties of the emitting gas across at least $600$~kpc to reproduce the observed roughly constant SB$_{\rm Ly\alpha}$ 
for the whole extent of the ELAN (Fig.~\ref{figlya}). We explain this modelling in what follows.

\begin{figure*}
\begin{center}
	\includegraphics[width=0.8\textwidth, clip]{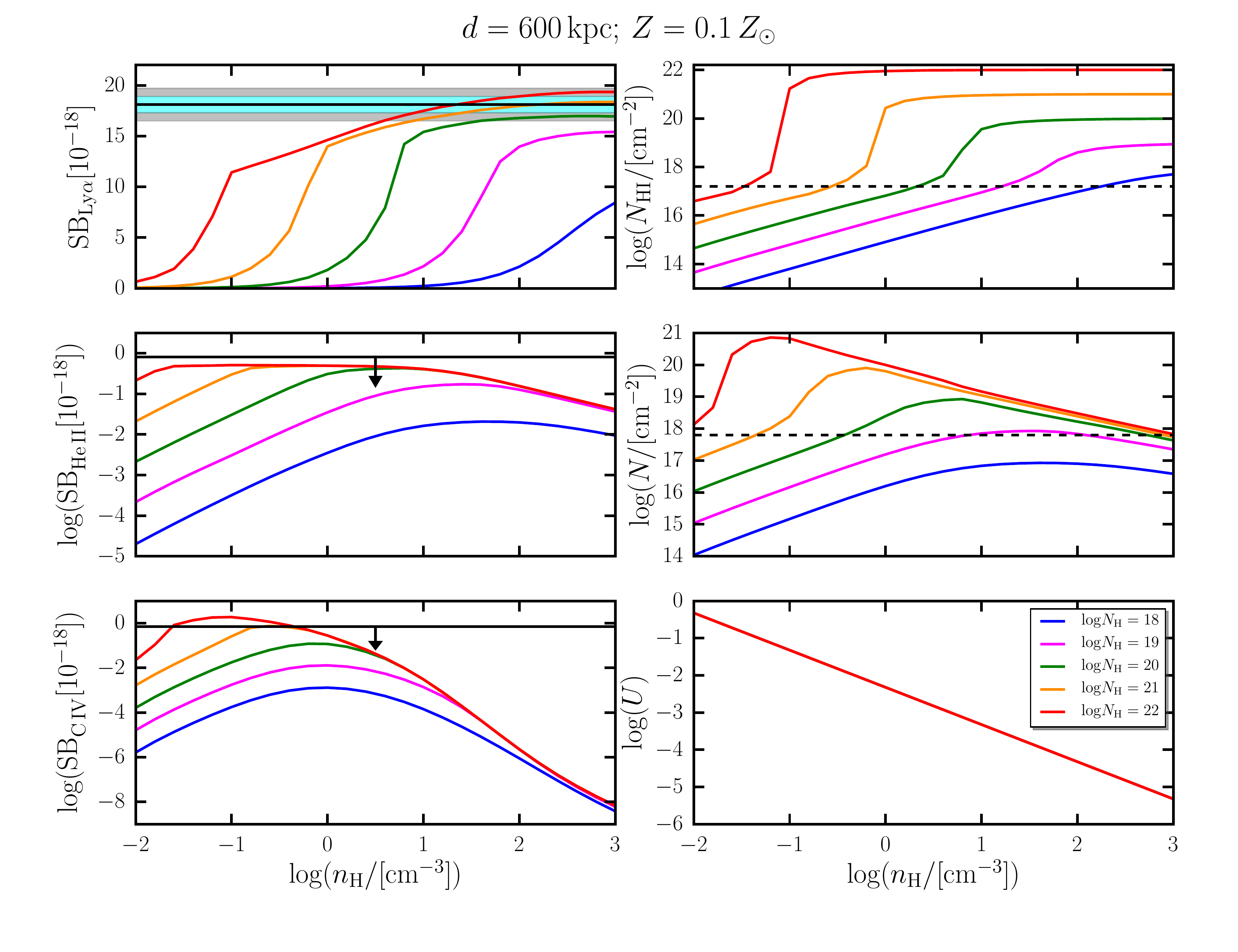}
\end{center}
\caption{{\bf Photoionization modelling of the maximum within the NW redshifted emission assumed to be at $600$~kpc, $Z=0.1$~Z$_{\odot}$.} 
The calculation has been performed using the photoionization code Cloudy (\citealt{ferland13}). We show, as a function of $n_{\rm H}$, 
the predicted SB$_{\rm Ly\alpha}$ in units of $10^{-18} \cgssb$, the predicted column density of neutral hydrogen $N_{\rm HI}$, 
the predicted SB$_{\rm He\, II}$ in units of $10^{-18} \cgssb$, the predicted column density of singly ionized helium $N_{\rm He\, II}$, 
the predicted SB$_{\rm C\, IV}$ in units of $10^{-18} \cgssb$, and the ionization parameter $U$, respectively in {\bf (A)}, {\bf (B)},  
{\bf (C)}, {\bf (D)}, {\bf (E)}, and {\bf (F)}. 
The solid horizontal lines show our measurement for  SB$_{\rm Ly\alpha}$ (together with shaded regions for the 1 and 2$\sigma$ uncertainties) 
and the upper limits for SB$_{\rm He\, II}$ and SB$_{\rm C\, IV}$.
The horizontal dashed lines indicate the theoretical threshold dividing the optically thin regime from the optically thick case for that element. 
For neutral hydrogen is at $N_{\rm HI}=10^{17.2}$~cm$^{-2}$, while for helium is at  $N_{\rm He\, II}=10^{17.8}$~cm$^{-2}$. 
If the quasar \qso\ is illuminating the emitting gas within a large-scale structure along the line-of-sight, optically thick 
models with $Z=0.1$~Z$_{\odot}$, $n_{\rm H}\gtrsim 10$~cm$^{-3}$ and  $20\leq{\rm log}[N_{\rm H}/{\rm cm}^{-2}]\leq22$ would match our 
observational constraints for the maximum within the NW redshifted emission.}
\label{CloudySecond}
\end{figure*}

\begin{figure*}
\begin{center}
	\includegraphics[width=0.8\textwidth, clip]{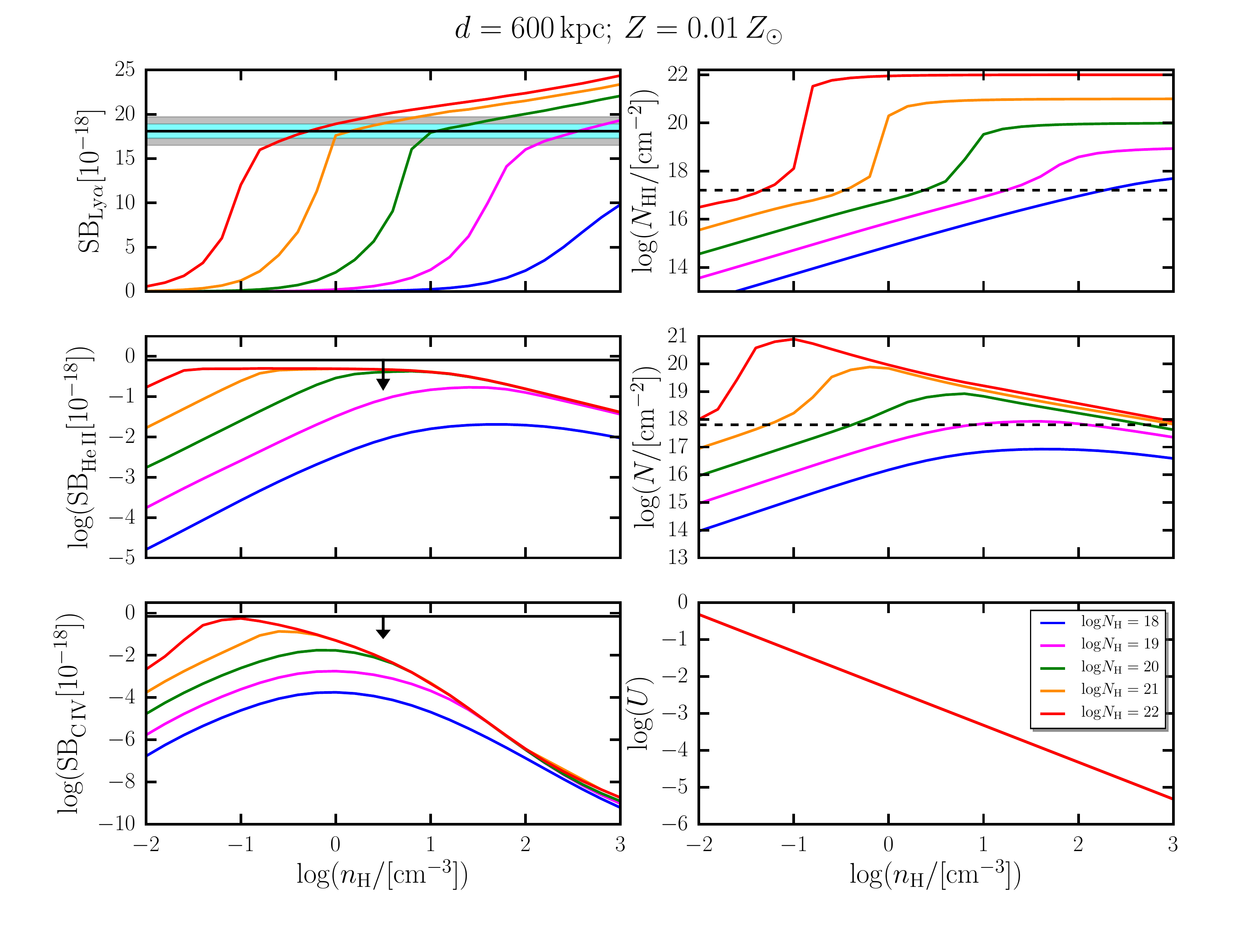}
\end{center}
\caption{{\bf Photoionization modelling of the maximum within the NW redshifted emission assumed to be at $600$~kpc, $Z=0.01$~Z$_{\odot}$.} 
The calculation has been performed using the photoionization code Cloudy (\citealt{ferland13}). We show, as a function of $n_{\rm H}$, 
the predicted SB$_{\rm Ly\alpha}$ in units of $10^{-18} \cgssb$, the predicted column density of neutral hydrogen $N_{\rm HI}$, 
the predicted SB$_{\rm He\, II}$ in units of $10^{-18} \cgssb$, the predicted column density of singly ionized helium $N_{\rm He\, II}$, 
the predicted SB$_{\rm C\, IV}$ in units of $10^{-18} \cgssb$, and the ionization parameter $U$, respectively in {\bf (A)}, {\bf (B)},  
{\bf (C)}, {\bf (D)}, {\bf (E)}, and {\bf (F)}. 
The solid horizontal lines show our measurement for  SB$_{\rm Ly\alpha}$ (together with shaded regions for the 1 and 2$\sigma$ uncertainties) 
and the upper limits for SB$_{\rm He\, II}$ and SB$_{\rm C\, IV}$.
The horizontal dashed lines indicate the theoretical threshold dividing the optically thin regime 
from the optically thick case for that element. For neutral hydrogen is at $N_{\rm HI}=10^{17.2}$~cm$^{-2}$, 
while for helium is at  $N_{\rm He\, II}=10^{17.8}$~cm$^{-2}$. If the quasar \qso\ is illuminating the emitting 
gas within a large-scale structure along the line-of-sight, optically thick models with $Z=0.01$~Z$_{\odot}$, 
$n_{\rm H}\gtrsim 0.3$~cm$^{-3}$ and  $19\leq{\rm log}[N_{\rm H}/{\rm cm}^{-2}]\leq22$ would match our observational constraints for the maximum within the NW redshifted emission.}
\label{CloudyThird}
\end{figure*}

Once again, we perform our calculations with the photoionization code Cloudy (\citealt{ferland13}) following the same assumptions 
as in section~\ref{sec:photo}. 
However, we now assume a distance of $600$~kpc (or $300$~km~s$^{-1}$ in the Hubble flow) for the NW emission.
In Fig.~\ref{CloudySecond} and Fig.~\ref{CloudyThird} we show the prediction for this set of new models for $Z=0.1$, 
and $0.01$~Z$_{\odot}$\footnote{We do not show the models for solar metallicity as they under-predict the observed SB$_{\rm Ly\alpha}$ of the NW clump. 
The model with SB$_{\rm Ly\alpha}$ closer to the observations under-predict them by a factor of 1.2, and has $n_{\rm H}=10^3$~cm$^{-3}$ and log$N_{\rm H}=22$.} 
and compare them to the same constraints used in section~\ref{sec:photo}. These models can reproduce the observed SB$_{\rm Ly\alpha}$ together with the 
upper limits in the \heii and \civ lines (see panel A and compare with panels B and C in both Fig.~\ref{CloudySecond} and Fig.~\ref{CloudyThird}). 
Specifically, the emitting gas in the NW part of the ELAN  is now predicted to be optically thick 
(log$[N_{\rm HI}/{\rm cm}^{-2}]\gg 17.2$, see panels B) and to have  $n_{\rm H}\gtrsim10$~cm$^{-3}$ and $20\leq{\rm log}[N_{\rm H}/{\rm cm}^{-2}]\leq 22$, or  $n_{\rm H}\gtrsim 0.3$~cm$^{-3}$ 
and $19\leq{\rm log}[N_{\rm H}/{\rm cm}^{-2}]\leq 22$, 
respectively for $Z=0.1$, and $0.01$~Z$_{\odot}$.

Even though these models are closer to the expectation for the  $n_{\rm H}$ and  $N_{\rm H}$ in the CGM and IGM, they suffer of at least two considerable issues.
First, the models that reproduce the observed \lya\ emission are completely neutral  (see panels B in Fig.~\ref{CloudySecond} and Fig.~\ref{CloudyThird}). With such a high column 
of neutral hydrogen (log$[N_{\rm HI}/{\rm cm}^{-2}]\gtrsim 19$, see panels B), \lya\ photons are expected to be heavily affected by resonant scattering effects. In particular, 
the bright NW emission should have a clear double peaked \lya\ line profile  (see e.g. Fig.~3 and 7 in \citealt{Cantalupo2005}).
On the contrary, as discussed in section~\ref{sec:scenShear} and appendix~\ref{sec:kubeviz}, the observed \lya\ emission at the location of the NW bright clump, 
and more in general in the whole extent of the ELAN, can be approximated by a single Gaussian at our spectral resolution (FWHM~$\approx 170$~km~s$^{-1}$ at $5000$~\AA).  

\begin{figure*}
\begin{center}
	\includegraphics[width=0.8\textwidth, clip]{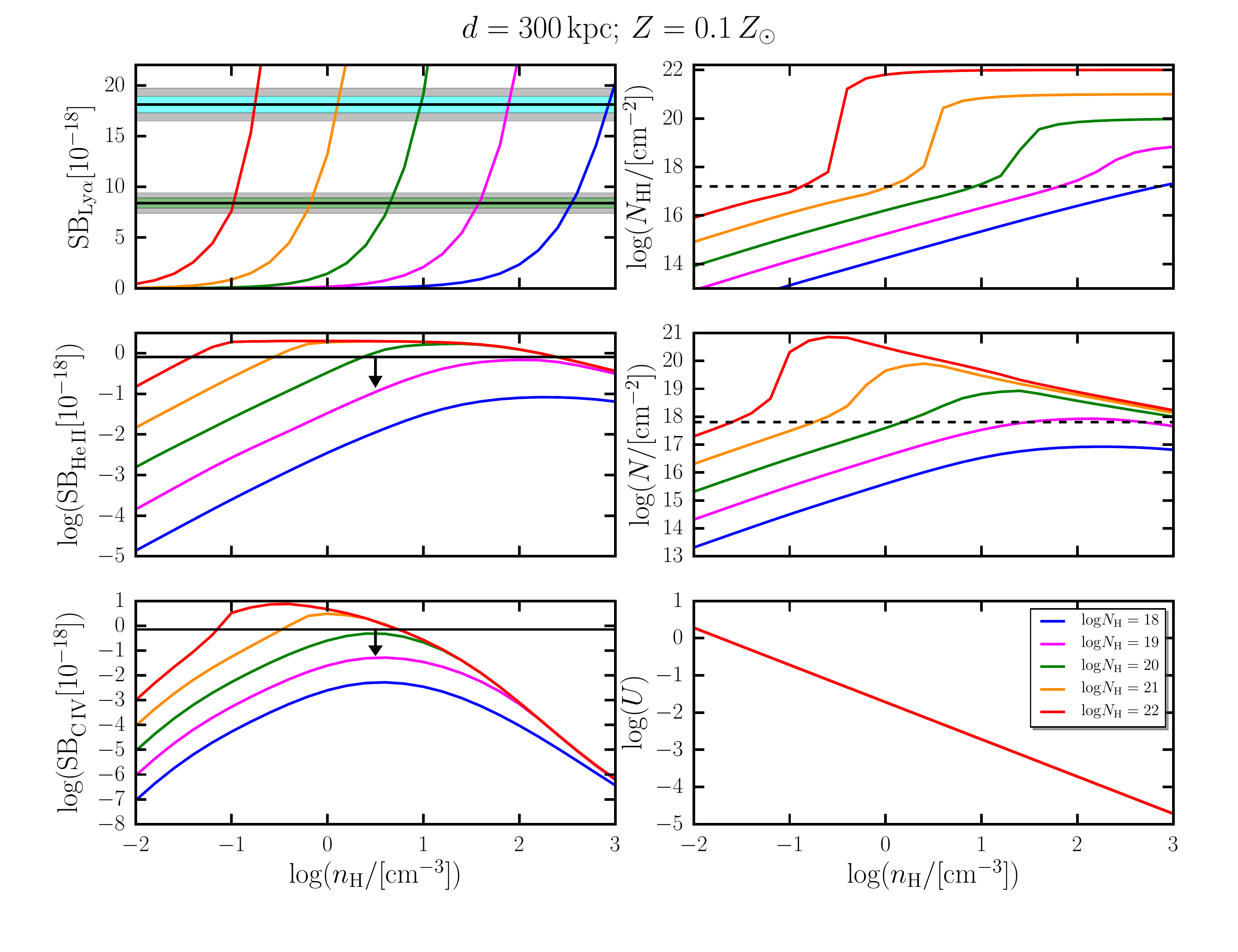}
\end{center}
\caption{{\bf Photoionization modelling of the maximum within the NW redshifted emission assumed to be at $300$~kpc, $Z=0.1$~Z$_{\odot}$.} 
The calculation has been performed using the photoionization code Cloudy (\citealt{ferland13}). We show, as a function of $n_{\rm H}$, the 
predicted SB$_{\rm Ly\alpha}$ in units of $10^{-18} \cgssb$, the predicted column density of neutral hydrogen $N_{\rm HI}$, the predicted 
SB$_{\rm He\, II}$ in units of $10^{-18} \cgssb$, the predicted column density of singly ionized helium $N_{\rm He\, II}$, 
the predicted SB$_{\rm C\, IV}$ in units of $10^{-18} \cgssb$, and the ionization parameter $U$, respectively in {\bf (A)}, {\bf (B)},  
{\bf (C)}, {\bf (D)}, {\bf (E)}, and {\bf (F)}. 
The solid horizontal lines show our measurement for  SB$_{\rm Ly\alpha}$, together with shaded regions for the 1 and 2$\sigma$ uncertainties (in cyan for 
the NW clump at $\approx120$~projected kpc; in green for the dashed region in panel E of Fig.~\ref{civheii}). We also show the upper limits for SB$_{\rm He\, II}$ and SB$_{\rm C\, IV}$.
The horizontal dashed lines indicate the theoretical threshold dividing the optically thin regime from the optically thick case for that element. 
For neutral hydrogen is at $N_{\rm HI}=10^{17.2}$~cm$^{-2}$, while for helium is at  $N_{\rm He\, II}=10^{17.8}$~cm$^{-2}$. If the quasar \qso\ 
is illuminating the emitting gas within a large-scale structure along the line-of-sight, optically thick models with $Z=0.1$~Z$_{\odot}$, $n_{\rm H}\gtrsim 100$~cm$^{-3}$ and 
${\rm log}[N_{\rm H}/{\rm cm}^{-2}]<19$ would match our observational constraints at $300$~kpc (or $\approx 144$~km~s$^{-1}$) within the Hubble flow.}
\label{CloudyFourth}
\end{figure*}

Secondly, if the velocity shifts (e.g. Fig.~\ref{figchords}) trace distances, i.e. a structure along the line-of-sight, there should be optically thin locations 
within the ELAN at smaller distances from the quasar \qso\ than the NW clump ($600$~kpc). In particular, given the known luminosity of the quasar \qso, all the illuminated 
gas at $\lesssim350$~kpc (or $\approx175$~km~s$^{-1}$ within the Hubble flow) will be optically thin.
Basically, most of the ELAN between the quasar and the NW clump has to be optically thin while showing roughly the same level of SB$_{\rm Ly\alpha}$ and similar 
non-detection in \heii and \civ as the optically thick gas at larger distances. 
In Fig.~\ref{CloudyFourth} we show the calculation performed with the Cloudy photoionization code (\citealt{ferland13}) for a distance of $300$~kpc 
(or $\approx144$~km~s$^{-1}$, or $\Delta z\approx0.002$). The location of this region at $\approx144$~km~s$^{-1}$ from the quasar systemic is highlighted in 
Fig.~\ref{civheii} by the red dashed contour.
This region has a slightly lower level of average emission (green shaded region in panel A of Fig.~\ref{CloudyFourth}) than the NW clump (cyan shaded 
region in panel A of Fig.~\ref{CloudyFourth}), i.e. SB$_{\rm Ly\alpha}=(0.84\pm0.05) \times 10^{-17} \cgssb$, but the same upper 
limits on \heii and \civ (as calculated in section~\ref{sec:heiiciv}).
It is once again clear that only models with very high $n_{\rm H}\gtrsim 100$~cm$^{-3}$ and low $N_{\rm H}<10^{19}$~cm$^{-2}$ can 
reproduce simultaneously the observed SB$_{\rm Ly\alpha}$, and the upper limits in \heii and \civ (see panel A and compare with panels B and C in Fig.~\ref{CloudyFourth}).
As already discussed in section~\ref{sec:photo}, such values are in contrast with current expectations for CGM and IGM gas, making it
implausible that the quasar \qso\ illuminates the ELAN. 

To solve these discrepancies, a scenario in which the gas is along the line-of-sight would require  the quasar to be heavily 
obscured in the direction of the emitting gas, so that the ELAN does not receive its hard ionizing photons.
Given that brighter AGN are expected to be able to disrupt the obscuring medium on small scales, and to have wider opening angles (\citealt{Lawrence1991,Kreimeyer2013}), 
the bright quasar \qso\ should radiate along the line-of-sight as it does towards us. 
This calculation therefore adds to the arguments against the ELAN being along the line-of-sight (presented in section~\ref{sec:scenShear}). 
However, only planned follow-up observations at high spectral resolution, and additional observations to verify the powering mechanisms within this ELAN could verify our conclusion.


\bsp	
\label{lastpage}
\end{document}